\documentclass[twocolumn,journal,pagenumbers, romanappendices]{IEEEtran}
\usepackage{listings}
\usepackage{amsmath}
\usepackage{tikz}
\usepackage{array}
\usepackage{mdwmath}
\usepackage{multirow}
\usepackage{mdwtab}
\usepackage{eqparbox}
\usepackage{amsfonts}
\usepackage{tikz}
\usepackage{array}
\usepackage{bbm}
\usepackage{mdwmath}
\usepackage{mdwtab}
\usepackage{eqparbox}
\usepackage{latexsym}
\usepackage{cite}
\usepackage{amssymb}
\usepackage{bm}
\usepackage{amssymb}
\usepackage{graphicx,color}
\usepackage{mathrsfs}
\usepackage{epsfig}
\usepackage{psfrag}
\usepackage{setspace}
\usepackage{algorithm}
\usepackage{algpseudocode}
\usepackage{stfloats}
\usepackage{mdwlist}

\newtheorem{theorem}{Theorem}
\newtheorem{definition}{Definition}
\newtheorem{proposition}{Proposition}
\newtheorem{assumption}{Assumption}
\newtheorem{corollary}{Corollary}
\newtheorem{remark}{Remark}

\usepackage{graphics}
\usepackage{epstopdf}
\usepackage[tight,footnotesize]{subfigure}
\usepackage{multirow,bigstrut,threeparttable}
\newcommand{\blue}{\textcolor[rgb]{0,0,0}}
\def \ra {\text{\rm a}}
\def \rb {\text{\rm b}}
\def \rc {\text{\rm c}}
\def \rV {\text{\rm V}}

\begin{document}
\title{Performance Limits and Geometric Properties\\of Array Localization}
\author{Yanjun~Han,~\IEEEmembership{Student Member,~IEEE},~Yuan~Shen,~\IEEEmembership{Member,~IEEE},~Xiao-Ping~Zhang,~\IEEEmembership{Senior Member,~IEEE},\\Moe Z. Win,~\IEEEmembership{Fellow,~IEEE},~and Huadong~Meng,~\IEEEmembership{Senior Member,~IEEE}%
\thanks{Manuscript received Month 00, 0000; revised Month 00, 0000; accepted Month 00, 0000. Date of current version Month 00, 0000. This research was supported, in part, by the National Natural Science Foundation of China (Grant No.\ 61501279), the Natural Sciences and Engineering Research Council of Canada (NSERC, No.\ RGPIN239031), the Office of Naval Research (Grant N00014-16-1-2141), and MIT Institute for Soldier Nanotechnologies. The material in this paper was presented in part at the 2014 IEEE International Conference on Acoustics, Speech, and Signal Processing, Florence, Italy.
}
\thanks{Y.~Han is with the Department of Electrical Engineering, Stanford University, Stanford, CA 94305, USA, and was with the Department of Electronic Engineering, Tsinghua University, Beijing 100084, China (email: yjhan@stanford.edu).} 
\thanks{Y.~Shen is with the Department of Electronic Engineering and Tsinghua National Laboratory for Information Science and Technology (TNList), Tsinghua University, Beijing 100084, China, and was with the Laboratory for Information and Decision Systems (LIDS), Massachusetts Institute of Technology, Cambridge, MA 02139, USA  (email: shenyuan\_ee@tsinghua.edu.cn).}
\thanks{X.-P.~Zhang is with the Department of Electrical and Computer Engineering, Ryerson University, Toronto, ON M5B 2K3, Canada (email: xzhang@ee.ryerson.ca).}
\thanks{M.~Z.~Win is with the Laboratory for Information and Decision Systems (LIDS), Massachusetts Institute of Technology, Cambridge, MA 02139, USA (e-mail: moewin@mit.edu).}
\thanks{H.~Meng is with California PATH, University of California, Berkeley, CA 94804, USA, and was with the Department of Electronic Engineering, Tsinghua University, Beijing 100084, China (email: hdmeng@berkeley.edu).}
\thanks{Communicated by O.~Simeone, Associate Editor for BBB.}
\thanks{Color versions of one or more of the figures in this paper are available online at http://ieeexplore.ieee.org.}
\thanks{Digital Object Identifier 10.1109/TIT.2016.xxxxxx}
}
\maketitle
\begin{abstract}
Location-aware networks are of great importance and interest in both {civil and military} applications. This paper determines the localization accuracy of an agent, which is equipped with an antenna array and localizes itself using wireless measurements with anchor nodes, in a far-field environment. In view of the Cram\'{e}r-Rao bound, we first derive the localization information for static scenarios and demonstrate that such information is a weighed sum of Fisher information matrices from each anchor-antenna measurement pair. Each matrix can be further decomposed into two parts: a distance part with intensity proportional to the squared baseband effective bandwidth of the transmitted signal and a direction part with intensity associated with the normalized anchor-antenna visual angle. Moreover, in dynamic scenarios, we show that the Doppler shift contributes additional direction information, with intensity determined by the agent velocity and the root mean squared time duration of the transmitted signal. In addition, two measures are proposed to evaluate the localization performance of wireless networks with different anchor-agent and array-antenna geometries, and both {formulae} and simulations are provided for typical anchor deployments and antenna arrays.
\end{abstract}


\newcounter{mytempeqncnt}

\section{Introduction}\label{sec_intro}
\IEEEPARstart{L}{ocalization} is of great importance with a wide variety of civil and military applications such as navigation, mobile network services, autonomous vehicles, social networking, and seeking and targeting people \cite{WinConMazSheGifDarChi:J11,GezTiaGiaKobMolPooSah:05, SheWin:J10a, SayTarKha:05, PahLiMak:02, CafStu:98, WymLieWin:J09, PatAshKypHerMosCor:05,  DarConFerGioWin:J09, KhaKarMou:09}. The global positioning system (GPS) is the most prominent technology to provide location-aware services, but its effectiveness is severely degraded in harsh environments, e.g., in buildings, urban canyons and undergrounds \cite{WinConMazSheGifDarChi:J11,GezTiaGiaKobMolPooSah:05, SheWin:J10a}. Localization using a wireless network is a feasible alternative to overcome the GPS limitation since location information can be obtained with the aid of a network that consists of anchor nodes with known position and agent nodes aiming to estimate self positions. 

Typically, a localization task is achieved by the radio communications between anchors and agents, which are equipping with a single antenna or an antenna array. By processing the received signals, relevant signal metrics can be extracted for localization, for example, time-of-arrival (TOA)\cite{TOA1,TOA2,JouDarWin:J08,ConGueDarDecWin:J12}, time-difference-of-arrival (TDOA)\cite{TDOA1,TDOA2,TDOA3}, angle-of-arrival (AOA)\cite{AOA2,study1, AOA1,AOA3}, and received signal strength (RSS)\cite{RSS1,RSS2,RSS3}. {Among these signal metrics, TOA and AOA are the two widely used in practice.} TOA is a time-based metric obtained via measuring the signal propagation time between the anchor and agent; then the distance measurements translate to location information by trilateration \cite{BasicIntro}. AOA is a metric characterizing the arriving direction of the signal at the agent, and it can be obtained using an array of antennas and spatial filtering; then the angle measurements translate to location information by triangulation\cite{AOA2}. Techniques using a combination of these signal metrics, such as many hybrid TOA/AOA systems, have also been studied in literature \cite{TDOA3,TOA-AOA}.

In practical scenarios, the transceived signals encounter non-ideal phenomena such as noise, fading, shadowing, multipath (signal reaches the receiver via multiple paths due to reflection) and non-line-of-sight (NLOS) propagation (the first arriving signal does not travel on a straight line)\cite{PatAshKypHerMosCor:05}, and therefore the location estimates are subject to uncertainty. In the interest of system design and operation, it is important to know the best attainable localization accuracy and the corresponding approaches to achieve such accuracy, which can be rephrased as obtaining the lower bound for localization errors and the achievability result in the language of information theory. For example, in designing the energy-efficient location-aware networks, attainable localization accuracy can be a meaningful performance objective to optimize \cite{SheDaiWin:J14, WanLeuHua:09, GarHaiCouLop:14, DaiSheWin:J15, DaiSheWin:J15a, GodPetPoo:11, JeoSimHaiKan:14}. To evaluate the localization performance in the presence of uncertainty, some studies consider specific systems that employ certain signal metrics extracted from the received waveforms, e.g., the time delay or the angle, and then determine the localization error based on the joint distribution of these metrics\cite{AOA2, study1}. However, the extracted metrics may discard useful information for localization, resulting in suboptimal localization performance. To address this issue, recent studies directly utilize the received waveforms \cite{TDOA3,SheWin:J10a} to exploit all relevant information and derive \emph{fundamental} limits for localization accuracy. For this purpose, the most commonly used tool is the Cram\'{e}r-Rao lower bound (CRLB) \cite{AOA2, SheWymWin:J10,study1,TDOA3,D2} due to its intriguing property in asymptotic statistics\cite{van2000asymptotic} in the sense of the H\'{a}jek convolution theorem \cite{hajek1970characterization} and the H\'{a}jek-Le Cam local asymptotic minimax theorem \cite{hajek1972local}, though some other bounds such as the Barankin bound\cite{Barakin} and Ziv-Zakai bound\cite{MonMazVitWin:J13} are also used.

{For general wideband systems, the fundamental limits of localization accuracy for a single agent has been obtained in \cite{SheWin:J10a} in terms of the CRLB, which are generalized to a cooperative framework with multiple agents \cite{SheWymWin:J10}. Moreover, it has been shown in \cite{SheWin:J10a} that AOA measurements obtained by wideband antenna arrays do not further improve position accuracy beyond that provided by TOA measurements, which implies that it suffices to use TOA measurements in wideband systems. 
However, the approaches developed for wideband systems are not applicable to commonly used modulation-based communication systems, because those systems modulate a baseband signal onto a carrier frequency with an \emph{unknown} initial phase\cite{RADAR,envelope,rice2000narrowband,bekkerman2006target}. Consequently, unlike wideband systems, not the entire passband signals can be used for TOA measurements due to the unknown phase in modulation-based communication systems. Nevertheless, in far-field environments, by using an antenna array, the carrier phases can be exploited for AOA measurements to improve the localization as widely recognized \cite{PatAshKypHerMosCor:05}.
These inherent differences from wideband systems call for a comprehensive investigation for the performance limits of localization accuracy in general wireless localization systems with antenna arrays.}


In dynamic scenarios that involve moving objects, the localization or tracking accuracy can be characterized by the Posterior CRLB (PCRLB) \cite{tichavsky1998posterior, meng2009computationally, SheMazWin:J12}. Since the derivation of PCRLB relies on multiple snapshots and measurable uncertainties of both observations and hidden states of time-varying locations, the instantaneous localization error of the mobile user is required as an input of PCRLB. To obtain the instantaneous localization error in the dynamic scenario, the Doppler shift may be favored as another source for localization in addition to the TOA and AOA measurements. Most existing research treats the Doppler shift as the frequency-shift solely on the carrier frequency and obtain the performance bounds accordingly\cite{envelope,rice2000narrowband,bekkerman2006target}, whereas the effects of Doppler shift on the baseband signal are simply neglected. Moreover, the accuracy limit of navigation in general wireless networks has only been obtained in the form of block matrices without considering Doppler shift \cite{SheMazWin:J12}. The effect of Doppler shift to the localization information still remains under-explored.

{Upon obtaining the localization accuracy, a natural question arises concerning the optimal geometry when deploying anchors and designing antenna arrays. A few studies have optimized the anchor-agent geometry by minimizing the condition number of the visibility matrix \cite{mckay1997geometry,hegazy2003sensor}, while some others focused on a CRLB-related cost\cite{caffery2000new,bishop2007optimality,martinez2006optimal,abel1990optimal}, where the anchor-agent geometry is optimized in an isotropic source localization. In \cite{godrich2010target}, the array-antenna geometry was also studied jointly with the anchor-agent geometry, and {it is proven} optimal to place anchors and antennas symmetrically on two circles, respectively. However, existing studies do not provide simple measures to compare two arbitrary geometric structures.}

{In this paper, we develop a general array localization system with Doppler shifts and use the CRLB to determine the performance limits of localization accuracy in a far-field environment. We also propose two measures to characterize the impact of the anchor-agent and array-antenna geometry on localization accuracy. In particular, we highlight the difference of our model and the wideband model in \cite{SheWin:J10a} by introducing the carrier frequency with unknown initial phases in the localization problem.} The main contributions are as follows.
\begin{itemize}
  \item We derive the performance limits of localization accuracy for a \emph{static} agent equipped with an antenna array in terms of the equivalent Fisher information matrix (EFIM). {The EFIM can be decomposed as} a weighed sum of measuring information matrices, where each matrix contains both distance and direction information with intensities determined by the corresponding anchor-antenna measurement pair. {Moreover, we show that the direction part} can provide dominant information for localization for narrowband signals.
  \item We derive the performance limits of localization accuracy for a \emph{moving} agent equipped with an antenna array in terms of the EFIM.  The Doppler shift is shown to contribute to the direction information with intensity associated with the root mean squared time duration of the transmitted signal, and its contribution to the direction information can be more significant than that of the antenna array.
  \item We propose two measures, i.e., the squared array aperture function and anchor geometric factors, to quantify the effects of anchor-agent geometry and array-antenna geometry on localization, respectively, and give the optimal geometric design {of the anchor and antenna locations}.
\end{itemize}

The rest of the paper is organized as follows. In Section \ref{sec_model}, we describe the system model and formulate the location estimation problem. In Section \ref{sec_EFIM}, we derive the squared position error bound (SPEB) using EFIM for a static agent, and Section \ref{sec_doppler} generalizes the results for a moving agent. {Based on the SPEB and the EFIM,} Section \ref{sec_geometry} quantifies the impact of the anchor-agent and array-antenna geometries {with some examples}. Numerical results are given in Section \ref{sec_numerical}, and conclusions are drawn in Section \ref{sec_conclusion}.

\begin{figure*}[!t]
\centering
\includegraphics[width=6in]{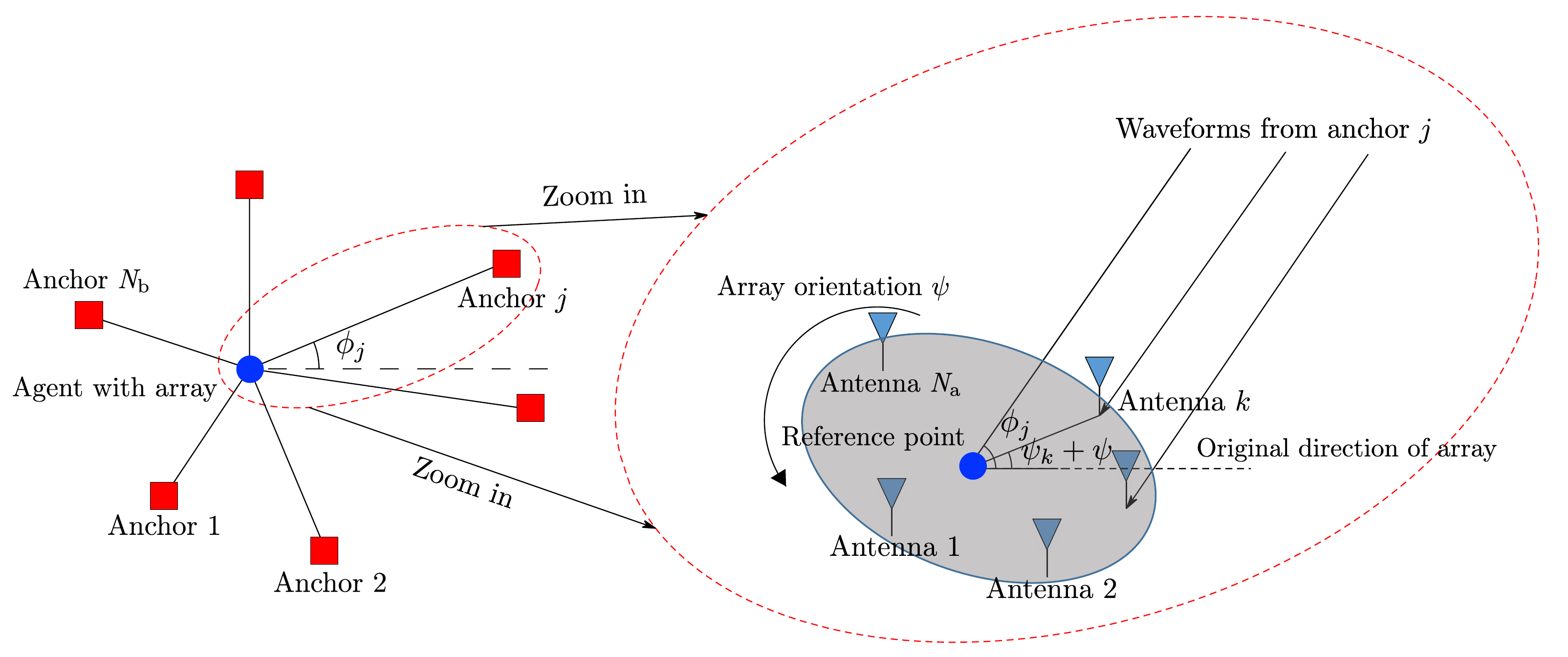}
\caption{\label{fig:scenario} System model: $N_{\text{b}}$ anchors with known position and one agent with an $N_{\text{a}}$-antenna array characterized by a reference point {and orientation $\psi$. The angle of the $k$-th antenna to the reference point is the sum of the original individual angle of this antenna $\psi_k$ and the array orientation $\psi$.}}
\label{Anchor}
\end{figure*}

\emph{Notation:} We use upper and lower case boldface to denote matrices and vectors, respectively; $f^*(x)$ and $f'(x)$ denote the complex conjugate and the first-order derivative of $f(x)$, respectively; $\Re\{z\}$ and $\Im\{z\}$ denote the real and imaginary part of a complex number $z$, respectively; $\mathbb{E}_{\bf x}\{\cdot\}$ denotes the expectation operator with respect to the random vector $\bf x$; ${\bf A}\succeq {\bf B}$ denotes {the L\"{o}wner semiorder of matrices which means} that ${\bf A}-{\bf B}$ is positive semi-definite; $\text{tr}\{{\bf A}\}$, ${\bf A}^{\text{T}}$, ${\bf A}^{\text{H}}$ and ${\bf A}^{-1}$ denote the trace, the transpose, the conjugate transpose and the inverse of matrix $\bf A$; $[\cdot]_{r_1:r_2,c_1:c_2}$ denotes a submatrix composed of the rows $r_1$ to $r_2$ and the columns $c_1$ to $c_2$ of its argument; $\|\cdot\|$ denotes the Euclidean norm of its argument; $x\ll y$ denotes that $y>0$ and $|x/y|$ is a negligible number far less than 1, and ${\bf A}\ll{\bf B}$ denotes that ${\bf x}^{\text{H}}{\bf Ax}\ll {\bf x}^{\text{H}}{\bf Bx}$ for any complex column vector $\bf x$ (implying that $\bf B$ is positive definite); $x\cong y$ denotes that $|x-y|\ll \max\{|x|,|y|\}$; $\mathbb{R}^n$ ($\mathbb{C}^n$) and $\mathbb{S}_{++}^n$ denote the $n$-dimensional real (complex) vector space and the set of all $n\times n$ complex positive definite matrices, respectively. 

The notations of frequently-used symbols are listed as follows.

\begin{basedescript}{\desclabelstyle{\pushlabel}\desclabelwidth{9.5em}}
  \item[$N_{\text{\rm b}}, N_{\text{\rm a}}$]{number of anchors and antennas}
\item[${\bf p}_j=(x_j,\,y_j)$] position of anchor $j$
  \item[${\bf p}=(x,\,y)$]{position of the reference point}
\item[${\bf p}_k^{\text{\rm Array}}\!=(x_k^{\text{\rm Array}},\,y_k^{\text{\rm Array}})$]{position of the $k$-th antenna}
  \item[$d_k, \psi_k$]{distance and direction of the $k$-th antenna to the reference point}

  \item[$\mathcal{N}_{\text{\rm L}}, \mathcal{N}_{\text{\rm NL}}$]{set of anchors with line-of-sight (LOS) and NLOS to agent}
  \item[$\psi, \psi_{\text{\rm d}}$]{array orientation and moving direction of the agent}
  {\item[$D_j, \phi_j, \tau_j$]{distance, direction and signal propagation time from anchor $j$ to the agent}}
  \item[$c, v, v_{\text{\rm r}}\triangleq v/c$]{propagation speed of the signal, speed and relative speed of the agent}
  \item[$\tau, {\bf p}_\tau$]{reference time and corresponding position of the reference point}
  {\item[$\xi_j$]{initial carrier phase of the transmitted signal from anchor $j$}}
    {\item[$L_{j}$]{number of multipath components (MPCs) from anchor $j$ to the agent}}
      \item[$\alpha_{j}^{(l)}, b_j^{(l)}, \gamma_j^{(l)}$]{channel gain, range bias and arrival-angle bias from anchor $j$ to the agent via $l$-th path}
    \item[$\tau_{jk}^{(l)}$]{signal propagation time from anchor $j$ to the $k$-th antenna via $l$-th path}
  \item[$[\,0,T_{\text{\rm ob}})$]{observation time interval}
  \item[$s(t), s_0(t)$]{the entire and the baseband signal}
  \item[$S(f), S_0(f)$]{Fourier Transform of $s(t), s_0(t)$}
  \item[$r_{jk}(t)$]{received waveform at antenna  $k$ from anchor $j$}
  {\item[$\beta, f_\rc,\gamma$]{effective baseband bandwidth, carrier frequency, and baseband-carrier correlation (BCC)}}
  \item[${\bf J}_{\bm{\theta}}, {\bf J}_{\text{\rm e}}(\bm{\theta})$]{Fisher information matrix (FIM) and EFIM w.r.t. parameter $\bm{\theta}$}
  {\item[${\bf J}_{\text{\rm r}}(\phi)$]{ranging direction matrix (RDM) with direction $\phi$}}
  \item[$N_0$]{spectral density of noise}
  \item[$\mathsf{SNR}_j^{(l)}$]{received signal-to-noise ratio (SNR) in $l$-th path from anchor $j$}
  \item[$\theta_{jk}^{\text{\rm V}},\omega_j$]{visual angle and its angular speed from antenna  $k$ to anchor $j$}
  \item[$\lambda_j,\chi_{j}$]{information intensity and path overlap coefficient (POC) from anchor $j$}
  \item[$t_{\text{\rm rms}}$]{root mean squared time duration}
  \item[$G(\theta)$]{squared array aperture function (SAAF)}
\end{basedescript}

\section{System Model}\label{sec_model}
This section presents a detailed description of the system models and {formulates} the location estimation problem. {Two scenarios are considered in this work: the static scenario in which the agent is stationary, and the dynamic scenario in which the agent is moving. 
}
\subsection{Static Scenario}\label{sec_model_system}
Consider a 2-D wireless network with $N_{\text{b}}$ anchors and one \emph{static}  agent equipped with a rigid antenna array consisting of $N_{\text{a}}$ elements (see Fig.~\ref{Anchor}). Anchors have perfect knowledge of their positions, denoted by ${\bf{p}}_j=(x_j,\,y_j)\in\mathbb{R}^2$, where $j\in\mathcal{N}_{\text{b}} = \{1,2,\ldots,N_{\text{b}}\}$ is the set of all anchors. The agent aims to estimate its self-position based on the received waveforms obtained by its $N_{\text{a}}$ array-antennas from all anchors, and ${\bf{p}}_k^{\text{\rm Array}}=(x_k^{\text{\rm Array}},\,y_k^{\text{\rm Array}})\in \mathbb{R}^2$ denotes the position of $k$-th antenna in the array where $k\in\{1,2,\ldots,N_{\text{a}}\}$. 

{The array rigidity implies that it} has exactly three degrees of freedom, i.e., translations and rotation, and hence it can be characterized by a predetermined reference point ${\bf{p}}=(x,y)$ and an orientation $\psi$. Then, by denoting the distance between the reference point and $k$-th antenna by $d_k$, and the direction (relative to oreintation) from the reference point to $k$-th antenna by $\psi_k$, we can express the position of each antenna as
\begin{align}\label{eq:array_structure}
  {\bf{p}}_k^{\text{\rm Array}} = {\bf{p}} + d_k\left[\begin{array}{c}
                                   \cos(\psi+\psi_k) \\
                                   \sin(\psi+\psi_k)
                                 \end{array}\right].
\end{align}

{This work focuses on far-field enviroments, where the distances between anchors and the agent are sufficiently larger than the array dimension so that (i) the angles from each anchor to all array-antennas are identical and (ii) the channel properties from each anchor to all array-antennas are identical, e.g., the same SNR. Moreover, the phase differences between received signals in adjacent antennas are assumed to be less than $2\pi$ so that there is no periodic phase ambiguity (i.e., the array element spacing is smaller than the signal wavelength). We write the propagation time delay and the direction from anchor $j$ to the agent (the reference point) as}
\begin{align}\label{eq:anchor_time_direction}
  \tau_j \triangleq \frac{D_j}{c} \triangleq \frac{\|\mathbf{p}_j - \mathbf{p}\|}{c},\qquad \phi_j \triangleq \arctan\frac{y-y_j}{x-x_j}
\end{align}
respectively, where $c$ is the propagation speed of the signal.

As for the signal model, we consider that anchor $j$ transmits a \emph{known} signal 
\begin{align}\label{eq:trans_sig}
	s(t)=s_0(t)\exp(j2\pi f_\rc t + \xi_j)
\end{align}
to the agent, where the signal is formed by the quadrature modulation that consists of the baseband signal (also called the complex envelope) $s_0(t)$, the carrier wave with central frequency $f_\rc$, and the initial carrier phase $\xi_j$.\footnote{The quadrature demodulation requires that the baseband signal $s_0(t)$ be bandlimited by $f_\rc$, i.e., $S_0(f)=0$ for all $|f|\ge f_\rc$, where $S_0(f)$ is the Fourier Transform of $s_0(t)$.} In practical modulation systems, the initial carrier phase $\xi_j$ is usually unknown, and hence we model $\xi_j$ as an unknown parameter in this work.

{Our channel model considers both multipath and NLOS propagation phenomena. Specifically, $\mathcal{N}_{\text{b}} = \mathcal{N}_{\text{L}} \cup \mathcal{N}_{\text{NL}}$, where $\mathcal{N}_{\text{L}}$ denotes the set of anchors providing LOS signals and $\mathcal{N}_{\text{NL}}$ for those providing NLOS signals. Together with the transmitted signal given in (\ref{eq:trans_sig}), the received waveform at $k$-th antenna from anchor $j$ can be written as \cite{SheWin:J10a,rice2000narrowband}
\begin{align}\label{eq:received_waveform}
  r_{jk}(t) &= \sum_{l=1}^{L_{j}}\alpha_{j}^{(l)}\cdot \sqrt{2}\Re\left\{s_0(t-\tau_{jk}^{(l)}) \right. \\
  & \times \left. \exp\big(j(2\pi f_\rc (t-\tau_{jk}^{(l)}) + \xi_j)\big)\right\}+z_{jk}(t), \;\; t\in[\,0,T_{\text{ob}}) \nonumber
\end{align}
where $\alpha_{j}^{(l)}\in \mathbb{R}$ and $\tau_{jk}^{(l)}$ are the amplitude and delay of the $l$-th path, respectively, and $L_{j}$ is the number of MPCs, $z_{jk}(t)$ represents the real observation noise modeled as additive white Gaussian noise (AWGN) with two-side power spectral density $N_0/2$, and $[\,0,T_{\text{ob}})$ is the observation time interval. {In far-field environments, by geometry (as shown in Fig.~\ref{fig:scenario}) the time delays can be written as}
\begin{align}\label{eq:time_delay_antenna}
    \tau_{jk}^{(l)} = \tau_j + \frac{-d_k\cos((\phi_j-\psi+\gamma_{j}^{(l)})-\psi_k)+b_{j}^{(l)}}{c}
\end{align}
where $\gamma_{j}^{(l)} \in [-\pi,\pi]$ and $b_{j}^{(l)}\geq 0$ are the arrival-angle bias and range bias of the $l$-th path, respectively. In particular, for the first path of a LOS signal, we have
\begin{align}
	b_{j}^{(1)} = 0 \quad \text{and} \quad \gamma_{j}^{(1)} = 0, \qquad j\in\mathcal{N}_{\text{L}}
\end{align}
and otherwise the range biases $b_{j}^{(l)}$ are positive and the arrival-angle biases can be between $-\pi$ to $\pi$.
{Note also that in far-field environments, the multipath parameters $L_j,\alpha_j^{(l)},b_j^{(l)}$ and $\gamma_j^{(l)}$ do not depend on the choice of the antenna element $k$.}

\begin{remark}
Intuitively speaking, when the initial phase $\xi_j$ is unknown, the time delays or TOA information $\tau_j$ is completely corrupted in the carrier phase, and hence \emph{only} the baseband part can be utilized for obtaining the distance information. 
Nevertheless, the phase differences between different antennas can cancel out the unknown parameter $\xi_j$, leading to useful information about $\phi_j$ but not $\tau_j$. Hence, the direction information can be retrieved from the carrier phases of antennas. 
This constitutes the key difference from the wideband model in \cite{SheWin:J10a} which assumes the initial carrier phases be precisely known and consequently
both the distance and direction information can be extracted from the carrier phases. 
\end{remark}



{Throughout this paper, we consider the case where there is no a priori knowledge about the parameters, i.e., all unknown parameters are deterministic and non-Bayesian approaches are used}.

\subsection{Dynamic Scenario}\label{sec_doppler_model}
Built upon the static scenario, we further consider a system model for the dynamic scenario in which the agent is moving at a constant speed $v$ along direction $\psi_{\text{\rm d}}$ throughout the observation time (see Fig.~\ref{Fig:Doppler_Struc}). Denote the position of reference point at reference time $\tau$ by ${\bf p}_{\tau}$, then the position of the $k$-th antenna at time $t$ can be written as
\begin{align}\label{eq:Dynamic_Array_Position}
  {\bf p}_k^{\text{Array}}(t) = {\bf{p}_\tau} + d_k\left[\begin{array}{c}
                                   \cos(\psi+\psi_k) \\
                                   \sin(\psi+\psi_k)
                                 \end{array}\right] + v(t-\tau)\left[\begin{array}{c}
                                   \cos\psi_{\text{\rm d}} \\
                                   \sin\psi_{\text{\rm d}}
                                 \end{array}\right].
\end{align}

\begin{figure}[!t]
  \centering
  \includegraphics[scale=0.5]{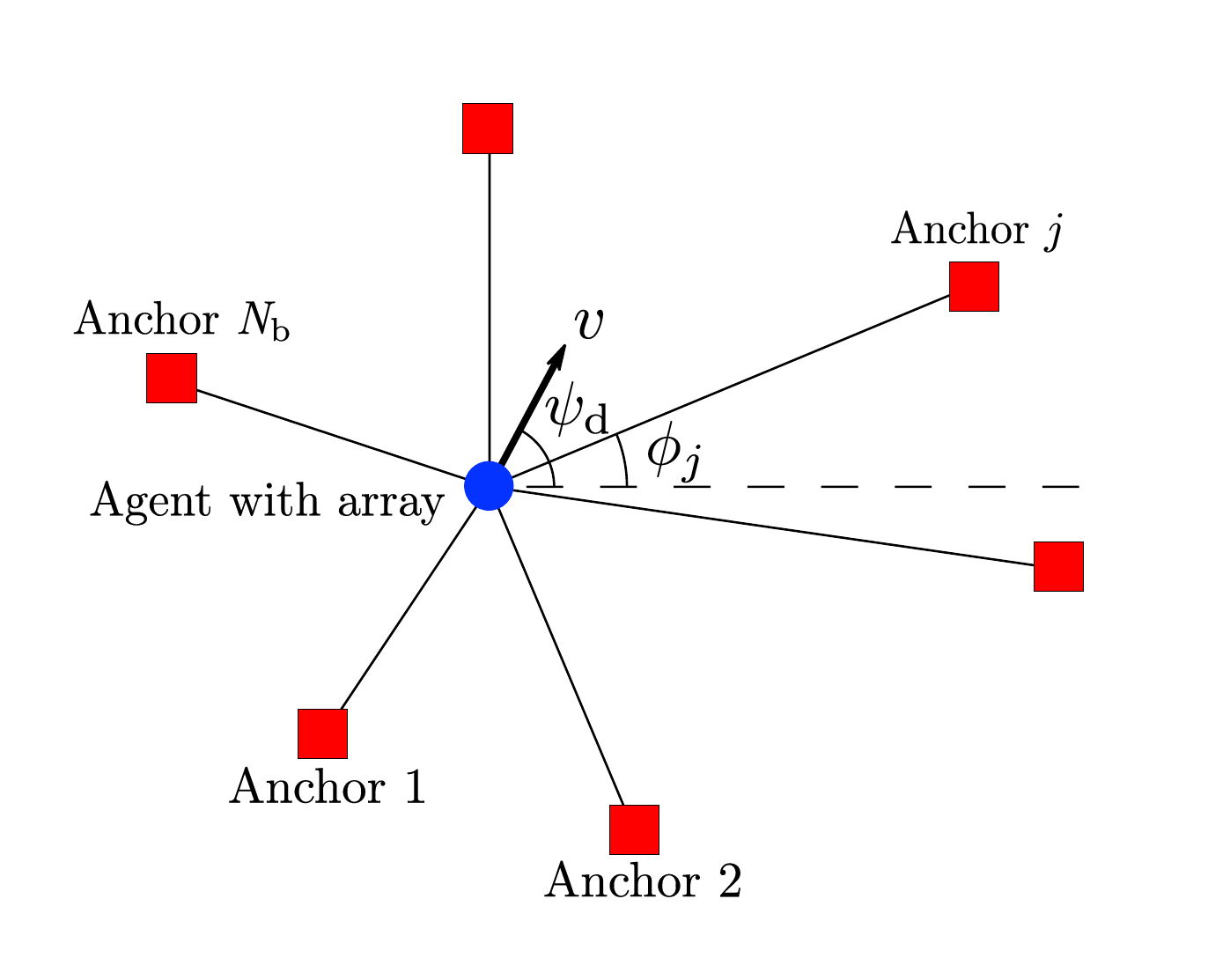}\\
  \caption{System model for {a moving agent}. The agent equipped with an antenna array moves at a constant speed $v$ along {the direction $\psi_{\text{d}}$}.}\label{Fig:Doppler_Struc}
\end{figure}

We still consider far-field environments where the angle from each anchor to all antennas and channel properties (e.g., fading gains and multipath delays) remain time-invariant throughout the observation time.\footnote{{This model is valid when $T_{\text{ob}}\ll T_{\text{co}}$, where $T_{\text{co}}$ is the channel coherence time. Since $T_{\text{co}}\propto {c}/{f_\rc v}$ \cite{Rappaport1996Wireless}, the preceding condition translates to $v\,T_{\text{\rm ob}}\ll {c}/{f_\rc}$, which holds for general practical settings, e.g., $v=30$\thinspace m/s, $f_\rc=2.5$\thinspace GHz and $T_{\text{\rm ob}}=1$\thinspace ms for an LTE example, or $f_\rc=1.8$\thinspace GHz, $T_{\text{\rm ob}}=0.6$\thinspace ms for a GSM example. Moreover, if we replace $T_{\text{\rm ob}}$ with the effective observation time, which is $\sim 1/B$ by the time-frequency duality (cf. Assumption \ref{assum_narrowband} for the definition of $B$), the previous condition can be written as ${v}/{c}\ll {B}/{f_\rc}$, which usually holds in practice.}} Then, similar to (\ref{eq:received_waveform}), the received waveform can be expressed as
{\begin{align}\label{eq:Dynamic_Chan}
r_{jk}(t) & = \sum_{l=1}^{L_{j}}\alpha_{j}^{(l)}\cdot \sqrt{2} \Re\left\{s_0\big(t-\tau_{jk}^{(l)}(t)\big) \right. \\ 
& \hspace{-4mm} \times \left.\exp\big(j(2\pi f_\rc(t-\tau_{jk}^{(l)}(t))+\xi_j)\big)\right\}+z_{jk}(t), \;\; t\in[\,0,T_{\text{ob}}) \nonumber
\end{align}
where the time-variant path delay by the Doppler effect is given by
\begin{align}\label{eq:Dynamic_Delay}
\tau_{jk}^{(l)}(t)
& = \frac{\|{\bf p}_j-{\bf p}_\tau\|-d_k\cos(\phi_j+\gamma_j^{(l)}-\psi-\psi_k)+b_j^{(l)}}{c(1-v_{\text{r}}\cos(\phi_j+\gamma_j^{(l)}-\psi_{\text{\rm d}}))} \nonumber\\
&\qquad -\frac{(t-\tau)v_{\text{r}}\,\cos(\phi_j+\gamma_j^{(l)}-\psi_{\text{\rm d}})}{1-v_{\text{r}}\cos(\phi_j+\gamma_j^{(l)}-\psi_{\text{\rm d}})} 
\end{align}
in which} $v_{\text{r}}\triangleq v/c$ denotes the relative speed.

{Different from the static scenario, we need to introduce two assumptions to simplify the expressions of the main results}.
\begin{assumption}[{Narrowband Signal}]\label{assum_narrowband}
  The baseband signal $s_0(t)$ is bandlimited by $B$, i.e., $S_0(f)=0$ for all $|f|>B$. Furthermore, $B\ll f_{\text{\rm c}}$.
\end{assumption}
\begin{remark}
{In the dynamic scenario we assume the narrowband signal, while in the static scenario only $B\le f_\rc$ is required for the quadrature demodulation.}
\end{remark}
\begin{assumption}[{Balanced Phase}]\label{assum_balanced}
The baseband signal $s_0(t)$ has a balanced phase, i.e.,
  \begin{align}\label{eq:Balanced_Phase_Assumption}
    \int_{-\infty}^{\infty} f|S_0(f)|^2\phi'(f)df = 0
  \end{align}
  where $S_0(f)=|S_0(f)|\exp(j\phi(f))$.
\end{assumption}
\begin{remark}
Assumption \ref{assum_balanced} holds for signals of the form $s_0(t)=\sum_n a_ng(t-nT)$, which is typical in communications given a white stationary ergodic source $\{a_n\}$ and the same filter $g(t)$ used in I--Q two--way modulation, or for signals with constant envelope modulation and random phase uniformly distributed in $[0,2\pi)$. {Moreover, note that Assumption \ref{assum_balanced} is only used for obtaining a simplified result, while the information structure for localization does not rely on this assumption.}
\end{remark}

\subsection{Location Estimation {and Error Bounds}}\label{sec_model_problem}
From a statistical inference perspective, a well-formulated estimation problem involves a parameter set, a statistical experiment, and the random variables generated by this experiment. According to the system setting, the parameter vector $\bm{\theta}$ to be estimated is given by
{\begin{align}\label{eq:parameter}
  \bm{\theta} =
\big[
  \begin{array}{cccccccc}
    {\bf{p}}^{{\text{T}}} & \psi & \psi_{\text{\rm d}} & v & \bm{\kappa}_1^{{\text{T}}} & \bm{\kappa}_2^{{\text{T}}} & \cdots & \bm{\kappa}_{N_{\text{b}}}^{{\text{T}}} \\
  \end{array}
\big]^{{\text{T}}}
\end{align}}
where
\begin{align}\label{eq:parameter_kappa_j}
  \bm{\kappa}_j = \big[
  \begin{array}{ccccc}
    {\xi_j} & \bm{\kappa}_{j}^{(1)\, \text{T}} & \bm{\kappa}_{j}^{(2)\, \text{T}} & \cdots & \bm{\kappa}_{j}^{(L_j)\, \text{T}}
  \end{array}
\big]^{{\text{T}}}
\end{align}
\begin{align}\label{eq:parameter_kappa_j_l}
\bm{\kappa}_{j}^{(l)} \triangleq \begin{cases}
      \big[
    \begin{array}{ccc}
         \text{Para}(\gamma_j^{(1)})&\text{Para}(b_{j}^{(1)})&\alpha_{j}^{(1)}
      \end{array}
      \big]^{{\text{T}}}&l=1,\\
      \big[
      \begin{array}{ccc}
         \text{\ \ \ \ } \gamma_{j}^{(l)} \text{\ \ \ \ }&\text{\ \ \ } b_{j}^{(l)} \text{\ \ \ \ \ }  & \alpha_{j}^{(l)}
      \end{array}
      \big]^{{\text{T}}}&l>1.
  \end{cases}
\end{align}
and $\text{Para}(x_j)$ denotes $\emptyset$ if $j\in\mathcal{N}_{\text{L}}$ and $x_j$ elsewhere. {The random variable ${\bf{r}}$ generated by our statistical experiment} is the vector representation of all the received waveforms ${\bf{r}}_{jk}$ obtained from the Karhunen-Loeve expansion of $r_{jk}(t)$, and this statistical experiment can be characterized into the log-likelihood function shown as
{
\begin{align}\label{eq:likelihood_function_real}
\begin{split}
&\ln f({\bf{r}}|\bm{\theta})  = -\frac{1}{N_0}\sum_{j=1}^{N_{\text{b}}}\sum_{k=1}^{N_{\text{a}}}
   \int_0^{T_{\text{ob}}}\Big|r_{jk}(t)- \sum_{l=1}^{L_j}\alpha_{j}^{(l)}\times \\ 
   &\quad \sqrt{2}\Re\left\{s_0\big(t-\tau_{jk}^{(l)}\big)\exp\big(j(2\pi f_\rc(t-\tau_{jk}^{(l)})+\xi_j)\big)\right\}\Big|^2dt
\end{split}
\end{align}
up to an additive constant.} Hence, the estimation problem is to estimate the parameter $\bm{\theta}$ from the observation $\bf r$ according to the known parameterized probability distribution in (\ref{eq:likelihood_function_real}). {Note that the received waveforms from different anchors can be perfectly separated at the agent due to some implicit multiple access mechanism, but we remark that our estimation problem and thus the error bounds do not depend on the specific mechanism.\footnote{For example, in both static and dynamic scenarios, for the time-division mechanism the likelihood function in (\ref{eq:likelihood_function_real}) remains the same, and for the frequency-division or the code-division mechanism it suffices to use different down-conversion frequencies $f_{\text{c},j}$ (for FDMA) or different baseband signals $s_{0,j}(t)$ (for CDMA) for the waveforms from different anchors.} 

Based on (\ref{eq:likelihood_function_real}), to derive an error bound for this estimation problem, we recall the notion of FIM defined as
\begin{align}\label{eq:fisher_info_J}
  {\bf{J}}_{\bm{\theta}} = \mathbb{E}_{{\bf{r}}}
  \left\{\bigg(\frac{\partial}{\partial{\bm{\theta}}}\ln f({\bf{r}|\bm{\theta}})\bigg)\left(\frac{\partial}{\partial{\bm{\theta}}}\ln f({\bf{r}|\bm{\theta}})\right)^{{\text{T}}}\right\}.
\end{align}
The well-known information inequality asserts that, for any unbiased estimator $\bm{\hat{\theta}}$ for $\bm{\theta}$, we have
$
  \mathbb{E}_{{\bf{r}}}\{(\bm{\hat{\theta}} - \bm{\theta})(\bm{\hat{\theta}} - \bm{\theta})^{{\text{T}}}\}
  \succeq {\bf{J}}_{\bm{\theta}}^{-1}
$\cite{vanTrees1968detection}. It follows that if $\hat{\bf p}$ is an unbiased estimator for ${\bf{p}}$, then
\begin{align}\label{eq:SPEB}
\mathbb{E}_{{\bf{r}}}\big\{\|{\bf{\hat{p}}} - {\bf{p}}\|^2\big\}
  \ge \text{tr}\big\{[{\bf{J}}_{\bm{\theta}}^{-1}]_{1:2,1:2}\big\}.
\end{align}
The right-hand side of (\ref{eq:SPEB}) is defined as the SPEB, cf. \cite[Def. 1]{SheWin:J10a}. To avoid inverting the FIM with large dimensions, we also adopt the notion of EFIM in \cite[Def. 2]{SheWin:J10a}, where the EFIM for the first $n$ components of $\bm{\theta}$ is defined as
$
      {\bf{J}}_{\text{\rm e}}(\bm{\theta}_{1:n}) = {\bf{A}} - {\bf{BC}}^{-1}{\bf{B}}^{{\text{T}}}
$,
where the original FIM for $\bm{\theta} \in \mathbb{R}^N$ is expressed as
 \begin{align}
      {\bf{J}}_{\bm{\theta}} = \left[
                               \begin{array}{cc}
                                 {\bf{A}}_{n\times n} & {\bf{B}}_{n\times(N-n)} \\
                                 {\bf{B}}_{(N-n)\times n}^{{\text{T}}} & {\bf{C}}_{(N-n)\times(N-n)} \\
                               \end{array}
                             \right].
\end{align}
The EFIM retains all the necessary information to derive the information inequality for the parameter vector $\bm{\theta}_{1:n}$, in the sense that $[{\bf{J}}_{\text{\rm e}}(\bm{\theta}_{1:n})]^{-1} = [{\bf{J}}_{\bm{\theta}}^{-1}]_{1:n,1:n}$ according to the Schur complement theory.}

\section{Localization Accuracy in the Static Scenario}\label{sec_EFIM}
This section determines the localization accuracy in terms of the SPEB and EFIM in the static scenario, {and highlights the role that the knowledge of the phase $\xi_j$ and array orientation $\psi$ plays in the reduction of localization errors. For notational convenience, we define ${\bf g}({\bf q}) \triangleq {\bf qq}^{\text{T}}$ and adopt the notion of RDM \cite[Def. 4]{SheWin:J10a} given by
  \begin{align}\label{eq:MDM}
  \begin{split}
    {\bf J}_{\text{\rm r}}(\phi) &\triangleq {\bf g}\left(\left[
                               \begin{array}{c}
                                 \cos\phi \\
                                 \sin\phi \\
                               \end{array}
                             \right]\right) = \left[
                               \begin{array}{c}
                                 \cos\phi \\
                                 \sin\phi \\
                               \end{array}
                             \right]
                             \left[
                               \begin{array}{c}
                                 \cos\phi \\
                                 \sin\phi \\
                               \end{array}
                             \right]^{\text{T}}.
    \end{split}
  \end{align}
  
\subsection{Equivalent Complex Passband Signal Model}
For the ease of FIM derivation, one may want to remove the $\Re\{\cdot\}$ operator and favor the following complex passband signal model
\begin{align}\label{eq:received_waveform_complex}
  \tilde{r}_{jk}(t) &= \sum_{l=1}^{L_{j}}\alpha_{j}^{(l)}s_0(t-\tau_{jk}^{(l)})\exp\big(j(2\pi f_\rc (t-\tau_{jk}^{(l)}) + \xi_j)\big) \nonumber \\
  &\qquad\qquad\qquad\quad \qquad +\tilde{z}_{jk}(t), \quad t\in[\,0,\,T_{\text{ob}})
\end{align}
where $\{\tilde{z}_{jk}(t),t\in[\,0,T_{\text{ob}})\}$ is the complex observation noise with both real and imaginary components following the same distribution as $\{{z}_{jk}(t),t\in[\,0,T_{\text{ob}})\}$, and all other parameters remain the same as those in (\ref{eq:received_waveform}). Then, the corresponding log-likelihood function becomes
\begin{align}\label{eq:likelihood_function}
   \ln f({\bf{\tilde{r}}}|\bm{\theta}) 
   & = -\frac{1}{N_0}\sum_{j=1}^{N_{\text{b}}}\sum_{k=1}^{N_{\text{a}}}
   \int_0^{T_{\text{ob}}}\Big|\tilde{r}_{jk}(t)- \sum_{l=1}^{L_{j}}\alpha_{j}^{(l)} \\ 
   & \quad \;\; \times s_0\big(t-\tau_{jk}^{(l)}\big)\exp\big(j(2\pi f_\rc(t-\tau_{jk}^{(l)})+\xi_j)\big)\Big|^2dt  \nonumber
\end{align}
up to an additive constant.
We next show that in the derivation of the FIM, the complex passband signal model given by (\ref{eq:received_waveform_complex}) and (\ref{eq:likelihood_function}) are equivalent to the real passband signal model in (\ref{eq:received_waveform}) and (\ref{eq:likelihood_function_real}).
\begin{proposition}[{Equivalent Passband Model}]\label{PRO_equivalence}
  If the baseband signal $s_0(t)$ is bandlimited by $f_\rc$, the log-likelihood functions (\ref{eq:likelihood_function_real}) and (\ref{eq:likelihood_function}) generate the same FIM.
\end{proposition}
\begin{IEEEproof}
  See Appendix \ref{PF_PRO_equivalence}.
\end{IEEEproof}
\begin{remark}
	In fact, when $B\le f_\rc$, we can prove a stronger result than Proposition \ref{PRO_equivalence}: the statistical experiments given by (\ref{eq:received_waveform}) and (\ref{eq:received_waveform_complex}) are equivalent in terms of a vanishing Le Cam's distance \cite{cam1986asymptotic}. As a result, for any loss function and any estimator $\hat{\bm{\theta}}_1$ for $\bm{\theta}$ in one model, there exists an estimator $\hat{\bm{\theta}}_2$ in the other model which has the identical risk as $\hat{\bm{\theta}}_1$ under any realization of the parameter $\bm{\theta}$. We omit the proof here, but point out that the key step is to prove that the random vector $\mathbf{r}$ obtained via (\ref{eq:received_waveform}) and $\tilde{\mathbf{{r}}}$ obtained via (\ref{eq:received_waveform_complex}) are mutual randomizations with the help of the Hilbert transform.
\end{remark}

We recall that $B\le f_\rc$ is a natural condition required by the quadrature demodulation. Hence, in the sequel we will stick to the complex observation model (\ref{eq:received_waveform_complex}) and the log-likelihood function (\ref{eq:likelihood_function}).

Before presenting the main results in following sections, we first define a few important metrics.
\begin{definition}[Effective Baseband Bandwidth \cite{vanTrees1968detection} and Baseband-Carrier Correlation]
The effective baseband bandwidth and the baseband-carrier correlation (BCC) of $s_0(t)$ are defined respectively as 
\begin{align}\label{eq:signal_bandwidth}
    \beta \triangleq \left(\frac{\int_{-\infty}^\infty f^2|S_0(f)|^2df}{\int_{-\infty}^\infty |S_0(f)|^2df}\right)^{\frac{1}{2}}
\end{align}
and
\begin{align}
      \gamma \triangleq \frac{\int_{-\infty}^\infty f|S_0(f)|^2df}{\left(\int_{-\infty}^\infty |S_0(f)|^2df\right)^{\frac{1}{2}}\left(\int_{-\infty}^\infty f^2|S_0(f)|^2df\right)^{\frac{1}{2}}} \,.
 \end{align}
\end{definition}

\begin{definition}[{Squared Array Aperture Function}]\label{def_SAAF}
  The squared array aperture function (SAAF) for an array is defined as
    \begin{align}\label{eq:SAAF}
G(\theta) \triangleq \frac{1}{N_{\text{\rm a}}^2}\sum_{1\le k<l\le N_{\text{\rm a}}}\big(d_k\sin(\theta-\psi_k)-d_l\sin(\theta-\psi_l)\big)^2.
\end{align}
\end{definition}
\begin{remark}
The SAAF $G(\theta)$ is the effective array aperture observed from incident the angle $\theta$, and fully quantifies the effect of array-antenna geometry on localization, as will be shown in Section \ref{sec_geometry}.
\end{remark}

\subsection{Case with Known Array Orientation and Initial Phase}
We first consider the case where both the array orientation $\psi$ and the initial phase $\xi_j$ are known. This scenario reduces to the wideband case studied in \cite{SheWin:J10a} in the far-field environment. 
The results are given in the following theorem.
\begin{theorem}[{Full-knowledge Static EFIM}]\label{TH_static_EFIM_full}
  When both the array orientation and the initial phase are known, the EFIM for the position is
  \begin{align}\label{eq:EFIM_ori_known_static_full}
  \begin{split}
    {\bf{J}}_{\text{\rm e}}({\bf{p}}) = \sum_{j\in \mathcal{N}_{\text{\rm L}}}\sum_{k=1}^{N_{\text{\rm a}}}\lambda_j(\beta^2+f_{\text{\rm c}}^2+2\gamma\beta f_\text{\rm c}){\bf J}_{\text{\rm r}}(\phi_j + \theta_{jk}^\text{\rm V})
  \end{split}
  \end{align}
  where $\theta_{jk}^{\text{\rm V}}$ is the \emph{visual angle} expressed as
  \begin{align}\label{eq:visual_angle}
    \theta_{jk}^{\text{\rm V}} \triangleq \frac{d_k\sin(\phi_j-\psi-\psi_k)}{D_j}
  \end{align}
  and
  \begin{align}\label{eq:information_intensity}
  \lambda_{j} \triangleq \frac{8\pi^2 \mathsf{SNR}_{j}^{(1)}(1-\chi_{j})}{c^2}
\end{align}
with the path-overlap coefficient (POC) $\chi_{j}\in[0,1]$ defined in \cite[Thm. 1]{SheWin:J10a} and the SNR $\mathsf{SNR}_{j}^{(l)}$ given by
  \begin{align}\label{eq:SNR}
    \mathsf{SNR}_{j}^{(l)} \triangleq \frac{|\alpha_{j}^{(l)}|^2}{N_0}\int_{-\infty}^\infty |S_0(f)|^2df \,.
  \end{align}
\end{theorem}
\begin{IEEEproof}
  See Appendix \ref{PF_TH_static_EFIM_full}.
\end{IEEEproof}}
{Theorem \ref{TH_static_EFIM_full} implies that in the full knowledge case, the EFIM for the position is a weighed sum of the RDM from each anchor-antenna pair, with direction $\phi_j+\theta_{jk}^\rV$ (i.e., from anchor $j$ to the $k$-th antenna, cf. Fig. \ref{fig:fisherinfo}) and intensity $\lambda_j(\beta^2+f_\rc^2+2\gamma\beta f_\rc)$. Hence, each anchor-antenna pair provides distance information for localization, which sums up to the overall localization information. We also have the following observations}.
{  \begin{itemize}
  \item The support of the intensity $\lambda_j$ is $\mathcal{N}_{\text{L}}$, which means that the anchors providing NLOS signal are not useful for localization, for the actual distance and direction are completely corrupted by the first range bias $b_j^{(1)}$ and first arrival-angle bias $\gamma_j^{(1)}$, respectively.
  \item The intensity $\lambda_j$ depends on the SNR of the first path and the POC $\chi_j$, which characterizes the effect of multipath propagation for localization. It is shown in \cite{SheWin:J10a} that $\chi_{j}$ is determined by the first contiguous cluster (cf. \cite[Def. 3]{SheWin:J10a}) of the received waveform and does not depend on the path amplitudes $\alpha_j^{(l)}$, and $\chi_{j}=0$ when the signal of the first coming path from anchor $j$ does not overlap with those of other paths. {Moreover, the POC $\chi_j$ is solely determined by the autocorrelation function of the baseband signal $s_0(t)$, the carrier frequency $f_\rc$, and channel parameters $\alpha_j^{(l)}, b_j^{(l)}$ and $\gamma_j^{(l)}, 1\le l\le L_j$.}
  \item The $\beta^2+f_\rc^2+2\gamma\beta f_\rc$ term is the squared effective bandwidth of the entire signal $s(t)$ \cite{SheWin:J10a}, which means that the entire bandwidth can be utilized for localization in the full knowledge case.
\end{itemize}}

{Hence, the localization performance for the full knowledge case reduces to the wideband case \cite{SheWin:J10a}, and AOA measurements obtained by antenna arrays do not further improve position accuracy beyond that provided by TOA measurements}.

\subsection{Case with Known Array Orientation {but Unknown Initial Phase}}\label{sec_EFIM_static_known}
{We now turn to the case in which the orientation $\psi$ is known but not the initial phase $\xi_j$. {Theorem \ref{TH_static_EFIM} derives the corresponding localization information.
\begin{theorem}[{Orientation-known Static EFIM}]\label{TH_static_EFIM}
 When the array orientation is known but the initial phase is unknown, the EFIM for the position is
  \begin{align}\label{eq:EFIM_ori_known_static_theta}
    & {\bf{J}}_{\text{\rm e}}({\bf{p}})  = \sum_{j\in \mathcal{N}_{\text{\rm L}}}\lambda_j\Big((1-\gamma^2)\beta^2\sum_{k=1}^{N_{\text{\rm a}}} {\bf J}_{\text{\rm r}}(\phi_j+\theta_{jk}^{\text{\rm V}}) \\
    & \qquad \qquad + \frac{(\gamma\beta+f_\text{\rm c})^2}{N_\ra}\sum_{1\le k<l\le N_\ra}(\theta_{jk}^{\text{\rm V}}-\theta_{jl}^{\text{\rm V}})^2{\bf J}_{\text{\rm r}}\big(\phi_j+\frac{\pi}{2}\big)\Big) \nonumber
  \end{align}
which yields an equivalent expression in terms of the SAAF $G(\theta)$ as
  \begin{align}\label{eq:EFIM_ori_known_static}
    {\bf{J}}_{\text{\rm e}}({\bf{p}}) & = \sum_{j\in \mathcal{N}_{\text{\rm L}}}\lambda_j\Big((1-\gamma^2)\beta^2\sum_{k=1}^{N_{\text{\rm a}}} {\bf J}_{\text{\rm r}}(\phi_j+\theta_{jk}^{\text{\rm V}}) \\& \qquad + \frac{N_\text{\rm a}(\gamma\beta+f_\text{\rm c})^2G(\phi_j-\psi)}{D_j^2}{\bf J}_{\text{\rm r}}\big(\phi_j+\frac{\pi}{2}\big)\Big). \nonumber
  \end{align}
\end{theorem}}
\begin{IEEEproof}
See Appendix \ref{PF_TH_static_EFIM}.
\end{IEEEproof}

%

{In far-field environments, we have $\theta_{jk}^\rV\ll1$ and thus the following approximation for the EFIM given by (\ref{eq:EFIM_ori_known_static}).
\begin{corollary}
  If $\theta_{jk}^\rV\ll1$, the EFIM for the position in Theorem \ref{TH_static_EFIM} can be approximated as
  \begin{align}\label{eq:EFIM_approx}
    {\bf{J}}_{\text{\rm e}}({\bf{p}}) 
    &= \sum_{j\in \mathcal{N}_{\text{\rm L}}}\lambda_j\Big((1-\gamma^2)\beta^2\sum_{k=1}^{N_{\text{\rm a}}} {\bf J}_{\text{\rm r}}(\phi_j)\\
    &\qquad  + \frac{N_\text{\rm a}(\gamma\beta+f_\text{\rm c})^2G(\phi_j-\psi)}{D_j^2}{\bf J}_{\text{\rm r}}\big(\phi_j+\frac{\pi}{2}\big)\Big). \nonumber
  \end{align}
\end{corollary}

Note that the expression given by (\ref{eq:EFIM_approx}) does not depend on the reference point $\bf p$, and hence we can have the following alternative expression of (\ref{eq:EFIM_approx}) when the array center is chosen as the reference point.
\begin{corollary}\label{cor_EFIM_appro}
When the array center is chosen as the reference point, the EFIM in (\ref{eq:EFIM_approx}) becomes
  \begin{align}\label{eq:EFIM_appro}
    {\bf{J}}_{\text{\rm e}}({\bf{p}}) &= \sum_{j\in \mathcal{N}_{\text{\rm L}}}\sum_{k=1}^{N_{\text{\rm a}}}\lambda_j\big((1-\gamma^2)\beta^2{\bf J}_{\text{\rm r}}(\phi_j)\\
    & \qquad\qquad\qquad + (\gamma\beta+f_\rc)^2(\theta_{jk}^{\text{\rm V}})^2{\bf J}_{\text{\rm r}}(\phi_j+\frac{\pi}{2})\big). \nonumber
  \end{align}
\end{corollary}}

Based on the expression of EFIM given in (\ref{eq:EFIM_appro}), some observations and insights can be drawn as follows.
\subsubsection{Distance Information}
the  ${\bf J}_{\text{\rm r}}(\phi_j)$ term is the measuring information for distance, {with intensity proportional to $(1-\gamma^2)\beta^2$} and  direction along the radial angle $\phi_j$ to anchor $j$, i.e., from anchor $j$ to the reference point (see Fig. \ref{fig:fisherinfo}). Hence, this term is the information from TOA and provides localization information with direction \emph{towards} the agent, {and only the baseband signal can contribute to TOA information.}
 
 \begin{figure*}[!b]
 	\normalsize
 	\setcounter{mytempeqncnt}{\value{equation}}
 	\setcounter{equation}{33}
 	\hrulefill
	\vspace*{1mm}
 	\begin{align}\label{eq:EFIM_position_un}
 	{\bf{J}}_{\text{\rm e}}^{\text{\rm un}}({\bf{p}}) 
	&= {\bf J}_{\text{\rm e}}({\bf p})  - \frac{N_\ra f_\rc^2}{\sum_{j\in\mathcal{N}_{\text{\rm L}}}\lambda_jG(\phi_j-\psi)}{\bf g}
	\bigg(\sum_{j\in\mathcal{N}_{\text{\rm L}}}\frac{\lambda_j G(\phi_j-\psi)}{D_j}\bigg[
 	\begin{array}{c}
 	-\sin\phi_j \\
 	\cos\phi_j \\
 	\end{array}
 	\bigg]\bigg) \nonumber \\
 	&= \sum_{j\in\mathcal{N}_{\text{\rm L}}} N_\ra\lambda_j\beta^2{\bf J}_{\text{\rm r}}(\phi_j)+N_\ra f_\rc^2\sum_{i,j\in\mathcal{N}_{\text{\rm L}}}\frac{\lambda_i\lambda_jG(\phi_i-\psi)G(\phi_j-\psi)}
 	{2\sum_{k\in\mathcal{N}_{\text{\rm L}}}\lambda_kG(\phi_k-\psi)}{\bf g}\bigg(\frac{1}{D_i}\bigg[
 	\begin{array}{c}
 	-\sin\phi_i \\
 	\cos\phi_i \\
 	\end{array}
 	\bigg]
 	- \frac{1}{D_j}\bigg[
 	\begin{array}{c}
 	-\sin\phi_j \\
 	\cos\phi_j \\
 	\end{array}
 	\bigg]
 	\bigg)
 	\end{align}
 	\setcounter{equation}{\value{mytempeqncnt}}
 	\vspace*{4pt}
 \end{figure*}

 \begin{figure}[!t]
  \centering
  \includegraphics[scale=0.45]{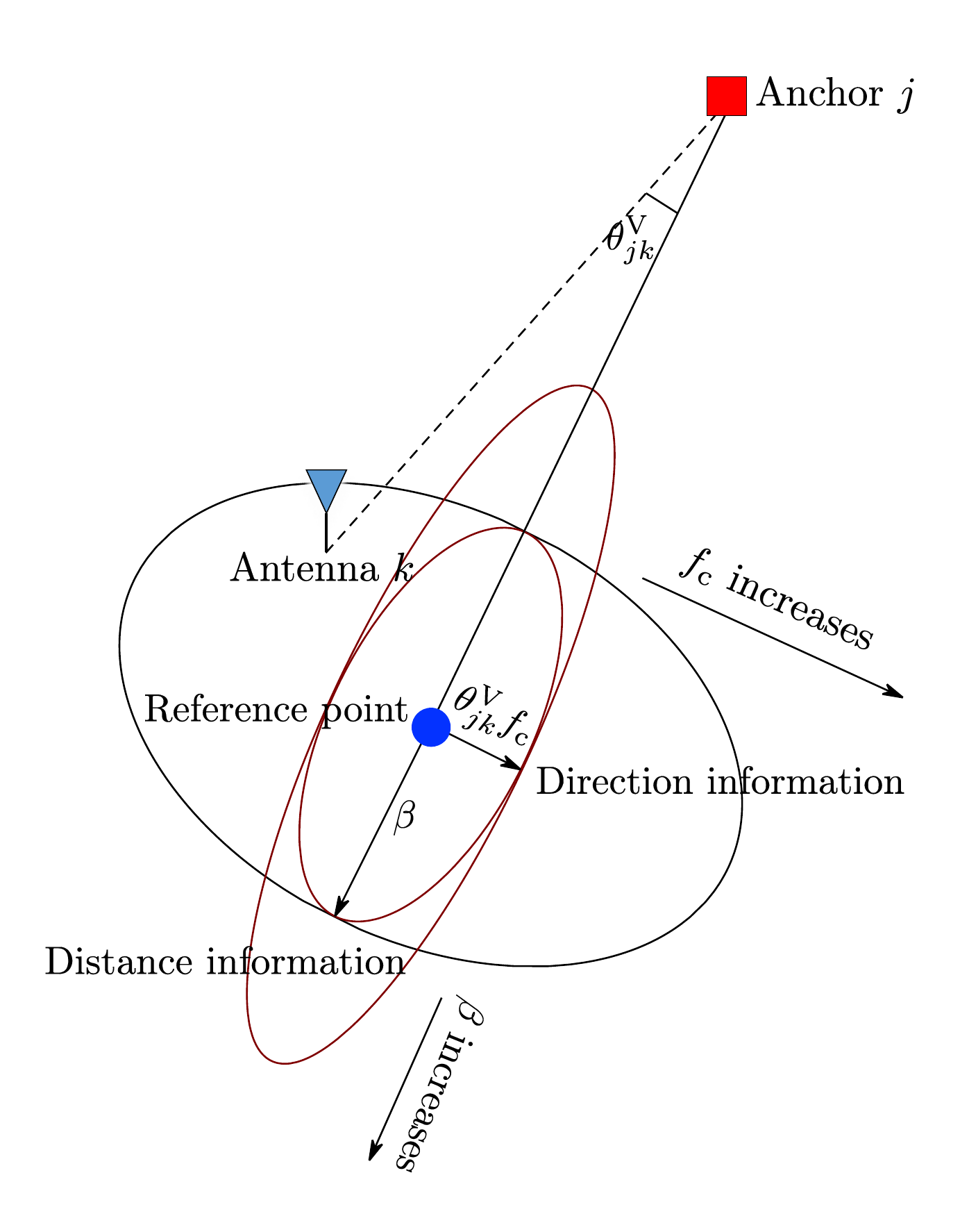}
  \vspace*{-3mm}
  \caption{Illustration for the localization information ellipse formed by distance information and direction information {in the case of $\gamma=0$}. The ellipse differs in shape as $\beta$ or $f_{\text{\rm c}}$ varies.}\label{fig:fisherinfo}
\end{figure}
 
\subsubsection{Direction Information}
the ${\bf J}_{\text{\rm r}}(\phi_j+\frac{\pi}{2})$ term corresponds to the measuring information for direction, i.e., from AOA, and has a tangent direction to anchor $j$ (the direction perpendicular to that connecting anchor $j$ and the reference point; see Fig. \ref{fig:fisherinfo}). The intensity of this term consists of two parts. The first part $\theta_{jk}^{\text{\rm V}}$ is the \emph{visual angle} of the effective aperture for $k$-th antenna observed from anchor $j$, and the second part $\gamma\beta+f_\rc$ is the {\emph{effective carrier frequency}}. As a system-level interpretation, assuming $\gamma=0$ for simplicity, the intensity scaled by $c^{-2}$, i.e., $(\theta_{jk}^{\text{\rm V}}f_\rc/c)^2$, is the squared  visual angle normalized by the wavelength. {Hence, the AOA information can be retrieved from the effective carrier frequency.}

\subsubsection{Geometric Interpretation}
the EFIM ${\bf{J}}_{\text{\rm e}}({\bf{p}})$ is a weighed sum of measuring information from each anchor-antenna pair, where each pair  provides information in two orthogonal directions summing up to the entire localization information (depicted as an ellipse in Fig. \ref{fig:fisherinfo}). {Note that 
\begin{align}
\beta^2+f_\rc^2+2\gamma\beta f_\rc=(1-\gamma^2)\beta^2+(\gamma\beta+f_\rc)^2
\end{align}
and we conclude by Theorem \ref{TH_static_EFIM_full} and \ref{TH_static_EFIM} that, in wideband cases \cite{SheWin:J10a} the effective bandwidth of the entire signal can be used for obtaining both distance and direction information, while in our model the overall bandwidth is decomposed into two parts, i.e., baseband signal for distance information and carrier frequency for direction information. In particular, AOA information can make significant contributions to localization accuracy beyond that obtained by TOA measurements in our model}.

{Moreover, the direction of the major axis of ellipse depends on whether $\theta_{jk}^{\text{\rm V}}f_{\text{\rm c}}\gtrless\beta$. In traditional TOA systems, $\beta$ is comparable to $f_{\text{\rm c}}$, and hence the distance information dominates since $\theta_{jk}^{\text{\rm V}}\ll1$ (due to far-field environments). In contrast, in traditional AOA systems, $f_{\text{\rm c}}\gg\beta$, and hence the direction information dominates. In practice, since $d_k\cong c/f_{\text{\rm c}}$ holds, the criterion becomes $c\gtrless \beta D_j$. For example, when $\beta\cong1$\thinspace MHz, the dominance of the direction information requires $D_j\ll 300$m, conforming to the fact that AOA information is more effective in short-distance localization}.

{Now we discuss some properties of the BCC $\gamma\in[-1,1]$, which characterizes the extent how close is the baseband signal $s_0(t)$ to a single-frequency signal. For example, $\gamma=\pm 1$ implies $s_0(t)=\exp(\pm j2\pi \beta t+\phi)$, and thus the entire baseband signal contributes to the AOA information and we cannot extract any TOA information from $s(t)=s_0(t)\exp(j2\pi f_\rc t)$. On the other hand, when $\gamma=0$ (equivalent to $\int f|S_0(f)|^2df=0$), the entire baseband signal can be utilized for obtaining the TOA information. 

Without loss of generality, we will assume $\gamma=0$ in the following sections, since otherwise we can always substitute the baseband signal and carrier frequency by $\tilde{s}_0(t) = s_0(t)\exp(-j2\pi\gamma\beta t)$ and $\tilde{f}_\rc = f_\rc + \gamma\beta$.}

\subsection{Case with Unknown Array Orientation {and Initial Phase}}\label{sec_EFIM_static_unknown}
We next consider the case in which the array orientation $\psi$ also becomes an unknown parameter to be estimated. Similar to Theorem \ref{TH_static_EFIM}, the EFIM for the position and orientation is derived accordingly.
\begin{theorem}[{Orientation-unknown Static EFIM}]\label{TH_static_EFIM_un}
When neither the initial phase nor the array orientation is known, the EFIM for the position and orientation is
\begin{align}\label{eq:EFIM_ori_unknown_static}
      {\bf{J}}_{\text{\rm e}}(\{{\bf{p}},\psi\}) 
      & = \sum_{j\in \mathcal{N}_{\text{\rm L}}}  \sum_{k=1}^{N_{\text{\rm a}}} \lambda_j \beta^2{\bf g}\left(\left[
                               \begin{array}{c}
                                 \cos(\phi_j + \theta_{jk}^{\text{\rm V}})\\
                                 \sin(\phi_j + \theta_{jk}^{\text{\rm V}})\\
                                 -D_j\theta_{jk}^{\text{\rm V}}\\
                               \end{array}
                             \right]\right)
                              \nonumber \\
                              & \qquad + \frac{\lambda_j N_\ra f_\text{\rm c}^2G(\phi_j-\psi)}{D_j^2}{\bf g}\left(\left[
                               \begin{array}{c}
                                 -\sin\phi_j \\
                                 \cos\phi_j \\
                                 -D_j\\
                               \end{array}
                             \right]\right).
  \end{align}
\end{theorem}
\begin{IEEEproof}
  See Appendix \ref{PF_TH_static_EFIM_un}.
\end{IEEEproof}

\begin{corollary}\label{cor_EFIM_ori_unknown}
{If $\theta_{jk}^\rV \ll  1$}, the EFIM for the position in the orientation-unknown case can be approximated as (\ref{eq:EFIM_position_un}), shown at the bottom of \blue{this} page, where ${\bf J}_{\text{\rm e}}({\bf p})$ is given by (\ref{eq:EFIM_approx}).
\end{corollary}
\begin{IEEEproof}
 Equations (\ref{eq:EFIM_position_un}) follows directly from (\ref{eq:EFIM_ori_unknown_static}), $\theta_{jk}^\rV\ll1$, and the definition of EFIM.
\end{IEEEproof}
\addtocounter{equation}{1}

Theorem \ref{TH_static_EFIM_un} claims that the EFIM for the position and orientation is also a weighed matrix sum of measuring information from each anchor-antenna pair, {and thus the overall localization information in the orientation-unknown case possesses a similar structure. Moreover, since ${\bf J}_{\text{\rm e}}({\bf p})-{\bf{J}}_{\text{\rm e}}^{\text{\rm un}}({\bf{p}}) $ is positive semi-definite}, the unknown orientation degrades the localization accuracy. 

Note that the approximated EFIM ${\bf{J}}_{\text{\rm e}}^{\text{\rm un}}({\bf{p}})$ for the orientation-unknown case still does not depend on the reference point $\bf p$, which seems contradictory to the wideband case in \cite{SheWin:J10a}. However, we remark that the invariance of ${\bf{J}}_{\text{\rm e}}^{\text{\rm un}}({\bf{p}})$ on $\bf p$ in far-field enviroments is due to the fact that the AOA information does not rely on $\bf p$, and the $\theta_{jk}^\rV$ term can be neglected in the TOA information.

\section{Localization Accuracy in the Dynamic Scenario}\label{sec_doppler}
As is shown in the preceding section, the time-invariant delays between anchor-antenna pairs are all the sources of localization information in the static scenario. In this section, we turn to the dynamic scenario where the Doppler effect can be utilized {in addition to} the TOA and AOA measurements for localization.

\subsection{Case with Known Orientation and Velocity}\label{sec_doppler_EFIM}
We first consider the scenario in which both the velocity and orientation of the antenna array are known. {This scenario is relevant in practice, as the agent can obtain its velocity and orientation locally, e.g., by a compass and accelerometer. Note that in the dynamic scenario, Assumptions \ref{assum_narrowband} and \ref{assum_balanced} are needed to simplify the expressions of the EFIM, as shown in the next theorem}.
 \begin{theorem}[{Orientation- and Velocity-known Moving EFIM}]\label{TH_dynamic_EFIM}
  The EFIM for the position is
    \begin{align}
    {\bf J}_{\text{\rm e}}({\bf p}_\tau) = \sum_{j\in\mathcal{N}_{\text{\rm L}}} & \Big[A_{1j}{\bf J}_{\text{\rm r}}(\phi_j) + A_{2j}{\bf J}_{\text{\rm r}}(\phi_j+\frac{\pi}{2}) \nonumber \\ 
    &  \quad + A_{3j}\big({\bf J}_{\text{\rm r}}(\phi_j+\frac{\pi}{4})-{\bf J}_{\text{\rm r}}(\phi_j-\frac{\pi}{4})\big) \Big]
  \end{align}
where $A_{1j},A_{2j},A_{3j}$ are given by (\ref{eq:A_1j_final}), (\ref{eq:Phi22_exp}) and (\ref{eq:A_3j}), respectively. Furthermore, under Assumption \ref{assum_narrowband} and \ref{assum_balanced}, the EFIM can be approximated as
\begin{align}\label{eq:Dynamic_position_EFIM}
      {\bf J}_{\text{\rm e}}({\bf p}_\tau) &\cong \sum_{j\in\mathcal{N}_{\text{\rm L}}}
      \frac{\lambda_jN_\ra}{1-v_{\text{\rm r}}\cos(\phi_j-\psi_{\text{\rm d}})}\Big[\,\beta^2{\bf J}_{\text{\rm r}}(\phi_j)  \nonumber\\
      & \qquad+ f_{\text{\rm c}}^2\Big(\frac{G(\phi_j-\psi)}{D_j^2} + \omega_j^2t_{\text{\rm rms}}^2\Big){\bf J}_{\text{\rm r}}(\phi_j+\frac{\pi}{2})\,\Big]
  \end{align}
  where $t_{\text{\rm rms}}$ is the root mean squared time duration of the baseband signal $s_0(t)$ defined as
  \begin{align}\label{eq:t_eff}
    t_{\text{\rm rms}} \triangleq \frac{\iint_{t_1<t_2} (t_2-t_1)^2|s_0(t_1)|^2|s_0(t_2)|^2dt_1dt_2}{\left(\int |s_0(t)|^2dt\right)^2}
  \end{align}
  and $\omega_j$ is the angular speed of the visual angle given by
  \begin{align}\label{eq:Dynamic_position_omega}
    \omega_j \triangleq \frac{v\sin(\psi_{\text{\rm d}} - \phi_j)}{D_j}.
  \end{align}
\end{theorem}
\begin{IEEEproof}
  See Appendix \ref{PF_TH_dynamic_EFIM}.
\end{IEEEproof}

\begin{remark}
By the proof details given in the appendix, we can show that the EFIM in (\ref{eq:Dynamic_position_EFIM}) is a tight approximation up to a multiplicative approximation error less than $\left(1+{3.17B}/{f_\rc}\right)^2-1$.
\end{remark}

Due to the invariance of ${\bf J}_{\text{\rm e}}({\bf p}_\tau)$ on the reference point $\bf p$ and the reference time $\tau$, similar to Corollary \ref{cor_EFIM_appro}, we can choose the array center to be the reference point and $\tau=0$ to be the reference time.
\begin{corollary}\label{cor_EFIM_doppler}
When the array center is chosen as the reference point and $\tau=0$, the EFIM in (\ref{eq:Dynamic_position_EFIM}) becomes
    \begin{align}
      {\bf J}_{\text{\rm e}}({\bf p}_\tau) &\cong \sum_{j\in\mathcal{N}_{\text{\rm L}}}\sum_{k=1}^{N_\ra}
      \frac{\lambda_j}{1-v_{\text{\rm r}}\cos(\phi_j-\psi_{\text{\rm d}})}\Big[\,\beta^2{\bf J}_{\text{\rm r}}(\phi_j) \nonumber \\
      &\;\;  \qquad+ f_{\text{\rm c}}^2\left((\theta_{jk}^\rV)^2 + (\omega_jt_{\text{\rm rms}})^2\right){\bf J}_{\text{\rm r}}(\phi_j+\frac{\pi}{2})\,\Big].
  \end{align}
\end{corollary}
\begin{figure}[!t]
  \centering
  \includegraphics[scale=0.45]{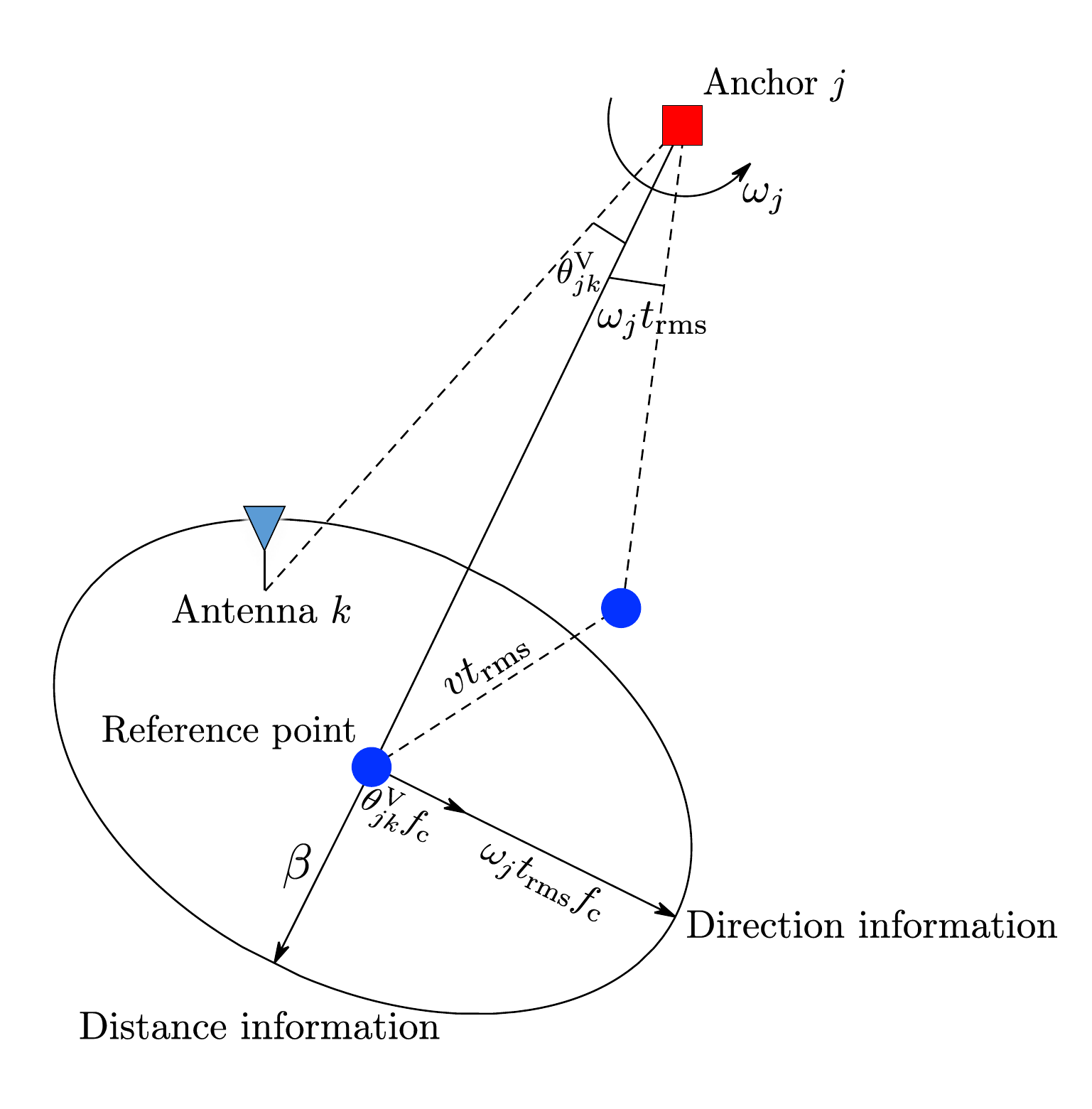}\\
  \caption{Illustration for the localization information ellipse formed by distance information and direction information. The direction information comes from both AOA measurement and the Doppler shift.}\label{Fig:Doppler_Info}
\end{figure}

\begin{figure*}[!b]
	\normalsize
	\hrulefill
		\vspace*{1mm}
	\begin{align}\label{eq:Dynamic_EFIM_un}
{\bf{J}}_{\text{\rm e}}(\{{\bf{p}}_\tau,\psi,\psi_{\text{\rm d}}\}) 
\cong \sum_{j\in \mathcal{N}_{\text{\rm L}}}N_\ra \lambda_j\left(\beta^2 {\bf g}\left(\left[
\begin{array}{c}
\cos\phi_j \\
\sin\phi_j \\
0 \\
0 \\
\end{array}
\right]
\right) + \frac{f_\rc^2G(\phi_j-\psi)}{D_j^2}{\bf g}\left(\left[
\begin{array}{c}
-\sin\phi_j \\
\cos\phi_j \\
-D_j \\
0 \\
\end{array}
\right]\right) + f_\rc^2\omega_j^2t_{\text{\rm rms}}^2{\bf g}\left(\left[
\begin{array}{c}
-\sin\phi_j \\
\cos\phi_j\\
0 \\
-D_j \\
\end{array}
\right]
\right)
\right)
\end{align}
	\hrulefill
		\vspace*{1mm}
  \begin{align}\label{eq:EFIM_doppler_un}
  {\bf{J}}_{\text{\rm e}}^{\text{\rm un}}({\bf{p}}) 
  =N_\ra \Bigg(\sum_{j\in\mathcal{N}_{\text{\rm L}}}\lambda_j\beta^2{\bf J}_{\text{\rm r}}(\phi_j)+\frac{1}{2}f_{\text{\rm c}}^2
  \sum_{i,j\in\mathcal{N}_{\text{\rm L}}}
  & \lambda_i\lambda_j\bigg(\frac{G(\phi_i-\psi)G(\phi_j-\psi)}{\sum_{k\in\mathcal{N}_{\text{\rm L}}}\lambda_kG(\phi_k-\psi)}
+\frac{v^2t_{\text{\rm rms}}^2\sin^2(\phi_i-\psi_{\text{\rm d}})\sin^2(\phi_j-\psi_{\text{\rm d}})}
  {\sum_{k\in\mathcal{N}_{\text{\rm L}}}\lambda_k\sin^2(\phi_k-\psi_{\text{\rm d}})}\bigg) \nonumber \\
  &
  \times
  {\bf g}\bigg(\frac{1}{D_i}\bigg[
  \begin{array}{c}
  -\sin\phi_i \\
  \cos\phi_i \\
  \end{array}
  \bigg]
  - \frac{1}{D_j}\bigg[
  \begin{array}{c}
  -\sin\phi_j \\
  \cos\phi_j \\
  \end{array}
  \bigg]
  \bigg)
  \Bigg)
  \end{align}
  \hrulefill
  		\vspace*{1mm}
  	\begin{align}\label{eq:Dynamic_EFIM_unun}
  	\begin{split}
  	{\bf{J}}_{\text{\rm e}}(\{{\bf{p}}_\tau,\psi,\psi_{\text{\rm d}},v\}) 
  	\cong \sum_{j\in \mathcal{N}_{\text{\rm L}}}N_\ra \lambda_j\left(\beta^2 {\bf g}\left(\left[
  	\begin{array}{c}
  	\cos\phi_j \\
  	\sin\phi_j \\
  	0 \\
  	0 \\
  	0 \\
  	\end{array}
  	\right]
  	\right) + \frac{f_\rc^2G(\phi_j-\psi)}{D_j^2}{\bf g}\left(\left[
  	\begin{array}{c}
  	-\sin\phi_j \\
  	\cos\phi_j \\
  	-D_j \\
  	0 \\
  	0 \\
  	\end{array}
  	\right]\right) + f_\rc^2\omega_j^2t_{\text{\rm rms}}^2{\bf g}\left(\left[
  	\begin{array}{c}
  	-\sin\phi_j \\
  	\cos\phi_j\\
  	0 \\
  	-D_j \\
  	\frac{\cos(\phi_j-\psi_{\text{\rm d}})}{\omega_j}\\
  	\end{array}
  	\right]
  	\right)
  	\right)
  	\end{split}
  	\end{align}
	\vspace*{4pt}
\end{figure*}

Some observations on the effect of Doppler shift can be drawn from Theorem \ref{TH_dynamic_EFIM} {and Corollay \ref{cor_EFIM_doppler}} as follows.
   \subsubsection{Intensity Effect}
   Compared with Theorem \ref{TH_static_EFIM}, there is a new coefficient $(1-v_{\text{\rm r}}\cos(\phi_j-\psi_{\text{\rm d}}))^{-1}$ on the information intensity, {which we refer to as the intensity effect of the Doppler shift}. Intuitively, the Doppler shift enlarges both the baseband bandwidth and carrier frequency by $(1-v_{\text{\rm r}}\cos(\phi_j-\psi_{\text{\rm d}}))^{-1}$ times, whereas the the SNR is reduced by $(1-v_{\text{\rm r}}\cos(\phi_j-\psi_{\text{\rm d}}))$ times. Hence, this new coefficient is obtained by the fact that the EFIM for the position is proportional to SNR times the squared bandwidth. Note that whether the intensity effect does help or harm to localization depends on the array orientation, but this effect is negligible since $v_{\text{r}}=v/c\ll1$. {With a slight abuse of notations, we will still denote $\lambda_j (1-v_{\text{\rm r}}\cos(\phi_j-\psi_{\text{\rm d}}))^{-1}$ by $\lambda_j$ in the following}. 
   \subsubsection{Direction Effect}
  The second effect of Doppler shift on localization is the direction effect, i.e., it provides additional direction information with intensity {$f_{\text{c}}^2\omega_j^2t_{\text{rms}}^2$. This direction information originates from the dependence of the Doppler shift on the direction of the anchor, and faster speed is preferred for accumulating more direction information}. Hence, as shown in Fig. \ref{Fig:Doppler_Info}, the localization information can be decomposed into the distance and direction information, where the latter consists of two parts from AOA measurement and Doppler shifts, respectively. {In particular, the Doppler shifts do not affect the distance information for localization.}
  \subsubsection{Geometric Interpretation}
  The {geometric} interpretations for the new variables $\omega_j$ and $t_{\text{rms}}$ are illustrated in Fig. \ref{Fig:Doppler_Info}. The variable $t_{\text{rms}}$ can be interpreted as the equivalent moving time of the agent from its initial position to the final position, and $\omega_jt_{\text{rms}}$ can be viewed as a synthetic aperture formed by the moving agent, which has the same effect as the real array aperture. Hence, similar to the interpretation that $\theta_{jk}^{\text{\rm V}}$ is the visual angle between the reference point and $k$-th antenna at any fixed time, $\omega_j$ is the angular speed of the visual angle formed by the reference points at different time. {Moreover, the overall localization information for direction is simply the sum of the synthetic aperture and the real array aperture. In practice, it is likely that the synthetic aperture formed by the moving agent during the observation time is larger than the real array aperture, and thus the Doppler effect can provide considerably more direction information for localization in the dynamic scenarios. Note that since $t_{\text{rms}}$ does not depend on the reference time $\tau$, the localization accuracy for the agent position remains the same at any time, which is consistent to our intuition under the known-velocity scenario. }

\subsection{Case with Unknown Orientation and Velocity}\label{sec_doppler_EFIM_un}
This subsection will address scenarios in which the array orientation, agent speed or its moving direction (possibly all) is unknown to the agent. {We first consider the case where the agent knows its speed but not the array orientation or the moving direction, which is practically relevant as the speed is the only local quantity invariant with translation or rotation among all parameters.}
{\begin{theorem}[{Orientation- and Direction-unknown Speed-known Moving EFIM}]\label{TH_dynamic_EFIM_un}
    Under Assumption \ref{assum_narrowband} and \ref{assum_balanced}, when the agent speed is known but both the array orientation and the moving direction are unknown, the EFIM for the position, orientation and moving direction is given by (\ref{eq:Dynamic_EFIM_un}), shown at the bottom of \blue{this} page.
\end{theorem}
\begin{IEEEproof}
  Similar to the proof of Theorem \ref{TH_static_EFIM_un}.
\end{IEEEproof}}

Analogous to Corollary \ref{cor_EFIM_ori_unknown}, the EFIM for the position can be derived based on (\ref{eq:Dynamic_EFIM_un}) as follows.
{\begin{corollary}\label{cor_EFIM_doppler_un}
Given the conditions of Theorem \ref{TH_dynamic_EFIM_un}, the EFIM for the position is given by (\ref{eq:EFIM_doppler_un}), shown at the bottom of \blue{this} page.
\end{corollary}}

The EFIM given by (\ref{eq:Dynamic_EFIM_un}) in Theorem \ref{TH_dynamic_EFIM_un} implies that the estimation of parameters $\psi$ and $\psi_{\text{\rm d}}$ does not affect each other given all other parameters. Moreover, comparing the EFIM (\ref{eq:EFIM_doppler_un}) for the dynamic scenario with (\ref{eq:EFIM_position_un}) for the static scenario shows that there is just an additional Doppler term with intensity proportional to the squared time duration $t_{\text{\rm rms}}^2$, and consequently, the information decomposition shown in Fig. \ref{Fig:Doppler_Info} remains the same for the dynamic scenario. Similar to the static scenario, the EFIM ${\bf{J}}_{\text{\rm e}}^{\text{\rm un}}({\bf{p}})$ for the dynamic scenario does not depend on the reference point $\bf p$ or the reference time $\tau$ in far-field enviroments.

We close this section by presenting the EFIM when the agent knows nothing about the orientation or velocity. The results are summarized in the following theorem.
{\begin{theorem}[{Orientation- and Velocity-unknown Moving EFIM}]\label{TH_dynamic_EFIM_unun}
  {Under Assumption \ref{assum_narrowband} and \ref{assum_balanced}}, when neither the array orientation nor the agent velocity is known, the EFIM for ${\bf p}, \psi, \psi_{\text{\rm d}}$ and $v$ is given by (\ref{eq:Dynamic_EFIM_unun}), shown at the bottom of \blue{this page}.
\end{theorem}}
\begin{IEEEproof}
  Similar to the proof of Theorem \ref{TH_static_EFIM_un}.
\end{IEEEproof}

{Compared with Theorem \ref{TH_dynamic_EFIM_un}, Theorem \ref{TH_dynamic_EFIM_unun} shows that the unknown speed only contaminates the Doppler term in the localization information, while the Doppler shift still contributes additional information to localization compared with the static scenario.}

\section{Geometric Properties in Localization}\label{sec_geometry}
In the preceding sections, we have shown that the EFIM for the position consists of three parts, i.e., the localization information provided by TOA, Doppler shift, and AOA. Note that all these parts depend on the geometric structure of the anchors and agent, and the AOA information is further dependent on the geometric structure of the antenna array. We call these geometric structures as {anchor-agent geometry and array-antenna geometry}, respectively, and characterize their effects on the localization accuracy in this section.

\subsection{Effects of Array-antenna Geometry}\label{sec_geometry_array}
We first investigate the array-antenna geometry as it only affects the AOA information. Based on the EFIMs for the position given in (\ref{eq:Dynamic_position_EFIM}) and (\ref{eq:EFIM_doppler_un}), the array-antenna geometry affects the localization information through the SAAF $G(\theta)$ given in (\ref{eq:SAAF}). In fact, in the view of traditional AOA localization, $G(\theta)$ is exactly the squared array aperture given the incident angle $\theta$ of the waveform. Moreover, {$G(\theta)$ is uniquely determined by the location of all antennas and fully quantifies the effect of array-antenna geometry on localization}, i.e., different array geometries with the same $G(\theta)$ have an identical performance on localization accuracy.

Based on the EFIM (\ref{eq:EFIM_doppler_un}) in the case where we know the speed but do not know the array orientation and moving direction, some useful observations can be drawn accordingly:

 \begin{itemize}
  \item The EFIM consists of a distance component and a direction component, where the array-antenna geometry, the array orientation, and the Doppler shift only affect the direction component;
  \item Compared with the EFIM in (\ref{eq:Dynamic_position_EFIM}), {the missing knowledge about the array orientation and agent moving direction causes the information loss only in direction component};
  \item {For two types of arrays, if $G_1(\theta) \ge G_2(\theta), \forall\, \theta$, the first array will yield a larger EFIM than the second array in terms of the L\"{o}wner semiorder $\succeq$. Hence, roughly speaking, arrays with larger $G(\theta)$ provide better localization accuracy. In particular}, joint signal processing among antennas is required for obtaining AOA information since $G(\theta)=0$ when $N_{\text{a}}=1$;
  \item The necessary conditions for an array localization system (i.e., non-singular ${\bf{J}}_{\text{e}}({\bf{p}})$) under various network parameters are given in Table \ref{table_example}, which shows the minimum number of anchors and antennas required for localization. {We remark that} the trilateration and triangulation principles also prove the sufficiency of these conditions disregarding the global \emph{ambiguity} (i.e., two possible locations of the agent).\footnote{For example, two anchors can localize the agent using only the baseband signals, and the ambiguity phenomenon can be overcome by some prior knowledge.} Note that the restrictions on antenna number are relaxed in the dynamic scenario because joint processing is not required to obtain the Doppler information.
\end{itemize}

\begin{table*}[!t]
  \centering
  \renewcommand{\arraystretch}{1.3}
  \caption{Requirements for the number of anchors and antennas}
\scriptsize
  \begin{threeparttable}
    \begin{tabular}{|c|c|c|c|c|c|}
    \hline
    \multirow{2}[4]{*}{} & \multicolumn{3}{c|}{$v=0$} & \multicolumn{2}{c|}{$v>0$\tnote{1}} \bigstrut\\
\cline{2-6}          & $f_{\text{c}}=0$ & $\beta=0$ & $\beta>0,f_{\text{c}}>0$ & $\beta=0$ & $\beta>0,f_{\text{c}}>0$ \bigstrut\\
    \hline
    Known orientation and moving direction & $\ge$2 anchors & $\ge$2 anchors \& $\ge$2 antennas & $\ge$2 anchors or $\ge$2 antennas & $\ge$2 anchors & No restriction \bigstrut\\
    \hline
    Unknown orientation and moving direction & $\ge$2 anchors & $\ge$3 anchors \& $\ge$2 antennas & $\ge$2 anchors & $\ge$3 anchors & $\ge$2 anchors \bigstrut\\
    \hline
    \end{tabular}%
    \begin{tablenotes}
      \item[1] Due to the requirement that $\beta\ll f_{\text{c}}$ in the moving model, there is no $f_{\text{c}}=0$ case when $v>0$ in this table.
    \end{tablenotes}
    \end{threeparttable}
  \label{table_example}%
\end{table*}%

{In summary, large SAAFs are preferred for localization, and it requires a large array diameter, a diversified antenna geometry, a proper incident angle, and at least two antennas. Since the antenna geometry is of interest in the array design, we consider the optimal array-antenna geometry given the antenna number and the array diameter.}
\begin{definition}[{Array Diameter}]\label{def_array_diameter}
  The diameter of an array is {the diameter of the smallest circle which can fully cover} the array, i.e.,
  \begin{align}\label{eq:GP_2_D4}
    D \triangleq 2\cdot\inf_{{\bf{p}}_{\text{\rm c}}\in\mathbb{R}^2}\sup_{1\le k\le N_{\text{\rm a}}}\|{\bf{p}}_{\text{\rm c}}-{\bf{p}}_k^{\text{\rm Array}}\|.
  \end{align}
\end{definition}

Before stepping further, we introduce two special types of arrays first.
\begin{definition}[{UOA and UCOA}]\label{def_UOA}
  The uniformly oriented array (UOA) is an array with
$  
    \widetilde{\bf xy} = 0, \widetilde{{\bf x}^2} = \widetilde{{\bf y}^2},
$  
  where ${\bf x}, {\bf y} \in \mathbb{R}^{N_{\text{\rm a}}}$ contain the $x$- and $y$-coordinates, respectively, of all antennas in order, and we adopt the notation
  \begin{align}\label{eq:oriented_2}
    \widetilde{\mathbf{vw}} \triangleq \frac{1}{N}\sum_{k=1}^{N}v_kw_k-\frac{1}{N^2}\bigg(\sum_{k=1}^{N}v_k\bigg)\bigg(\sum_{k=1}^{N}w_k\bigg)
  \end{align}
     for any ${\bf v}, {\bf w}  \in \mathbb{R}^{N_{\text{\rm a}}}$, and $\widetilde{{\bf v}^2}$ is abbreviated for $\widetilde{\bf vv}$. In addition, a UOA is called a uniformly circular oriented array (UCOA), if all antennas lie on a circle centered at the array coordinate center.
\end{definition}

\begin{figure*}[!b]
	\normalsize
	\setcounter{mytempeqncnt}{\value{equation}}
	\setcounter{equation}{49}
	\hrulefill
	\vspace*{1mm}
\begin{align}\label{eq:4_1_1_3}
	{\bf{J}}_{\text{e}}^{\text{ULA}}({\bf{p}}) 
	=N_{\text{a}}\Bigg(\sum_{j\in\mathcal{N}_{\text{L}}}\lambda_j\beta^2{\bf J}_{\text{\rm r}}(\phi_j)+\frac{1}{2}f_{\text{c}}^2
	&\sum_{i,j\in\mathcal{N}_{\text{L}}} \lambda_i\lambda_j
	\bigg(\frac{(N_{\text{a}}+1)D^2}{12(N_{\text{a}}-1)}\cdot\frac{\sin^2(\phi_i-\psi)\sin^2(\phi_j-\psi)}
	{\sum_{k\in\mathcal{N}_{\text{L}}}\lambda_k\sin^2(\phi_k-\psi)}\\
	&
	+\frac{v^2t_{\text{\rm rms}}^2\sin^2(\phi_i-\psi_{\text{\rm d}})\sin^2(\phi_j-\psi_{\text{\rm d}})}
	{\sum_{k\in\mathcal{N}_{\text{L}}}\lambda_k\sin^2(\phi_k-\psi_{\text{\rm d}})}\bigg)
	{\bf g}\bigg(\frac{1}{D_i}\bigg[
	\begin{array}{c}
	-\sin\phi_i \\
	\cos\phi_i \\
	\end{array}
	\bigg]
	- \frac{1}{D_j}\bigg[
	\begin{array}{c}
	-\sin\phi_j \\
	\cos\phi_j \\
	\end{array}
	\bigg]
	\bigg)
	\Bigg) \nonumber
\end{align}
\hrulefill
\vspace*{1mm}
\setcounter{equation}{51}
\begin{align}\label{eq:4_1_2_3}
	{\bf{J}}_{\text{e}}^{\text{UCA}}({\bf{p}}) 
	=N_{\text{a}}\Bigg(\sum_{j\in\mathcal{N}_{\text{L}}}\lambda_j\beta^2{\bf J}_{\text{\rm r}}(\phi_j)+\frac{1}{2}f_{\text{c}}^2\sum_{i,j\in\mathcal{N}_{\text{L}}} &\lambda_i\lambda_j
	\bigg(\frac{D^2}
	{8\sum_{k\in\mathcal{N}_{\text{L}}}\lambda_k}+\frac{v^2t_{\text{\rm rms}}^2\sin^2(\phi_i-\psi_{\text{\rm d}})\sin^2(\phi_j-\psi_{\text{\rm d}})}
	{\sum_{k\in\mathcal{N}_{\text{L}}}\lambda_k\sin^2(\phi_k-\psi_{\text{\rm d}})}\bigg) \nonumber \\
	& \times
	{\bf g}\bigg(\frac{1}{D_i}\bigg[
	\begin{array}{c}
	-\sin\phi_i \\
	\cos\phi_i \\
	\end{array}
	\bigg]
	- \frac{1}{D_j}\bigg[
	\begin{array}{c}
	-\sin\phi_j \\
	\cos\phi_j \\
	\end{array}
	\bigg]
	\bigg)
	\Bigg)
\end{align}
	\setcounter{equation}{\value{mytempeqncnt}}
	\vspace*{4pt}
\end{figure*}

By the preceding definition, we can rewrite the SAAF as
\begin{align}\label{eq:SAAF_equiv}
\begin{split}
  G(\theta) = \widetilde{{\bf x}^2}\sin^2\theta + \widetilde{{\bf y}^2}\cos^2\theta- \widetilde{\bf xy}\sin2\theta
\end{split}
\end{align}
and thus the SAAF for UOA and UCOA does not depend on the incident angle $\theta$, which will be abbreviated as $G_{\text{UOA}}$ and $G_{\text{UCOA}}$, respectively. Hence, UOAs possess such a symmetry that the intensity of the direction information obtained by a UOA is invariant with its orientation. UCOAs are more symmetric and has the largest average SAAF, as shown in the following proposition.
\begin{proposition}[{Largest Average SAAF}]\label{Pro_Geo}
  For any array with diameter $D$, UCOAs possess the largest SAAF, i.e.,
  \begin{align}\label{eq:4_1_3_3}
    \frac{1}{2\pi}\int_0^{2\pi} G(\theta) d\theta \le \frac{D^2}{8} = G_{\text{\rm UCOA}}.
  \end{align}
\end{proposition}
\begin{IEEEproof}
  See Appendix \ref{PF_Pro_Geo}.
\end{IEEEproof}


To obtain the optimal array-antenna geometry, we consider two widely-used criteria, i.e., the lowest \emph{average} SPEB and the lowest \emph{worst-case} SPEB, which correspond to the Bayes estimator under the non-informative prior and the minimax estimator in statistics, respectively. 
These two criteria characterize the average and worst-case performance of the array localization system over different array orientations.
The next theorem shows that, when both the array orientation and its velocity are known, the UCOA meets these two optimality criteria and thus is suggested to be used in practice.
\begin{theorem}[{Optimal Array-antenna Geometry}]\label{TH_array_geo}
   When both the orientation and velocity are known, UCOA has the lowest {average and worst-case} SPEB over all arrays with an identical diameter, i.e.,
 \begin{align}
     \text{\rm SPEB}^{\text{\rm UCOA}} \le \frac{1}{2\pi}\int_0^{2\pi}\text{\rm SPEB}(\psi)d\psi \le \sup_{\psi\in[0,2\pi]}\text{\rm SPEB}(\psi)
  \end{align}
where $\text{\rm SPEB}^{\text{\rm UCOA}}$ denotes the orientation-invariant SPEB obtained by UCOA.
\end{theorem}
\begin{IEEEproof}
  See Appendix \ref{PF_TH_array_geo}.
\end{IEEEproof}

{We next focus on two specific examples, i.e., we restrict the array-antenna geometry to two simple but practically useful cases: uniform linear array (ULA) and uniform circular array (UCA).} A summary of all special arrays is listed in Fig. \ref{Fig.ArrGeo}.

\subsubsection{ULA}
In ULA, all antennas are placed on a line with
\begin{align}\label{eq:4_1_1_1}
  {\bf{p}}_{k}^{\text{\rm Array}}(\psi) = {\bf{p}} + \frac{D}{N_{\text{a}}-1}\cdot\Big(k-\frac{N_{\text{a}}+1}{2}
  \Big)\left[
           \begin{array}{c}
             \cos\psi \\
             \sin\psi \\
           \end{array}
         \right]
\end{align}
where ${\bf{p}}$ is the {array} coordinate center and $D$ is the array diameter. Then the squared array aperture function can be easily derived as
\begin{align}\label{eq:4_1_1_2}
  G_{\text{ULA}}(\theta) &= \frac{1}{N_{\text{a}}^2}\bigg(\frac{D}{N_{\text{a}}-1}\bigg)^2\sum_{1\le k<l\le N_{\text{a}}}(k-l)^2\sin^2\theta \nonumber \\
  &= \frac{N_{\text{a}}+1}{12(N_{\text{a}}-1)}D^2\sin^2\theta \,.
\end{align}

Hence $G_{\text{ULA}}(\theta)$ is determined by the antenna number, the array diameter and the incident angle. Specifically, $G_{\text{ULA}}(\theta)$  increases with the physical length of ULA and achieves its maximum when the incident angle is perpendicular to the array orientation.
The EFIM for the position with unknown array orientation and moving direction is given by (\ref{eq:4_1_1_3}), shown at the bottom of  \blue{this page}.
\addtocounter{equation}{1}

\subsubsection{UCA}
In UCA, all antennas form a regular polygon {with}
\begin{align}\label{eq:4_1_2_1}
  {\bf{p}}_{k}^{\text{\rm Array}}(\psi) = {\bf{p}} + \frac{D}{2}
  \left[
           \begin{array}{c}
             \cos(\psi+\frac{2k\pi}{N_{\text{a}}}) \\
             \sin(\psi+\frac{2k\pi}{N_{\text{a}}}) \\
           \end{array}
         \right] \,.
\end{align}
To distinguish UCA with ULA, we assume that $N_{\text{a}}\ge3$. The SAAF is
$
  G_{\text{UCA}}(\theta) = {D^2}/{8}
$
based on Proposition \ref{Pro_Geo} and the fact that UCA is a special case of UCOA. Unlike ULA, the SAAF of UCA is determined purely by the array diameter and is invariant with $\theta$. The expression for ${\bf{J}}_{\text{e}}^{\text{UCA}}({\bf{p}})$ is given by (\ref{eq:4_1_2_3}), shown at the bottom of \blue{this page.}
\addtocounter{equation}{1}
\subsubsection{Comparison}
Comparing the localization accuracy using ULA and UCA based directly on (\ref{eq:4_1_1_3}) and (\ref{eq:4_1_2_3}) is complicated, so the alternative method is to compare their SAAFs. Given an identical diameter for these arrays, it can be shown that
\begin{align}\label{eq:4_1_3_1}
  \frac{G_{\text{ULA}}(\theta)}{G_{\text{UCA}}(\theta)} = \frac{2(N_{\text{a}}+1)}{3(N_{\text{a}}-1)}\sin^2\theta.
\end{align}

If $N_{\text{a}}\ge5$, then $G_{\text{UCA}}(\theta)\ge G_{\text{ULA}}(\theta)$ for all incident direction $\theta$, which indicates that UCA always outperforms ULA {when there are more than four antennas in the array}. If $N_{\text{a}}=3,4$, then whether $G_{\text{UCA}}(\theta)\ge G_{\text{ULA}}(\theta)$ or not depends on the incident direction $\theta$. However, note that
\begin{align}\label{eq:4_1_3_2}
  \frac{1}{2\pi}\int_0^{2\pi} G_{\text{ULA}}(\theta)d\theta = \frac{N_{\text{a}}+1}{24(N_{\text{a}}-1)}D^2 \le G_{\text{UCA}}(\theta)
\end{align}
i.e., the expected $G_{\text{ULA}}(\theta)$ does not exceed $G_{\text{UCA}}(\theta)$. In other words, UCA outperforms ULA in {average given} that the incident direction is uniformly distributed on $[0,2\pi)$. This result confirms Proposition \ref{Pro_Geo}.

\begin{figure}[!t]
        \subfigure[UOA]{
            \includegraphics[scale=0.32]{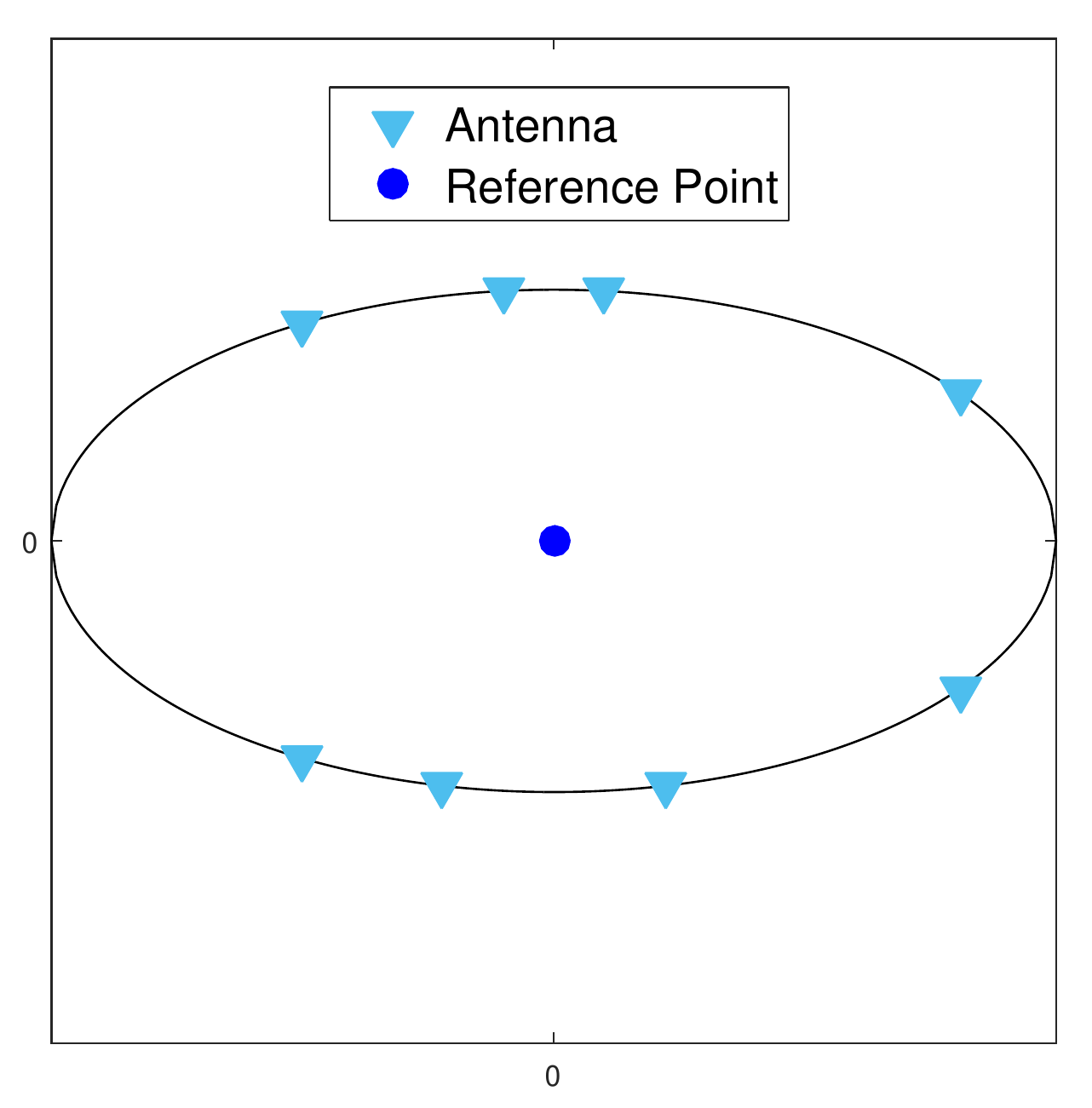}\\
            }
        \subfigure[UCOA]{
            \includegraphics[scale=0.32]{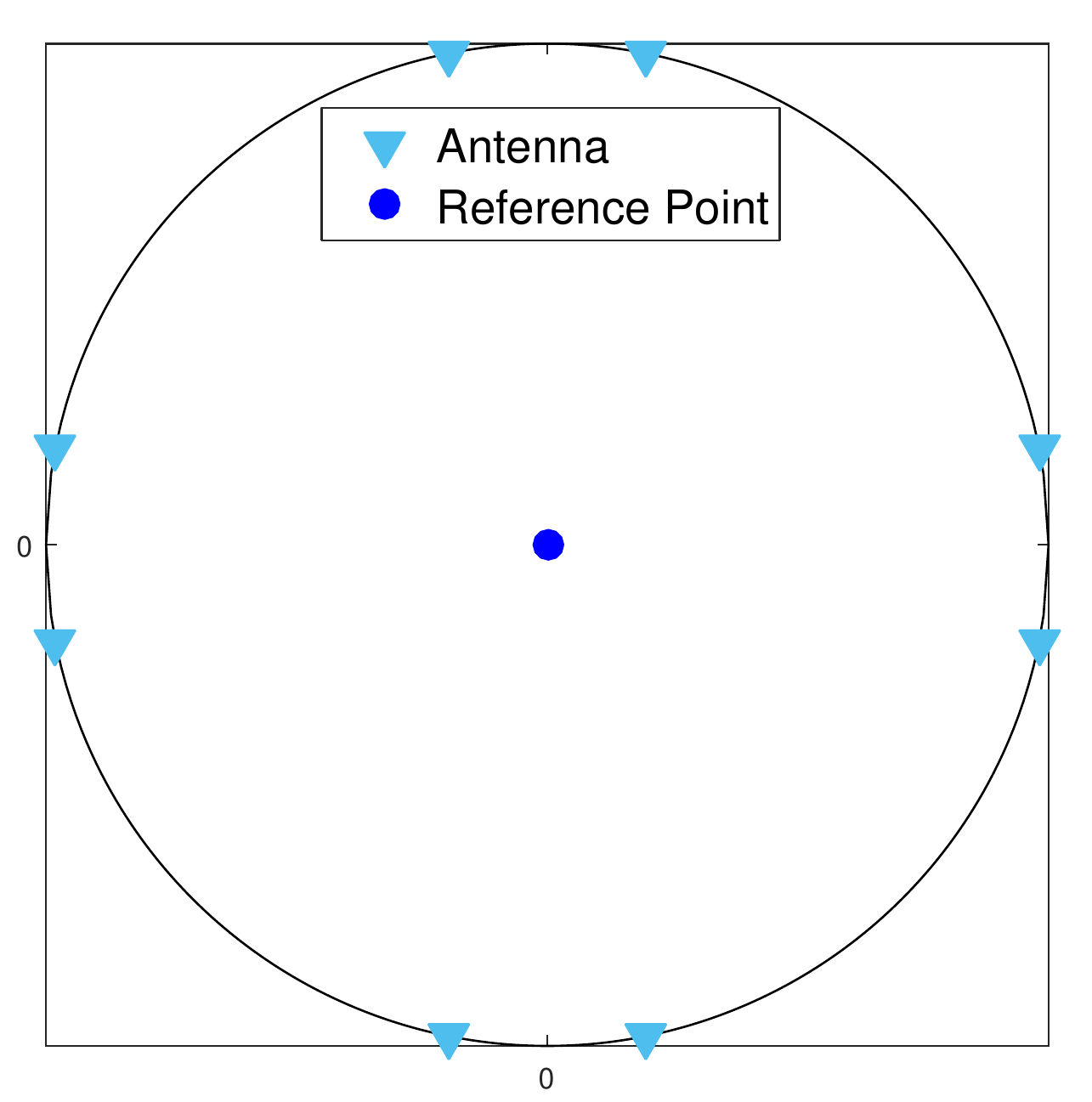}\\
            }
        \subfigure[ULA]{
            \includegraphics[scale=0.32]{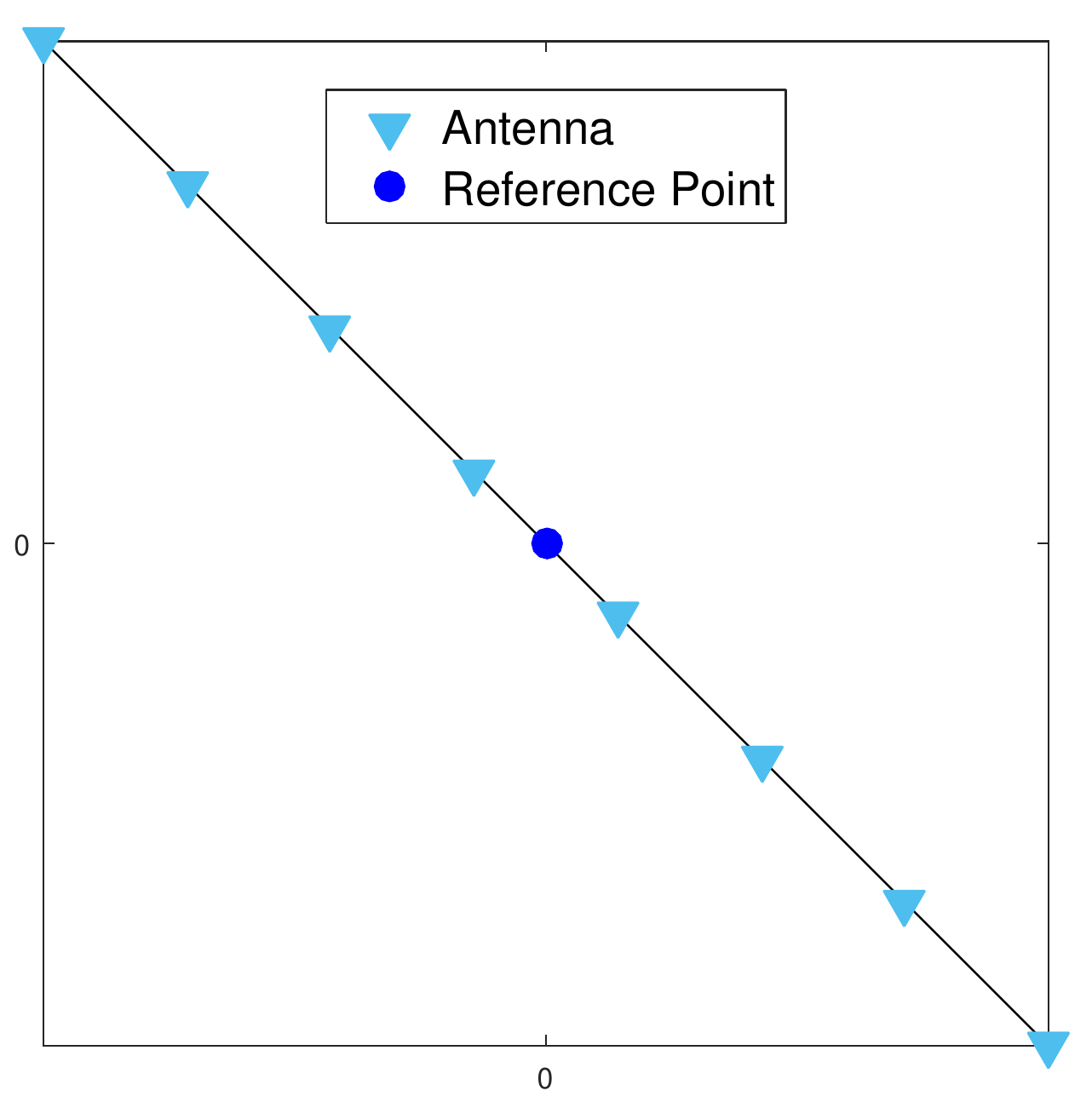}\\
            }
        \subfigure[UCA]{
            \includegraphics[scale=0.32]{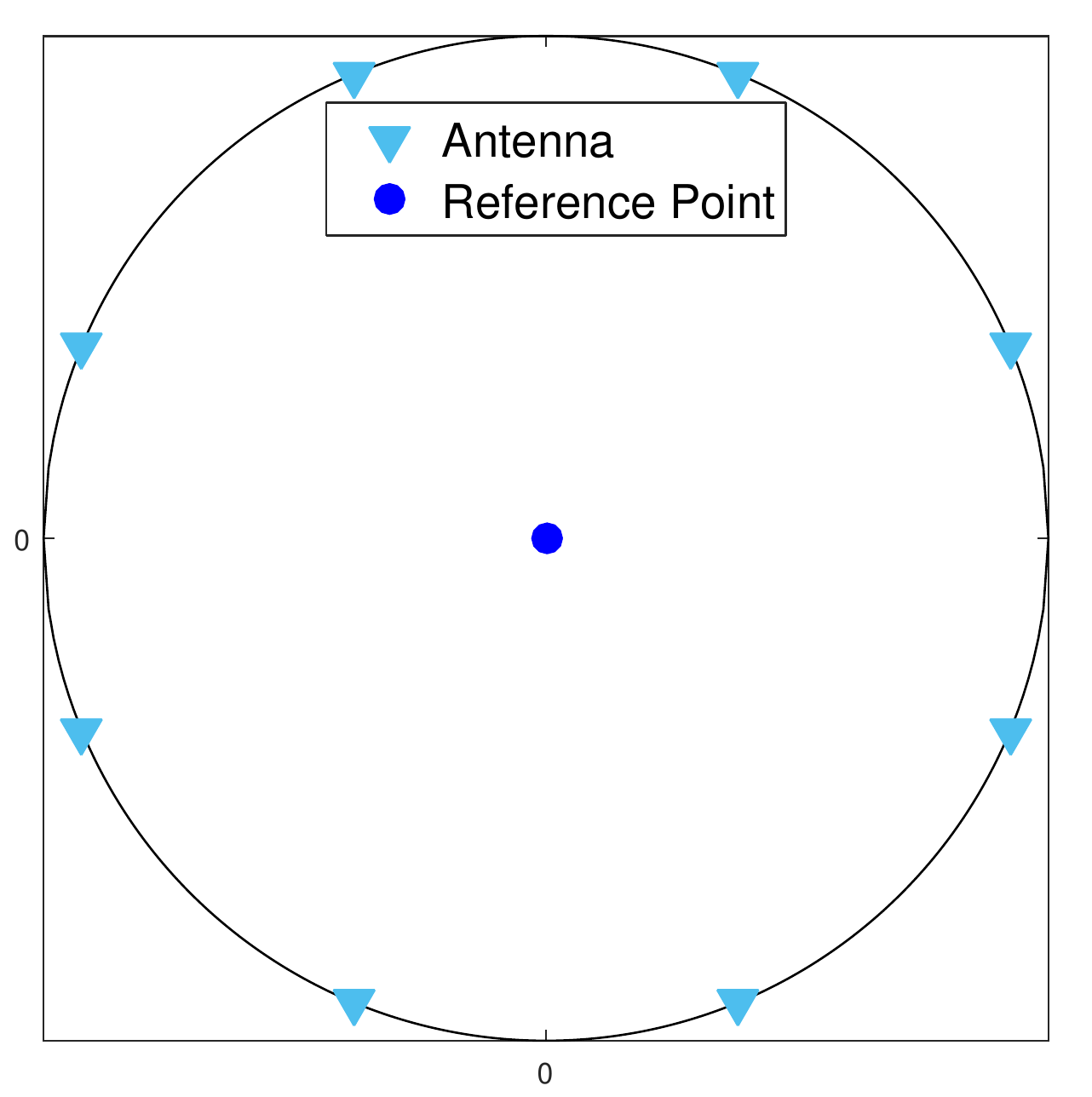}\\
            }
        \caption{Examples of special arrays, i.e., UOA, UCOA, ULA and UCA. Triangles represent the position of antennas, and the circle refers to the position of the reference point.}
        \label{Fig.ArrGeo}
\end{figure}

\subsection{Effects of Anchor-agent Geometry}\label{sec_geometry_anchor}
Once the array-antenna geometry is fixed, the anchor-agent geometry is a significant factor on the SPEB. In the following, we neglect the trivial fact that {a smaller $D_j$ is preferable and consider only the effect of} anchor-to-agent directions. Intuitively, if all $\phi_j$'s are close to each other, the localization information from the radial or tangent direction will be small (depending on the dominance between distance information and direction information), resulting in a low localization accuracy. On the contrary, if all $\phi_j$'s distribute uniformly on $[0,\pi]$, higher accuracy can be expected. {To find the optimal anchor-agent geometry}, we take the infimum of the SPEB over all possible $\phi_j$'s {while fixing other variables, e.g., distances and SNRs}. The result is summarized in the following theorem.
\begin{theorem}[{Optimal Anchor-agent Geometry}]\label{TH_anchor_geo}
  Given a UOA and a static scenario,
  the optimal anchor-agent geometry in the orientation-known case is given by
  \begin{align}\label{eq:geo_anc_1}
    \sum_{i\in \mathcal{N}_{\text{\rm L}}}u_i\exp(j2\phi_i) = 0
  \end{align}
 where $u_i \triangleq \lambda_i(\beta^2 - f_{\text{\rm c}}^2{G_{\text{\rm UOA}}}/{D_i^2})$,
  while in the orientation-unknown case, the optimal anchor-agent geometry is given by both (\ref{eq:geo_anc_1}) and
  \begin{align}\label{eq:geo_anc_2}
    \sum_{i\in \mathcal{N}_{\text{\rm L}}}\frac{\lambda_i}{D_i}\exp(j\phi_i) = 0 \,.
  \end{align}
\end{theorem}
\begin{IEEEproof}
  See Appendix \ref{PF_TH_anchor_geo}.
\end{IEEEproof}
\begin{remark}
In light of Theorem \ref{TH_array_geo}, the UOA assumption is reasonable for the successive optimization first over the array-antenna geometry and then over the anchor-agent geometry. Note that we have also shown in the proof that the orientation-known SPEB is a strictly increasing function of $|\sum_{i\in \mathcal{N}_{\text{\rm L}}}u_i\exp(j2\phi_i)|$.
\end{remark}

We can draw the following observations from Theorem \ref{TH_anchor_geo}.
\begin{itemize}
	\item In the static orientation-known case, the optimal choice for $\phi_j$'s requires that all direction vectors with different intensities be offset by each other, cf. Fig. \ref{AnchorGeo} with $|\mathcal{N}_{\text{L}}|=3$. Hence, $\phi_j$'s should be \emph{diversified} for a high localization accuracy.
	\item A special scenario occurs in the static orientation-known case when $\beta D_i = f_{\text{\rm c}} \sqrt{G^{\text{UOA}}}$ for some $i$. Under this scenario, $\phi_i$ has no impact on SPEB as $u_i=0$, i.e., the distance and direction information exactly offset each other. Consequently, the measuring ellipse in Fig. \ref{fig:fisherinfo} becomes a circle, yielding an \emph{isotropic} localization with respect to anchor direction.
	\item In the static scenario when the orientation is unknown, we need an additional condition to offset the accuracy degradation caused by unknown orientation, which is generally achievable if there are more than three anchors because (\ref{eq:geo_anc_1}) and (\ref{eq:geo_anc_2}) impose four constraints in total.
\end{itemize}
\begin{figure}[!t]
\centering
\includegraphics[scale=0.4]{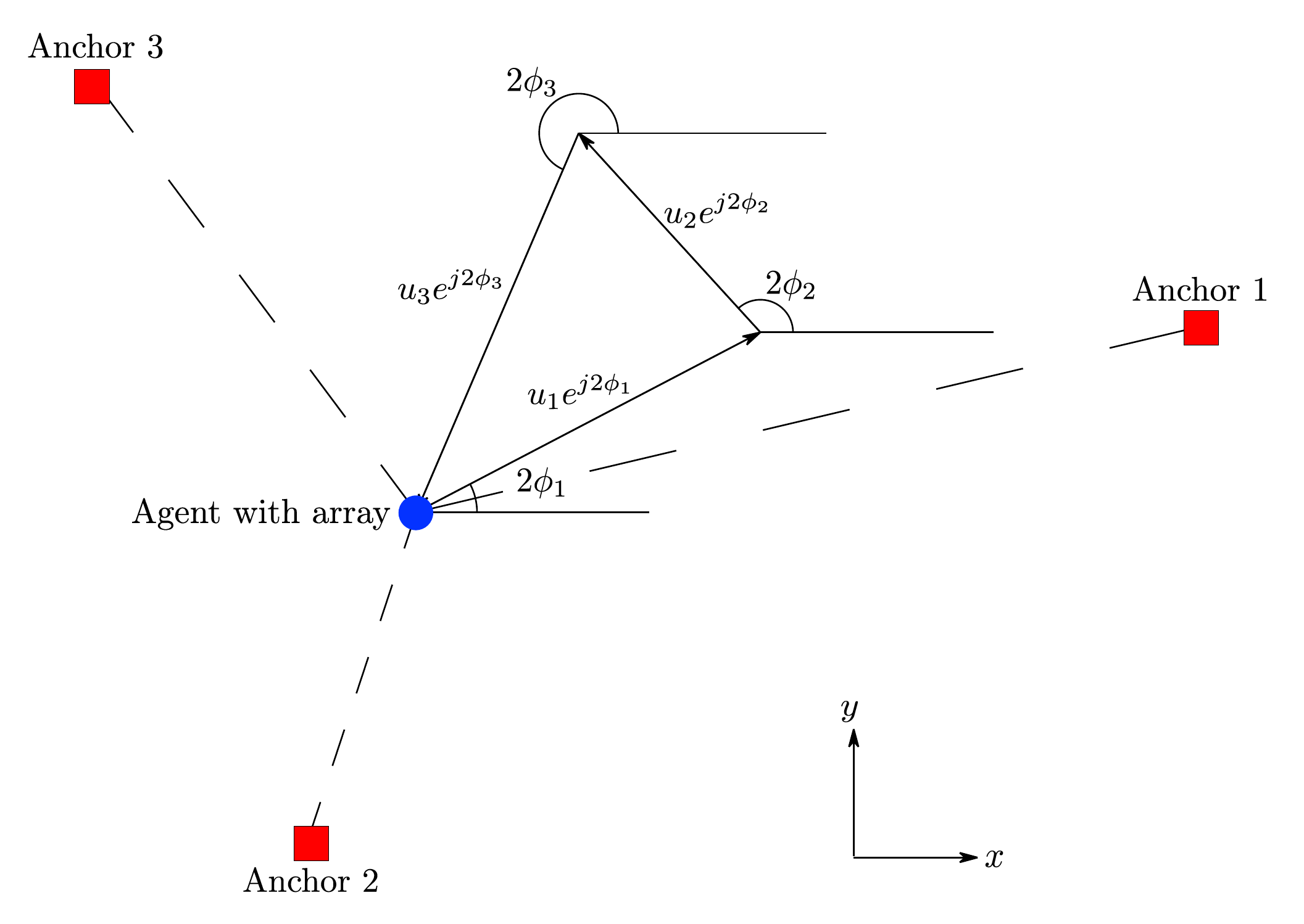}
\caption{The optimal anchor-agent geometry, where $\sum_{i\in\mathcal{N}_{\text{L}}} u_i\exp(j2\phi_i)=0$ with $|\mathcal{N}_{\text{L}}|=3$.}
\label{AnchorGeo}
\end{figure}

Note that Theorem \ref{TH_anchor_geo} not only provides an optimality criterion for anchor-agent geometry, but also suggests {two measures to characterize the anchor-agent geometry, i.e., the first- and the second-type anchor geometric factors
\begin{align}
\label{eq:GF_1}
\text{GF}_1 &\triangleq \Big|\sum_{i\in\mathcal{N}_{\text{L}}}u_i\exp(j2\phi_i)\Big|\\
\label{eq:GF_2}
\text{GF}_2 &\triangleq \Big|\sum_{i\in\mathcal{N}_{\text{L}}}\frac{\lambda_i}{D_i}\exp(j\phi_i)\Big|
\end{align}
where $\text{GF}_1$ \emph{fully} characterizes the effect of anchor-agent geometry on localization in the orientation-known case, while in the orientation-unknown case, the second-type anchor geometric factor $\text{GF}_2$ needs to be introduced to characterize the performance degradation caused by the unknown orientation}.

Now we draw some connections with other existing measures such as the geometric dilution of precision (GDOP) \cite{Zhu1992Calculation}. Consider an alternative expression for the orientation-known EFIM in (\ref{eq:EFIM_ori_known_static}) as ${\bf J}_{\text{e}}({\bf p}) = {\bf H^T}{\bf C}^{-1}{\bf H}$, where $\bf H$ is a $2|\mathcal{N}_{\text{L}}|\times 2$ matrix with $(2j-1,2j)$-th rows expressed as
\begin{align}\label{eq:GDOP_H}
\begin{split}
  [{\bf H}]_{2j-1:2j,1:2}&\triangleq \left[
                                \begin{array}{cc}
                                  \cos\theta_j & \sin\theta_j \\
                                  -D_j^{-1}\sin\theta_j & D_j^{-1}\cos\theta_j\\
                                \end{array}
                              \right]
  \end{split}
\end{align}
and $\bf C$ is a $2|\mathcal{N}_{\text{L}}| \times 2|\mathcal{N}_{\text{L}}|$ square matrix given by
\begin{align}
\label{eq:GDOP_C}
  {\bf C} &\triangleq \text{diag}(\bm{\Lambda}_1, \bm{\Lambda}_2, \cdots, \bm{\Lambda}_{|\mathcal{N}_{\text{L}}|})\\
  \bm{\Lambda}_j &\triangleq \frac{1}{N_{\text{a}}\lambda_j}\left[
            \begin{array}{cc}
              \beta^{-2} & 0 \\
              0 & (f_{\text{c}}^2G_{\text{UOA}})^{-1} \\
            \end{array}
          \right].
\end{align}

If we interpret $\bf C$ as the covariance matrix for all distance and direction metrics $(D_j,\phi_j)$, the standard GDOP approach \cite{wang2012efficient} can be applied to express the standard positioning error $\sigma_{\bf p}$ as
$
  \sigma_{\bf p} = \sqrt{\text{tr}\big\{\left({\bf H^T}{\bf C}^{-1}{\bf H}\right)^{-1}\big\}}
$,
which coincides with the root SPEB. Moreover, if all diagonal elements of $\bf C$, i.e., the measurement errors, are assumed to be equal (denoted by $\sigma_{\text{n}}^2$), then GDOP is defined as the ratio of standard positioning error to the measurement error
\begin{align}\label{eq:GDOP_expression}
  \text{GDOP} \triangleq \frac{\sigma_{\bf p}}{\sigma_{\text{n}}} = \sqrt{\text{tr}\big\{\left({\bf H^T}{\bf H}\right)^{-1}\big\}}.
\end{align}

By definition, the anchor-agent geometry with low GDOP is geometrically preferable. The minimization of GDOP in (\ref{eq:GDOP_expression}) yields 
$
  \sum_{i\in\mathcal{N}_{\text{L}}} \exp(j2\phi_i) = 0
$ \cite{wang2012efficient},
which is a special case for constant $u_i$ in (\ref{eq:geo_anc_1}). Hence, the measure $\text{GF}_1$ can be treated as a generalized version of the traditional GDOP in the orientation-known case, where different information intensities can be utilized to impose different weights on anchors. Moreover, in the static orientation-unknown case, we need also $\text{GF}_2$ to characterize the accuracy degradation caused by the unknown orientation{, which is not covered in the traditional GDOP approach.} In conclusion, by dropping the simplified assumption used by traditional GDOP that all measurement errors are equal, Theorem \ref{TH_anchor_geo} provides some generalized criteria for the optimal anchor-agent geometry, as well as two new measures to compare different anchor-agent geometry.
\begin{figure}[!t]
  \centering
  \includegraphics[width=3in]{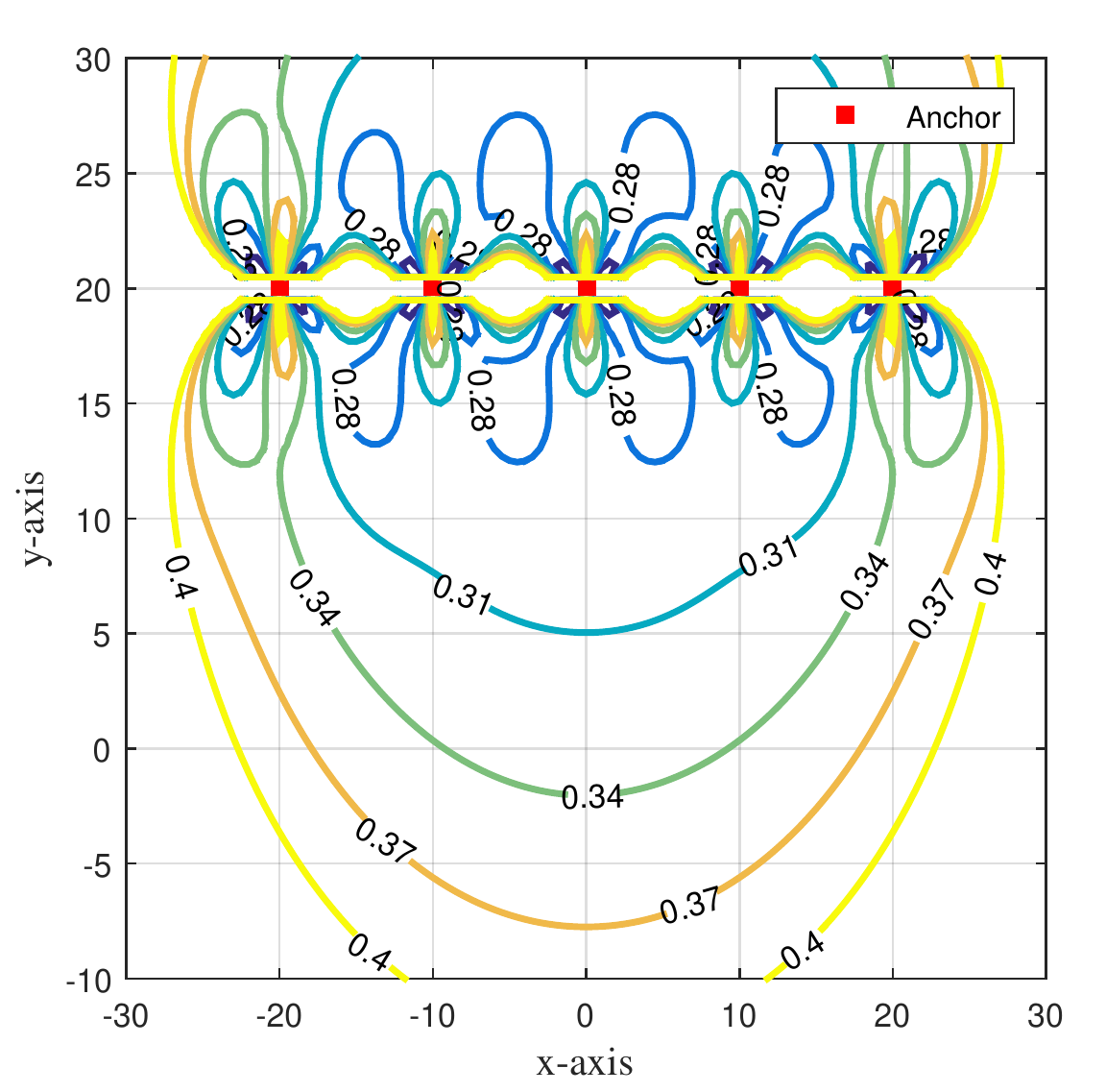}\\
  \caption{Root SPEB contours with five anchors placed on a line and static ULA with {$N_{\text{a}}=6,\beta=1\thinspace \text{MHz}, f_\rc=100\thinspace \text{MHz}$, $\mathsf{SNR}_j^{(1)}=30$\thinspace dB for all $j\in\mathcal{N}_{\text{b}}$. The array is \emph{parallel} to anchors and the units are all meters}.}\label{ExContour1}
\end{figure}

\section{Numerical Results}\label{sec_numerical}
This section presents numerical results evaluating the localization performance with respect to various system parameters, array-antenna geometry, and anchor-agent geometry.
\subsection{Determinants for SPEB}
We first check two determinants for SPEB, i.e., the absolute position and orientation which are both to be estimated. We place five identical anchors on $(-20\thinspace\text{m},20\thinspace\text{m})$,  $(-10\thinspace\text{m},20\thinspace\text{m})$, $\cdots$, $(20\thinspace\text{m},20\thinspace\text{m})$ and a ULA with diameter $D=0.5$\thinspace m and $N_{\text{a}}=6$. We consider a channel model with $\mathsf{SNR}_j^{(1)}=30$\thinspace dB for all $j\in\mathcal{N}_{\text{b}}$ and no multipath or NLOS, and a signal model with $\beta=1\thinspace \text{MHz}, f_\rc=100 \thinspace \text{MHz}$ and $\gamma=0$.} Given that the array diameter is small, it is reasonable to adopt static far-field environments here.

\begin{figure}[!t]
  \centering
  \includegraphics[width=3in]{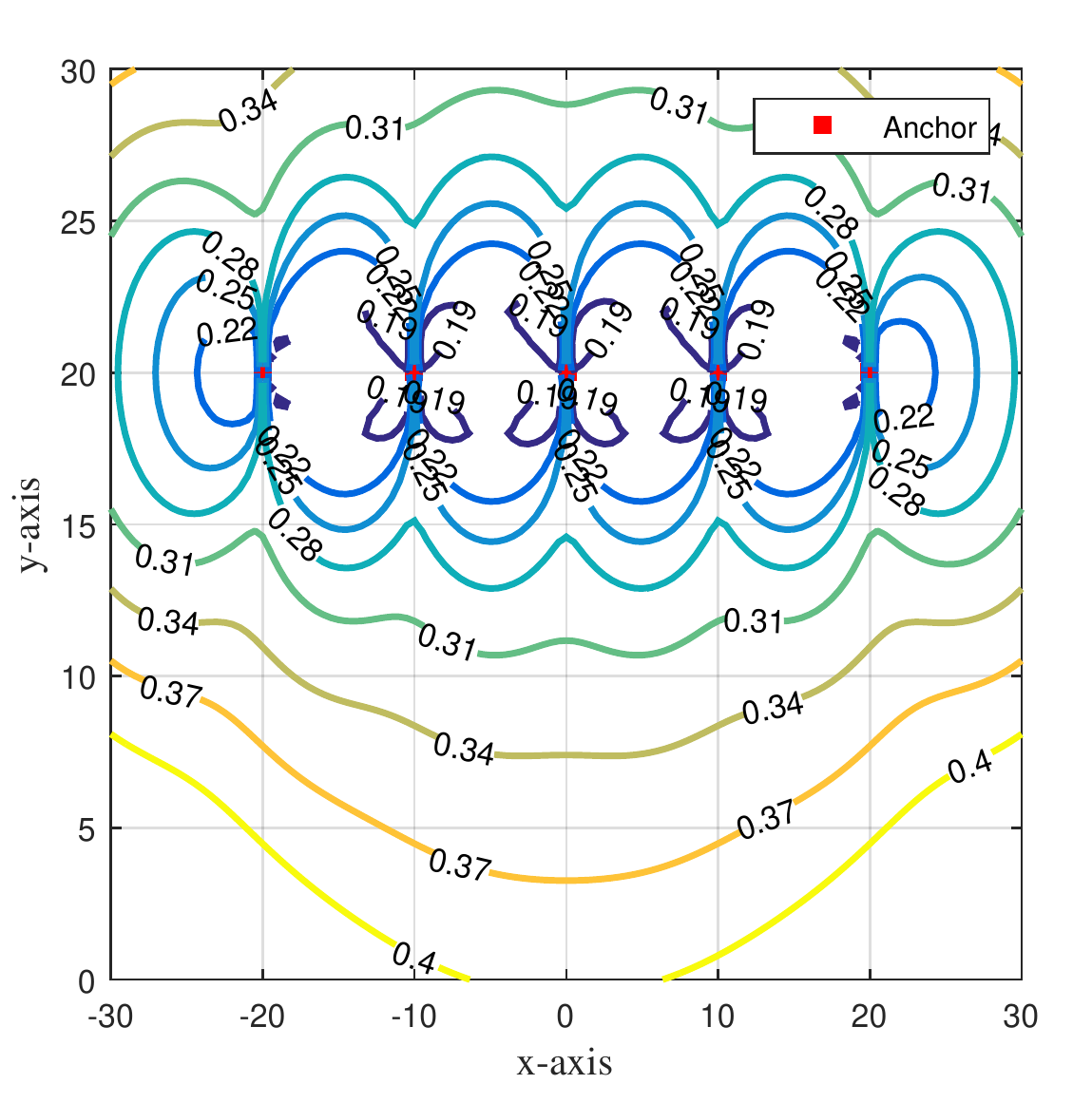}\\
  \caption{Root SPEB contours with five anchors placed on a line and static ULA with {$N_{\text{a}}=6,\beta=1\thinspace \text{MHz}, f_\rc=100\thinspace\text{MHz}$, $\mathsf{SNR}_j^{(1)}=30$\thinspace dB for all $j\in\mathcal{N}_{\text{b}}$. The array is \emph{perpendicular} to anchors and the units are all meters}.}\label{ExContour2}
\end{figure}
We consider two typical array orientations: in the first case, the array is parallel to anchors, i.e., antennas have identical $y$-coordinates, while in the second case, the array is perpendicular to anchors, i.e., antennas have identical $x$-coordinates. Any other case can be seen as a combination of the two.  Figs.~\ref{ExContour1} and \ref{ExContour2} show the root SPEB contours in these two cases with known array orientation.

When the array is parallel to the anchors, the line on which anchors are placed yields the lowest accuracy since SAAF is zero on that line. Moreover, there are valleys for root SPEB where the anchor-agent geometry and distance are balanced for localization. Meanwhile, when the array is perpendicular to the anchors, there are peaks when the array points to one of the anchors for similar SAAF reasons, but the line {which the anchors are placed on} no longer yields the lowest accuracy. Moreover, {in Fig. \ref{ExContour2} it seems that} arbitrarily high localization accuracy can be obtained by placing the array close to some anchor, but {we remark that the previous results cannot be applied here due to the contradiction to the far-field assumption}.

\begin{figure}[!t]
  \centering
  \includegraphics[width=3in]{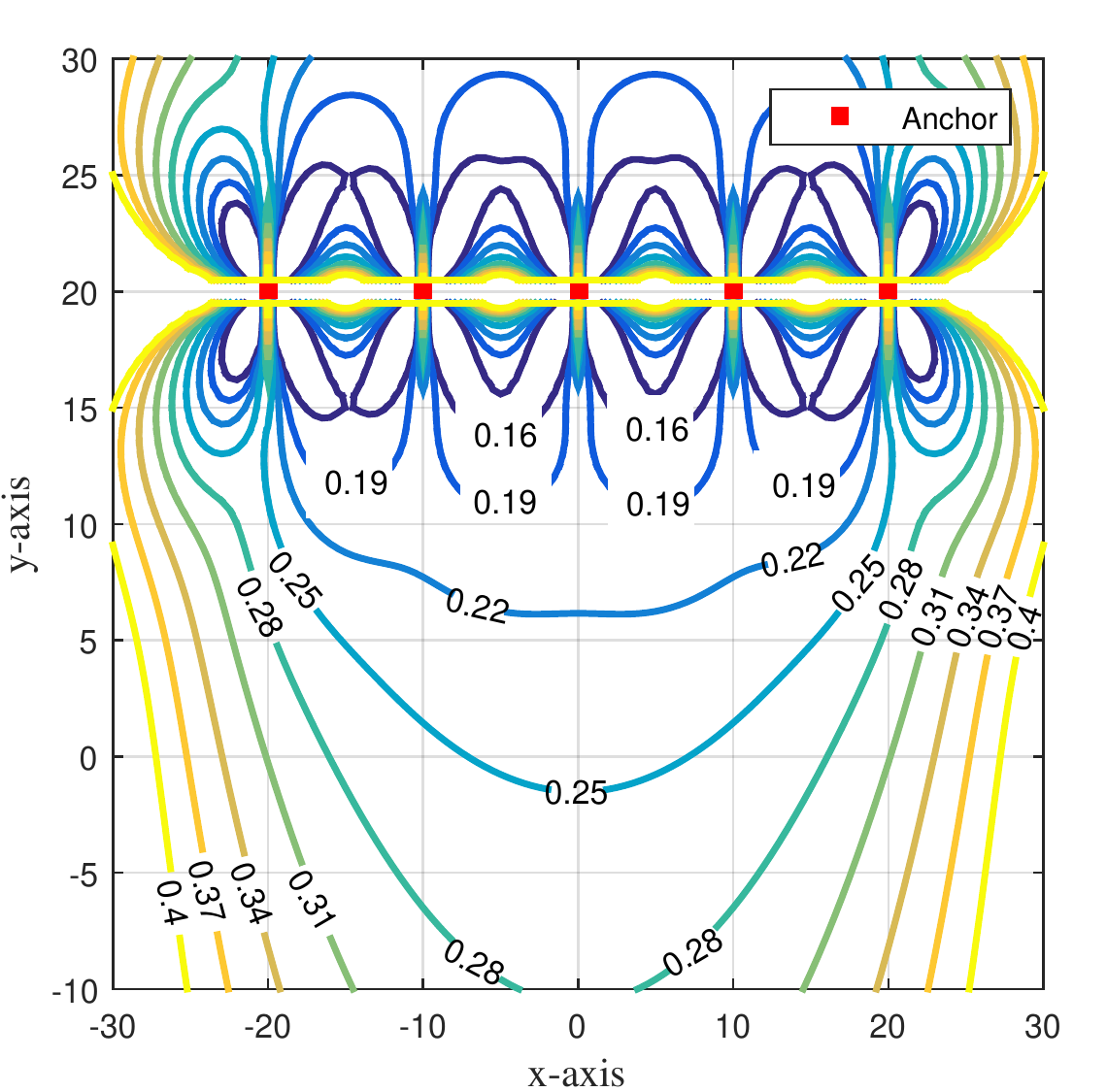}\\
  \caption{Root SPEB contours with five anchors placed on a line and moving ULA with $N_{\text{a}}=6$ and a known velocity {$v=30$\thinspace m/s along the array orientation. Other parameters are $\beta=1\thinspace\text{MHz}, f_\rc=100\thinspace\text{MHz}$, $t_{\text{rms}}=10$\thinspace ms and $\mathsf{SNR}_j^{(1)}=30$\thinspace dB for all $j\in\mathcal{N}_{\text{b}}$. The array is parallel to anchors and the units are all meters}.}\label{ExContour_Doppler}
\end{figure}
Now we turn to the {dynamic scenario where the agent is moving at $v=30$\thinspace m/s along its orientation, and the root mean squared time duration of the signal is $t_{\text{rms}}=10$\thinspace ms}. Fig. \ref{ExContour_Doppler} plots the corresponding root SPEB contours, where for simplicity only the first scenario that anchors are parallel to the array is considered here. Compared with Fig. \ref{ExContour1}, there is a remarkable gain in the localization {accuracy} due to the involvement of the Doppler shift.

Then we examine the effects of array orientation on SPEB by fixing the array position at ${\bf{p}}=(0,0)$ and changing its orientation $\psi$ from 0 (parallel to anchors) to $\pi/2$ (perpendicular to anchors) with {different $\beta = 10\thinspace \text{KHz}, 100\thinspace \text{KHz}, 1$\thinspace MHz and a constant $f_{\text{\rm c}}=100$\thinspace MHz. The array is assumed to be static. Fig.~\ref{ExOrient} illustrates the relationship between root SPEB and orientation in two cases, i.e., with known or unknown array orientation. 

We have the following observations from Fig.~\ref{ExOrient}}. When the baseband bandwidth is large, root SPEB is almost invariant with orientation, and the reverse conclusion holds for small $\beta$. This is attributed to the fact that with a large $\beta$, distance information is the main source of localization information which is invariant with array orientation and vice versa. In addition, the {dashdot} lines are lower than solid lines, indicating that unknown orientation will degrade localization accuracy. Moreover, when $\beta$ decreases with $f_{\text{\rm c}}$ fixed, the root SPEB in all three cases rise but tend to converge, indicating that {in contrast to the wideband cases,} AOA can provide information alone when there is no available TOA information, as we proved in Section \ref{sec_EFIM}.

\begin{figure}[!t]
  \centering
  \includegraphics[scale=0.6]{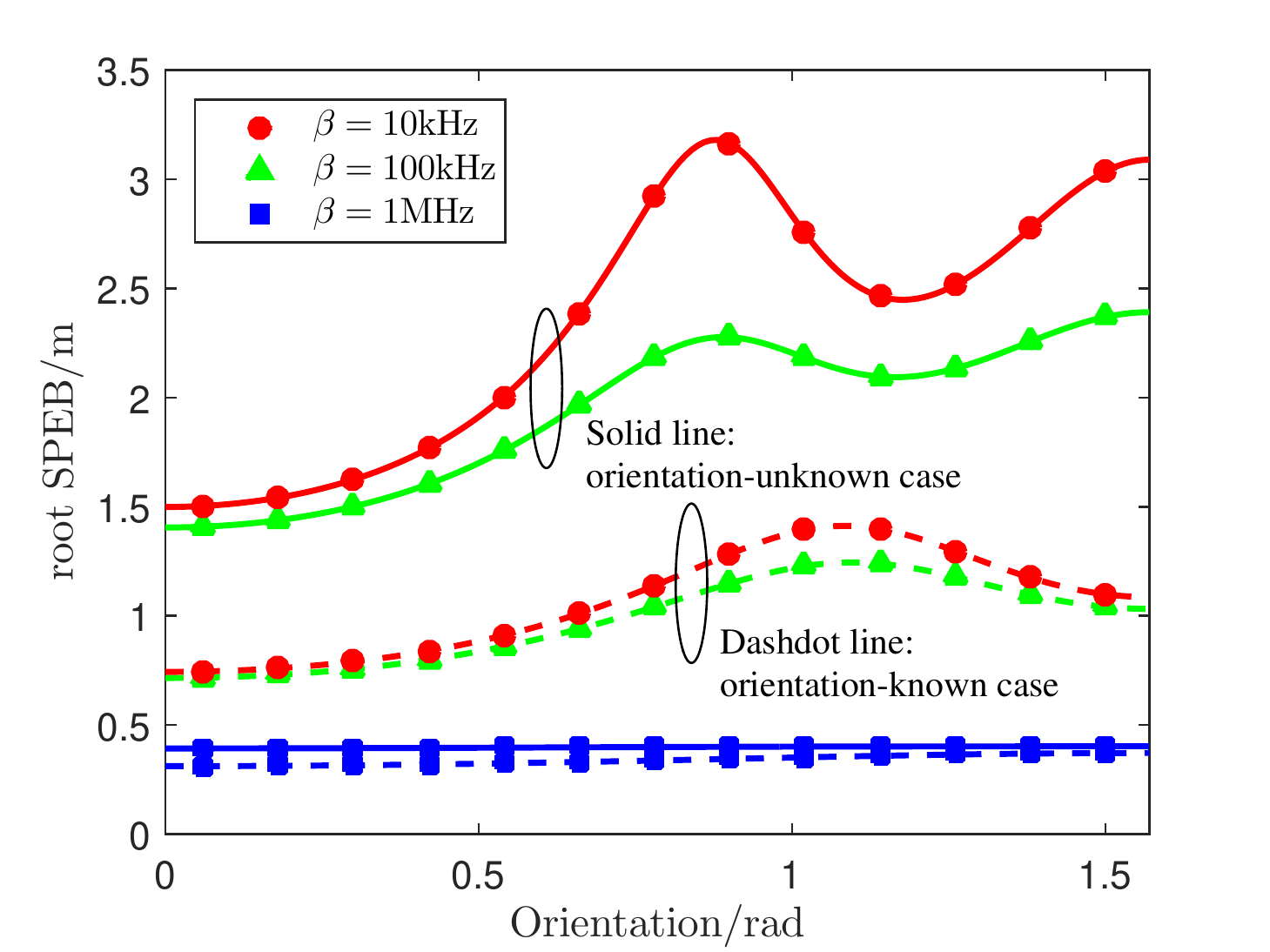}\\
  \caption{{Root SPEB of a static ULA as a function of orientation $\psi$ with fixed $f_{\text{\rm c}}=100$\thinspace MHz. The reference point is $(0,0)$, and $N_\rb=5, N_\ra=6,\mathsf{SNR}_j^{(1)}=30$\thinspace dB for all $j\in\mathcal{N}_{\text{b}}$. Solid line represents the orientation-unknown case, while dashdot line represents the orientation-known case. Curves with different line types correspond to different baseband bandwidth $\beta = 10\thinspace\text{KHz}, 100\thinspace\text{KHz}, 1$\thinspace MHz, respectively.}}\label{ExOrient}
\end{figure}

\subsection{Anchor-agent Geometry}
 The setting of our experiments is as follows. There are 4 anchors providing LOS signals with $f_{\text{c}}=200$\thinspace MHz, and the distance between each anchor and array is $D_j=50$\thinspace m with {different SNRs, i.e., 20\thinspace dB, 25\thinspace dB, 30\thinspace dB and 35\thinspace dB}. The anchor directions are randomly set according to a {uniform} distribution in $[\,0,\pi)$. As for the array, we set $N_{\text{a}}=6$ and consider a static UCA with diameter $D=1$\thinspace m, and denote the constant SAAF by $G_{\text{UCA}}$. In addition, $D_j\gg D$ entails the far-field environments here.

To investigate the relationship between SPEB and the anchor-agent geometry, we observe how SPEB varies with the measures provided in Theorem \ref{TH_anchor_geo}, i.e., two anchor geometric factors given by (\ref{eq:GF_1}) and (\ref{eq:GF_2}) {with a further normalization into the unit interval $[0,1]$.} Fig. \ref{ExAnchor1} and Fig. \ref{ExAnchor2} display this relationship in both orientation-known and orientation-unknown cases, respectively, under 10,000 Monte Carlo simulations.

\begin{figure}[!t]
  \centering
  \includegraphics[scale=0.6]{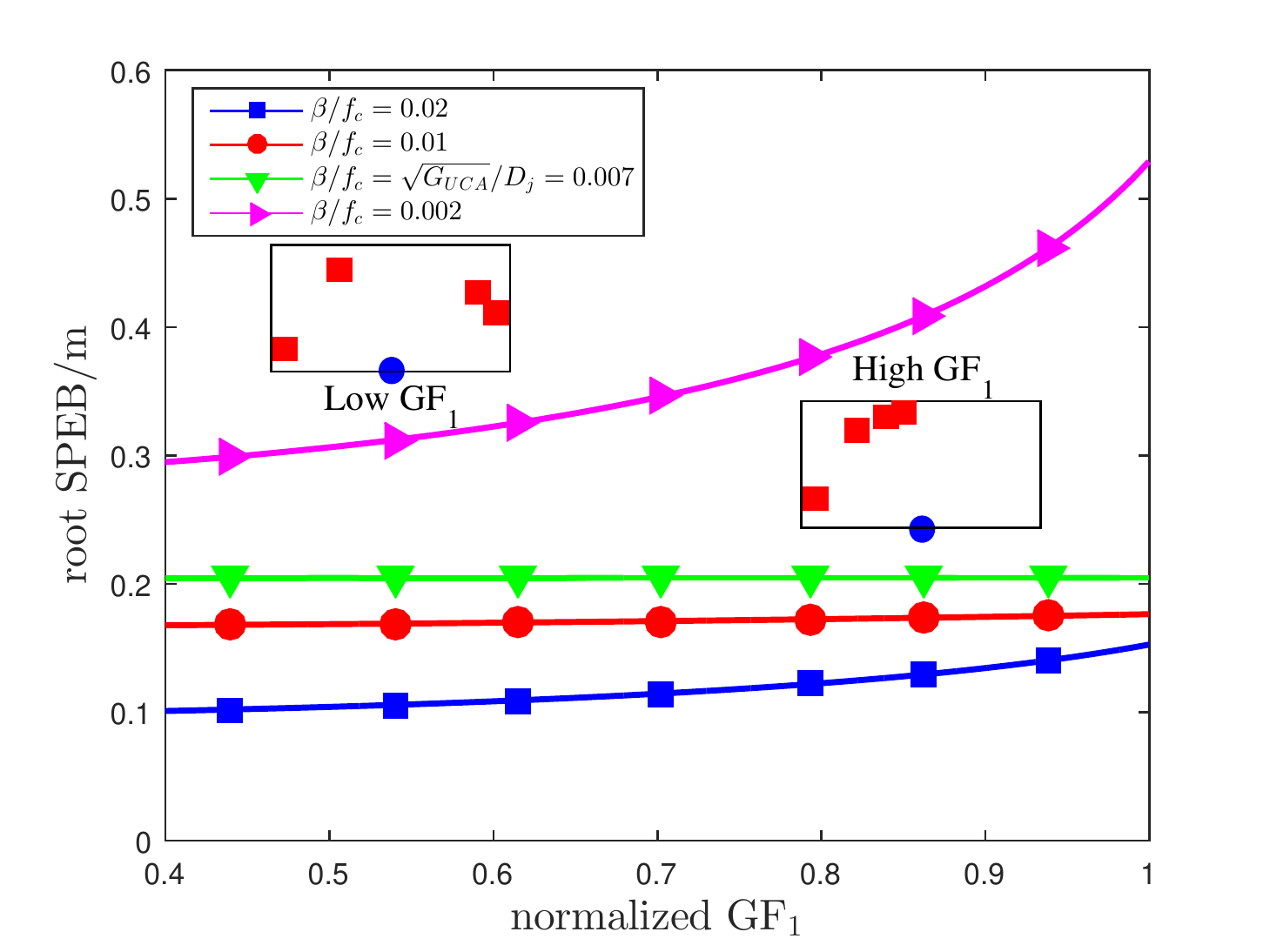}\\
  \caption{Orientation-known root SPEB as a function of the normalized first-type anchor geometric factor $\text{GF}_1$. Four curves with different line types correspond to different relative bandwidths $\beta/f_{\text{\rm c}} = 0.02, 0.01, 0.007, 0.002$, respectively, and small diagrams illustrate two specific examples of anchor distribution with small and large $\text{GF}_1$ (circle as the array location and squares as anchor locations).}\label{ExAnchor1}
\end{figure}

In Fig.~\ref{ExAnchor1}, root SPEBs are monotonically non-decreasing with $\text{GF}_1$, and the ascendent extent tends to zero when distance information and direction information offset each other, i.e., $\beta/f_{\text{\rm c}}= \sqrt{G^{\text{UCA}}}/D_j=0.007$. All these observations are consistent with Theorem \ref{TH_anchor_geo}. Hence, $\text{GF}_1$ is a sufficient indicator for the effect of anchor-agent geometry on SPEB. Moreover, as shown in {the two sub-diagrams} in Fig. \ref{ExAnchor1}, a large $\text{GF}_1$ requires anchor distributions with \emph{concentrated} directions, {but \emph{diversified} directions are preferred} to obtain a small $\text{GF}_1$ to be propitious to localization.

In Fig.~\ref{ExAnchor2}, we focus on the degradation of localization accuracy caused by unknown orientation. Since the root SPEB does not depend uniquely on $\text{GF}_2$ in the orientation-unknown case, we adopt an order-5 polynomial fit curve (the solid line in Fig. \ref{ExAnchor2}) for clarity. Note that this fit curve is sufficiently accurate, with a root MSE less than 0.005. It can be observed that the degradation tends to rise with large $\text{GF}_2$, hence $\text{GF}_2$ is a good indicator (though not sufficient) for the degradation phenomenon. As one can observe, the minimum orientation-unknown SPEB is achieved when both $\text{GF}_1$ and $\text{GF}_2$ are small, conforming again to the results in Theorem \ref{TH_anchor_geo}.

\begin{figure}[!t]
  \centering
  \includegraphics[scale=0.6]{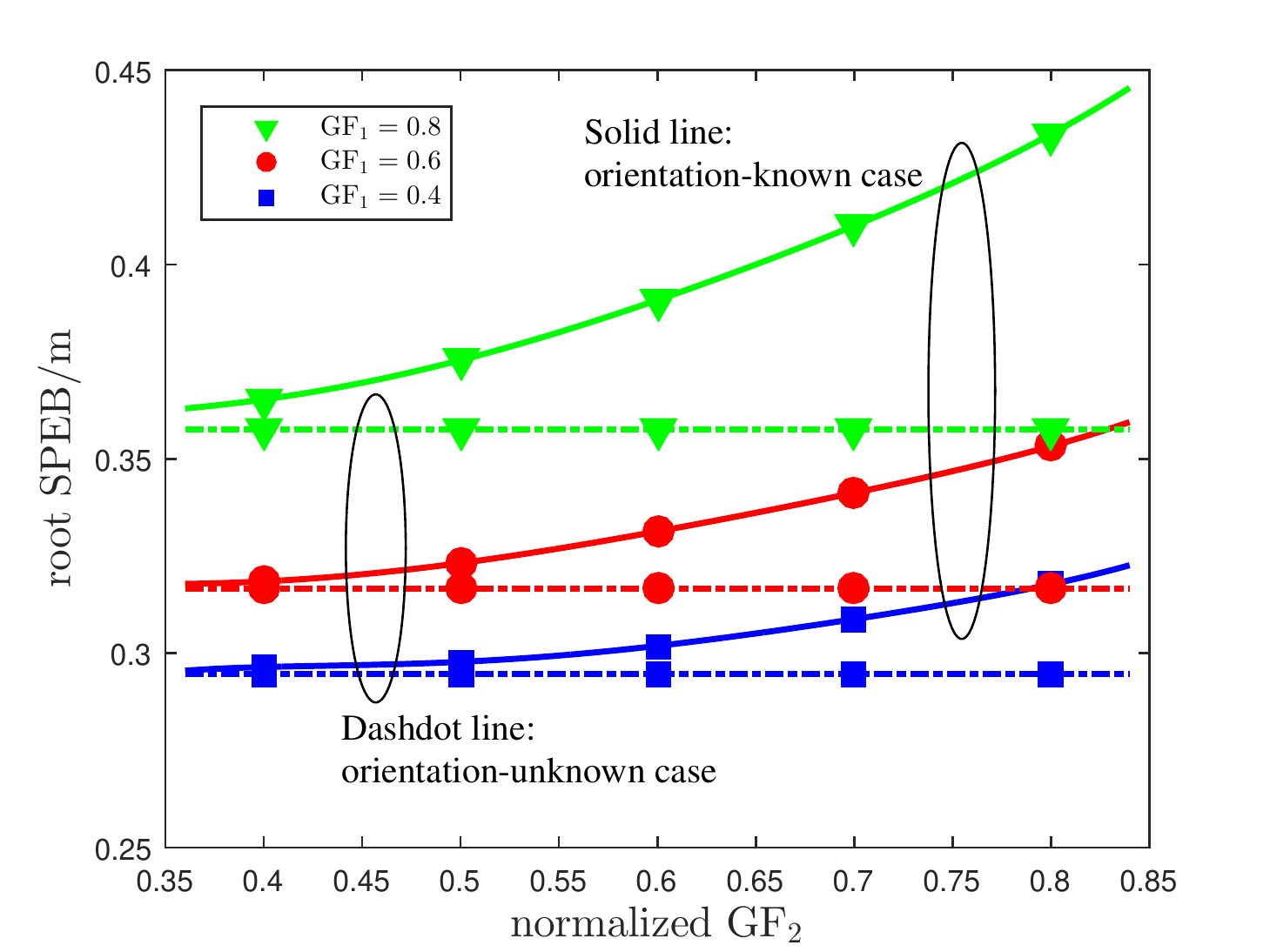}\\
  \caption{Orientation-known (dashdot line) and orientation-unknown (solid line) root SPEB as a function of normalized second-type anchor geometric factor $\text{GF}_2$. Curves with different markers correspond to $\text{GF}_1 = 0.4, 0.6, 0.8$, respectively. {The signal parameters are $\beta=40$\thinspace MHz and $f_\rc=200$\thinspace MHz}.}\label{ExAnchor2}
\end{figure}

\subsection{Array-antenna Geometry}

Finally, we consider two kinds of arrays, i.e., the ULA and UCA, to investigate {the effect of the array-antenna geometry on localization. For simplicity, we only explore the SPEB with unknown array orientation and velocity}. In addition to the same parameters in the experiment on anchor-agent geometry, we have four anchors evenly distributed instead of random locations and a moving agent at $v=30$\thinspace m/s instead of the static one. Fig. \ref{ExArray} shows the SPEB as a function of array orientation $\psi$ varying in $[\,0, 2\pi)$ and antenna number $N_{\text{a}}$ taking value in $\{3,6,12\}${, with baseband bandwidth $\beta=1$\thinspace MHz and carrier frequency $f_\rc=100$\thinspace MHz.}

We draw the following observations {from Fig. \ref{ExArray}}. Firstly, the SPEB for ULA is affected more significantly by array orientation than that for UCA, for UCA has a constant SAAF. Secondly, for both arrays, the SPEB decreases (approximately) to its half when the antenna number is doubled. Thirdly, UCA outperforms ULA uniformly in this example. All these observations are consistent with the theoretical comparison between their SAAFs in the preceding subsection.

\begin{figure}[!t]
  \centering
  \includegraphics[scale=0.6]{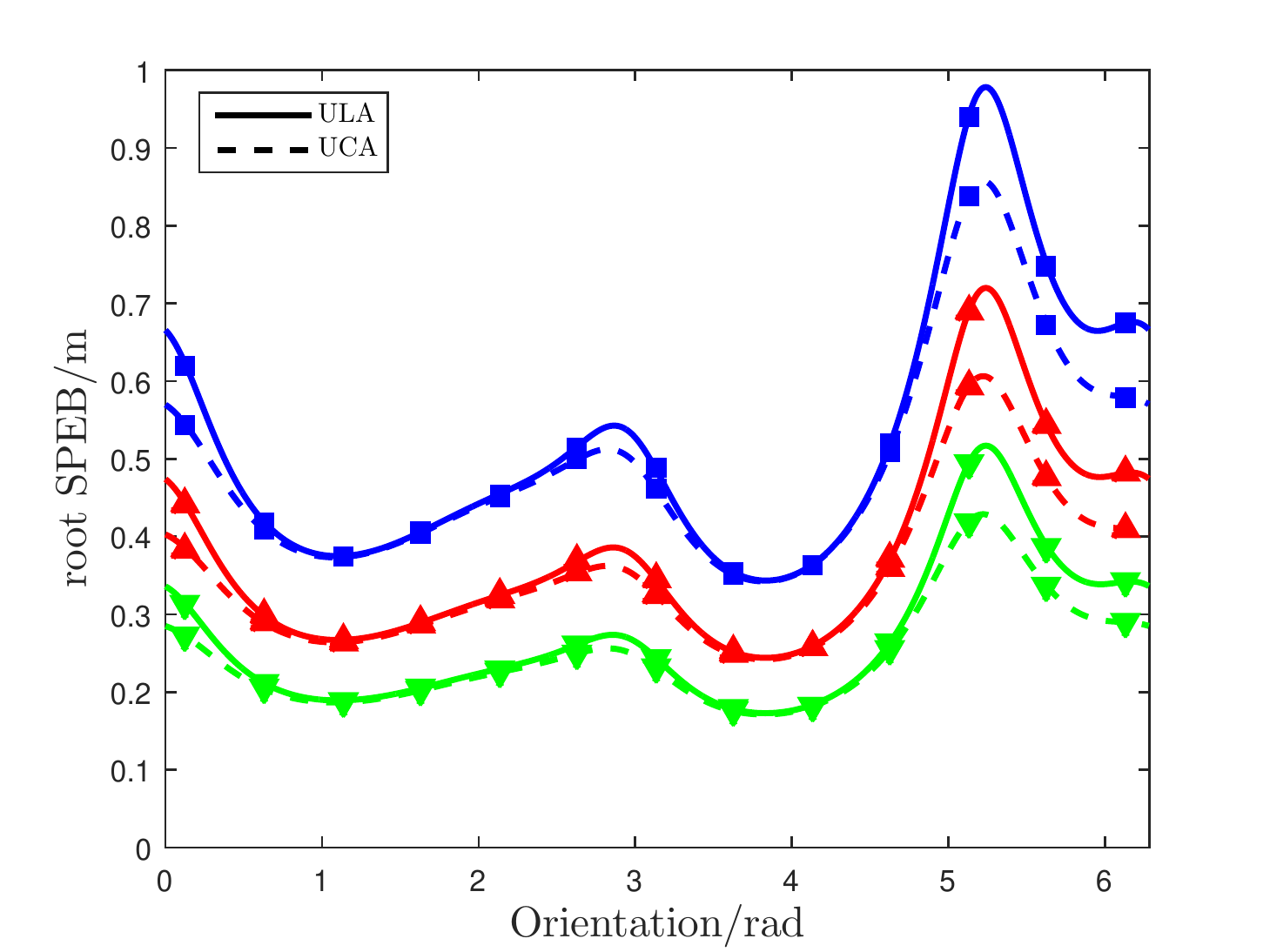}\\
  \caption{Orientation- and Velocity-unknown SPEB (solid line for ULA, dashdot line for UCA) as a function of array orientation $\psi$. {Four anchors transmit the signal with $\mathsf{SNR}=20$\thinspace dB, $25$\thinspace dB, $30$\thinspace dB and $35$\thinspace dB, respectively, and $\beta=1$\thinspace MHz, $f_\rc=100$\thinspace MHz.} Three curves with different line types correspond to different antenna numbers $N_{\text{a}}=3,6,12$, respectively （(from up to down).}\label{ExArray}
\end{figure}

\section{Conclusion}\label{sec_conclusion}
In this paper, we determined the performance limit of the array localization accuracy that exploits all relevant information in received waveforms, and showed that the localization information is a weighed sum of measuring information from each anchor-antenna measurement pair. In the static scenario, the measuring information can be decomposed into the TOA (distance) information with intensity proportional to the squared bandwidth of baseband signal along the radial direction, and the AOA (direction) information with intensity associated with visual angles and carrier frequency along the tangent direction. In the dynamic scenario, the Doppler shift contributes additional direction information with intensity determined by the speed of the agent and the root mean squared time duration of the transmitted signal, implying that the effect of agent mobility can be interpreted as an equivalent aperture synthesized along the course of moving. These results establish a complete physical interpretation for the structure of localization information in general localization networks.

Moreover, we proposed two measures, i.e., the squared array aperture function and anchor geometric factors, to quantify the impacts of network geometric structures on the localization performance. We proved that the UCOA and anchors with zero anchor geometric factors have the optimal localization performance over all kinds of anchor-agent and array-antenna geometries. These results can be used as guidelines for localization system design, as well as benchmarks for localization-aware networks and localization algorithms. 

{\section*{Acknowledgement}
We would like to thank O.~Simeone, the associate editor and the anonymous reviewers for their constructive suggestions, which significantly improve the quality of this paper.}

\appendices

\section{Derivation of EFIMs}
We will use the notation
\begin{align}\label{eq:fisher_info_notation}
  {\bf{F}}_{{\bf{z}}}({\bf{w}};{\bf{x}},{\bf{y}})\triangleq\mathbb{E}_{{\bf{z}}}
  \bigg\{\Big(\frac{\partial}{\partial{\bf{x}}}\ln f({\bf{w}})\Big)\Big(\frac{\partial}{\partial{\bf{y}}}\ln f({\bf{w}})\Big)^{{\text{T}}}\bigg\}
\end{align}
and
\begin{align}\label{eq:direc_vector}
    {\bf q}_{1j} =\left[
                \begin{array}{c}
                  \cos\phi_j \\
                  \sin\phi_j \\
                \end{array}
              \right], \qquad
               {\bf q}_{2j} = \left[
                \begin{array}{c}
                  -\sin\phi_j \\
                  \cos\phi_j \\
                \end{array}
              \right]
\end{align}
throughout this appendix.

\subsection{Proof of Proposition \ref{PRO_equivalence}}\label{PF_PRO_equivalence}
Denote by $\bf F$ and $\bf \tilde{F}$ the FIMs generated by the log-likelihood function (\ref{eq:likelihood_function_real}) and (\ref{eq:likelihood_function}), respectively, and further denote by ${\bf F}'$ the FIM obtained by the modified log-likelihood function where we replace all $\Re\{\cdot\}$ operators by $\Im\{\cdot\}$ in (\ref{eq:likelihood_function_real}). Since $|z|^2 = |\Re\{z\}|^2 + |\Im\{z\}|^2$ for any $z\in\mathbb{C}$, it is straightforward to show that ${\bf F}+{\bf F}' = 2{\bf \tilde{F}}$. Moreover, it can be easily shown that each entry of the difference ${\bf F}-{\bf F}'$ is a finite sum of the terms taking the following form
\begin{align}\label{eq:cross_terms}
  c\int t^ls_0^{(m)}(t)(s_0^{(n)}(t))^* \exp(j2\pi \cdot 2f_\rc t)dt
\end{align}
for some coefficient $c\in\mathbb{R}$ and nonnegative integers $l,m,n$. If $s_0(t)$ is bandlimited by $f_\rc$, then by the differential and convolutional properties of the Fourier Transform, $t^ls_0^{(m)}(t)(s_0^{(n)}(t))^*$ is bandlimited by $2f_\rc$ for any nonnegative integers $l,m,n$. As a result, the terms in (\ref{eq:cross_terms}) are all zero. Hence, ${\bf F}={\bf F}'$, and we conclude that ${\bf F}={\bf F}'={\bf \tilde{F}}$.
\subsection{Proof of Theorem \ref{TH_static_EFIM_full}}\label{PF_TH_static_EFIM_full}
Denote by $S(f)$ the Fourier Transform of $s(t)$, we have $S(f)=S_0(f-f_\text{\rm c})$. Applying the conclusion of \cite[Thm. 1]{SheWin:J10a}, the proof is completed by plugging the definition of $\beta$ and $\gamma$ into
\begin{align}
  &\int_{-\infty}^\infty f^2|S(f)|^2df = \int_{-\infty}^\infty f^2 |S_0(f)|^2df \nonumber\\
  &\qquad + f_\text{\rm c}^2 \int_{-\infty}^\infty |S_0(f)|^2df +
  2f_\text{\rm c} \int_{-\infty}^\infty f|S_0(f)|^2df
\end{align}
and the identity $\phi_{jk}=\phi_j + \theta_{jk}^{\text{\rm V}}$ (see Figure \ref{fig:fisherinfo}).

\begin{figure*}[!b]
	\normalsize
	\setcounter{mytempeqncnt}{\value{equation}}
	\setcounter{equation}{71}
	\hrulefill
	\vspace*{1mm}
	\begin{align}\label{eq:pf_Je_p}\nonumber
	{\bf J}_\text{\rm e}({\bf p}) 
	&=  {\bf{F}}_{{\bf{r}}}({\bf{r}}|{\bm{\theta}};{\bf{p}},{\bf{p}}) - \sum_{j=1}^{N_{\text{\rm b}}}  {\bf{F}}_{{\bf{r}}}({\bf{r}}|{\bm{\theta}};{\bf{p}},\xi_j) {\bf{F}}_{{\bf{r}}}({\bf{r}}|{\bm{\theta}};\xi_j,\xi_j)^{-1} {\bf{F}}_{{\bf{r}}}({\bf{r}}|{\bm{\theta}};{\bf{p}},\xi_j)^{\text{T}}\\\nonumber
	&= \sum_{j=1}^{N_{\text{\rm b}}}\sum_{k=1}^{N_\text{\rm a}}\frac{8\pi^2\mathsf{SNR}_j(\beta^2 + f_\text{\rm c}^2 + 2\gamma\beta f_\text{\rm c})}{c^2}{\bf J}_\text{\rm r}(\phi_j+\theta_{jk}^{\text{\rm V}}) \\\nonumber
	&\qquad\qquad \quad - \sum_{j=1}^{N_\text{\rm b}}\frac{8\pi^2\mathsf{SNR}_j(\gamma\beta+f_\text{\rm c})^2}{c^2N_\text{\rm a}}\bigg(N_\text{\rm a}{\bf q}_{1j}+\sum_{k=1}^{N_\text{\rm a}}\theta_{jk}^{\text{\rm V}}{\bf q}_{2j}\bigg)\bigg(N_\text{\rm a}{\bf q}_{1j}+\sum_{k=1}^{N_\text{\rm a}}\theta_{jk}^{\text{\rm V}}{\bf q}_{2j}\bigg)^{\text{T}}\\\nonumber
	&\cong \sum_{j=1}^{N_\text{\rm b}} \frac{8\pi^2\mathsf{SNR}_j}{c^2}\bigg[(1-\gamma^2)\beta^2\sum_{k=1}^{N_{\text{\rm a}}}{\bf J}_{\text{\rm r}}(\phi_j+\theta_{jk}^{\text{\rm V}}) + (\gamma\beta + f_\text{\rm c})^2\cdot \frac{\sum_{1\le k<l\le N_\text{\rm a}}(\theta_{jk}^{\text{\rm V}}-\theta_{jl}^{\text{\rm V}})^2}{N_\text{\rm a}}{\bf q}_{2j}{\bf q}_{2j}^{\text{T}}\bigg]\\
	&=  \sum_{j=1}^{N_\text{\rm b}} \lambda_j\bigg[(1-\gamma^2)\beta^2\sum_{k=1}^{N_{\text{\rm a}}}{\bf J}_{\text{\rm r}}(\phi_j+\theta_{jk}^{\text{\rm V}}) + \frac{N_\text{\rm a}(\gamma\beta + f_\text{\rm c})^2G(\phi_j-\psi)}{D_j^2}{\bf J}_{\text{\rm r}}\left(\phi_j+\frac{\pi}{2}\right)\bigg]
	\end{align}
	\setcounter{equation}{\value{mytempeqncnt}}
	\vspace*{1mm}
\end{figure*}

\subsection{Proof of Theorem \ref{TH_static_EFIM}}\label{PF_TH_static_EFIM}
We first consider the case where there is no multipath or NLOS phenomenon. By Theorem \ref{TH_static_EFIM_full} we know that
\begin{align}
 {\bf{F}}_{{\bf{r}}}({\bf{r}}|{\bm{\theta}};{\bf{p}},{\bf{p}}) = \sum_{j=1}^{N_{\text{\rm b}}}\sum_{k=1}^{N_\text{\rm a}}\lambda_j(\beta^2 + f_\text{\rm c}^2 + 2\gamma\beta f_\text{\rm c}){\bf J}_\text{\rm r}(\phi_j+\theta_{jk}^{\text{\rm V}}).
\end{align}

\begin{figure*}[!b]
	\normalsize
	\setcounter{mytempeqncnt}{\value{equation}}
	\setcounter{equation}{74}
	\hrulefill
	\vspace*{1mm}
	\begin{align}\label{eq:relation_p}
	[{\bf J}_\text{\rm e}(\{{\bf p},\psi\})]_{1:2,3} 
	&= {\bf F_r}({\bf r}|\bm{\theta};{\bf p},\psi) - \sum_{j\in \mathcal{N}_{\text{L}}}  {\bf{F}}_{{\bf{r}}}({\bf{r}}|{\bm{\theta}};{\bf{p}},\xi_j) {\bf{F}}_{{\bf{r}}}({\bf{r}}|{\bm{\theta}};\xi_j,\xi_j)^{-1} {\bf{F}}_{{\bf{r}}}({\bf{r}}|{\bm{\theta}};{\psi,\xi_j})^{\text{T}} \nonumber \\
	&= -\sum_{j\in\mathcal{N}_{\text{L}}} {\bf F_r}({\bf r}|\bm{\theta};{\bf p},\phi_j) + \sum_{i\in\mathcal{N}_{\text{L}}}\sum_{j\in\mathcal{N}_{\text{\rm L}}}{\bf{F}}_{{\bf{r}}}({\bf{r}}|{\bm{\theta}};{\bf{p}},\xi_j) {\bf{F}}_{{\bf{r}}}({\bf{r}}|{\bm{\theta}};\xi_j,\xi_j)^{-1} {\bf{F}}_{{\bf{r}}}({\bf{r}}|{\bm{\theta}};{\phi_i,\xi_j})^{\text{T}} \nonumber \\
	&= -\sum_{j\in\mathcal{N}_{\text{L}}} \lambda_jD_j\Big((\beta^2+f_\text{\rm c}^2)\sum_{k=1}^{N_\text{\rm a}}\left[\theta_{jk}^{\text{\rm V}}{\bf q}_{1j} + (\theta_{jk}^{\text{\rm V}})^2 {\bf q}_{2j}\right]- f_\text{\rm c}^2 \big[{\bf q}_{1j} + \frac{1}{N_\text{\rm a}}\sum_{k=1}^{N_\text{\rm a}}\theta_{jk}^{\text{\rm V}}{\bf q}_{2j}\big]\cdot \sum_{k=1}^{N_\text{\rm a}}\theta_{jk}^{\text{\rm V}}\Big) \nonumber \\
	&= -\sum_{j\in\mathcal{N}_{\text{L}}} \lambda_jD_j \bigg[\beta^2\sum_{k=1}^{N_\text{\rm a}}\theta_{jk}^{\text{V}}({\bf q}_{1j} + \theta_{jk}^{\text{V}}{\bf q}_{2j}) + \frac{N_\text{\rm a}f_\text{\rm c}^2G(\phi_j-\psi)}{D_j^2}{\bf q}_{2j}\bigg]
	\end{align}
	\hrulefill
	\vspace*{1mm}
	\begin{align}\label{eq:relation_varphi}\nonumber
	[{\bf J}_\text{\rm e}(\{{\bf p},\psi\})]_{3,3} &= {\bf F_r}({\bf r}|\bm{\theta};\psi,\psi) - \sum_{j\in \mathcal{N}_{\text{L}}}  {\bf{F}}_{{\bf{r}}}({\bf{r}}|{\bm{\theta}};\psi,\xi_j) {\bf{F}}_{{\bf{r}}}({\bf{r}}|{\bm{\theta}};\xi_j,\xi_j)^{-1}
	{\bf{F}}_{{\bf{r}}}({\bf{r}}|{\bm{\theta}};{\psi,\xi_j})^{\text{T}}\\\nonumber
	&= \sum_{j\in \mathcal{N}_{\text{L}}} {\bf F_r}({\bf r}|\bm{\theta};\phi_j,\phi_j) - \sum_{j\in \mathcal{N}_{\text{L}}}
	{\bf{F}}_{{\bf{r}}}({\bf{r}}|{\bm{\theta}};\phi_j,\xi_j) {\bf{F}}_{{\bf{r}}}({\bf{r}}|{\bm{\theta}};\xi_j,\xi_j)^{-1}
	{\bf{F}}_{{\bf{r}}}({\bf{r}}|{\bm{\theta}};{\phi_j,\xi_j})^{\text{T}}\\
	&= \sum_{j\in\mathcal{N}_{\text{L}}}\lambda_jD_j^2 \bigg[\beta^2\sum_{k=1}^{N_\text{\rm a}}(\theta_{jk}^{\text{V}})^2 + \frac{N_\text{\rm a}f_\text{\rm c}^2G(\phi_j-\psi)}{D_j^2}\bigg]
	\end{align}
	\setcounter{equation}{\value{mytempeqncnt}}
	\vspace*{4pt}
\end{figure*}

Now we proceed to compute ${\bf{F}}_{{\bf{r}}}({\bf{r}}|{\bm{\theta}};{\bf{p}},\xi_j)$ and ${\bf{F}}_{{\bf{r}}}({\bf{r}}|{\bm{\theta}};\xi_i,\xi_j)$. The details are as follows:
\begin{align}\nonumber
    &{\bf{F}}_{{\bf{r}}}({\bf{r}}|{\bm{\theta}};\tau_i,\xi_j)\\\nonumber
  &\qquad= \frac{2|\alpha_j|^2\delta_{ij}}{N_0}\Re\bigg\{\sum_{k=1}^{N_\text{\rm a}} \int (js_0^*)(s_0' + j2\pi f_\text{\rm c}s_0) dt \bigg\}\\\nonumber
  &\qquad= -2\delta_{ij}\cdot \Re\bigg\{\sum_{k=1}^{N_\text{\rm a}} 2\pi (\gamma\beta+f_\text{\rm c})\mathsf{SNR}_j\bigg\}\\
  &\qquad= -4\pi (\gamma\beta+f_\text{\rm c})\mathsf{SNR}_j\cdot N_\text{\rm a}\delta_{ij}
\end{align}
\begin{align}\nonumber
&{\bf{F}}_{{\bf{r}}}({\bf{r}}|{\bm{\theta}};\phi_i,\xi_j)\\\nonumber
  &\qquad= \frac{2|\alpha_j|^2\delta_{ij}\tau_j}{N_0}\Re\bigg\{\sum_{k=1}^{N_\text{\rm a}} \int (js_0^*)(s_0' + j2\pi f_\text{\rm c}s_0)\theta_{jk}^{\text{\rm V}} dt \bigg\}\\\nonumber
  &\qquad= -2\delta_{ij}\tau_j\cdot\Re\bigg\{\sum_{k=1}^{N_\text{\rm a}} 2\pi (\gamma\beta+f_\text{\rm c})\mathsf{SNR}_j\cdot \theta_{jk}^{\text{\rm V}}\bigg\} \\
  &\qquad= -4\pi\tau_j(\gamma\beta+f_\text{\rm c})\mathsf{SNR}_j\cdot \delta_{ij}\sum_{k=1}^{N_\text{\rm a}} \theta_{jk}^{\text{\rm V}}
\end{align}
\begin{align}\nonumber
&{\bf{F}}_{{\bf{r}}}({\bf{r}}|{\bm{\theta}};\xi_i,\xi_j) \\
&\qquad= \frac{2|\alpha_j|^2\delta_{ij}}{N_0}\sum_{k=1}^{N_\text{\rm a}}\int (js_0^*)(-js_0)dt = 2\, \mathsf{SNR}_j\cdot N_{\text a}\delta_{ij}
\end{align}
where $\delta_{ij}$ is the discrete Dirac function which equals 1 if $i=j$ and 0 otherwise. Combining these identities yields
\begin{align}\nonumber
    &{\bf{F}}_{{\bf{r}}}({\bf{r}}|{\bm{\theta}};{\bf{p}},\xi_j)\\\nonumber
     &\qquad= \sum_{i=1}^{N_\text{\rm b}}
    \Big[{\bf{F}}_{{\bf{r}}}({\bf{r}}|{\bm{\theta}};\tau_i,\xi_j)\cdot \frac{{\bf q}_{1i}}{c}+ {\bf{F}}_{{\bf{r}}}({\bf{r}}|{\bm{\theta}};\phi_i,\xi_j)\cdot \frac{{\bf q}_{2i}}{c\tau_i}\Big]\\
  &\qquad= -\frac{4\pi(\gamma\beta+f_\text{\rm c})\mathsf{SNR}_j}{c}\cdot\bigg(N_\text{\rm a}\,{\bf q}_{1j}+\sum_{k=1}^{N_\text{\rm a}}\theta_{jk}^\text{\rm V}{\bf q}_{2j}\bigg).
\end{align}

Then the EFIM is given by (\ref{eq:pf_Je_p}), shown at the bottom of \blue{the next} page, where we have used 
\addtocounter{equation}{1}
\begin{align}
{\bf J}_\text{\rm r}(\phi_j+\theta_{jk}^{\text{\rm V}}) 
\cong {\bf q}_{1j}{\bf q}_{1j}^{\text{T}} & + (\theta_{jk}^{\text{\rm V}})^2{\bf q}_{2j}{\bf q}_{2j}^{\text{T}}  \nonumber \\
& + \theta_{jk}^{\text{\rm V}}({\bf q}_{2j}{\bf q}_{1j}^{\text{T}} +{\bf q}_{1j}{\bf q}_{2j}^{\text{T}} )
\end{align}
in the third step by the far-field assumption $\theta_{jk}^{\text{\rm V}}\ll 1$.

In the general case with multipath and NLOS phenomena, we involve the path-overlap coefficient $\chi_j$ and the set of anchors $\mathcal{N}_{\text{\rm NL}}$ providing NLOS signals following the same way as that in the proof of \cite[Thm. 1]{SheWin:J10a}. The proof is complete.

\subsection{Proof of Theorem \ref{TH_static_EFIM_un}}\label{PF_TH_static_EFIM_un}
Since $\psi$ and $\phi_j$ appear in pairs in the log-likelihood function in (\ref{eq:likelihood_function}) through $\phi_j-\psi$, it can be obtained that
\begin{align}\label{eq:relation_gen}
  {\bf F_r}({\bf r}|\bm{\theta};{\bf x},\psi) = -\sum_{j\in\mathcal{N}_{\text{L}}} {\bf F_r}({\bf r}|\bm{\theta};{\bf x},\phi_j)
\end{align}
for any parameter $\bf x$. Hence, the entries of FIM for the position and orientation are given by (\ref{eq:relation_p}) and (\ref{eq:relation_varphi}), respectively, shown at the bottom of \blue{this} page. The combination of these two completes the proof.
\addtocounter{equation}{2}

\subsection{Proof of Theorem \ref{TH_dynamic_EFIM}}\label{PF_TH_dynamic_EFIM}
Without loss of generality we assume that $\tau=0$, otherwise $s_1(t)=s_0(t-\tau)$ can be used instead. Similar to the proof of Theorem \ref{TH_static_EFIM}, it suffices to consider the scenario without multipath or NLOS. Define $\delta_j \triangleq v_{\text{r}}\cos(\phi_j-\psi_{\text{\rm d}})$, and write
\begin{align}\nonumber
  {\bf J}_{\text{\rm e}}({\bf p}) \equiv \sum_{j=1}^{N_\text{\rm b}} &\left[A_{1j}{\bf q}_{1j}{\bf q}_{1j}^{\text{T}} + A_{2j}{\bf q}_{2j}{\bf q}_{2j}^{\text{T}}\right.\\
  &\left. \qquad + A_{3j}\left({\bf q}_{1j}{\bf q}_{2j}^{\text{T}}+{\bf q}_{2j}{\bf q}_{1j}^{\text{T}}\right)\right].
\end{align}

\begin{figure*}[!b]
	\normalsize
	\setcounter{mytempeqncnt}{\value{equation}}
	\setcounter{equation}{78}
	\hrulefill
	\vspace*{1mm}
	\begin{align}\label{eq:Phi22_exp}
	A_{2j} &= \frac{1}{(c\tau_j)^2}\left[{\bf F_r}({\bf r}|\bm{\theta};\phi_j,\phi_j) - {\bf F_r}({\bf r}|\bm{\theta};\phi_j,\xi_j){\bf F_r}({\bf r}|\bm{\theta};\xi_j,\xi_j)^{-1}{\bf F_r}({\bf r}|\bm{\theta};\phi_j,\xi_j)^{\text{T}}\right] \nonumber \\
	&= \frac{2|\alpha_j|^2}{c^2(1-\delta_j)N_0}\bigg[\sum_{k=1}^{N_\ra}\int |s'(t)(\theta_{jk}^\rV+\omega_jt)|^2dt
	- \frac{|\alpha_j|^2/N_0}{N_\ra\mathsf{SNR}_j}\bigg(\sum_{k=1}^{N_\ra}\Re\Big\{\int (js(t))^* \cdot s'(t)(\theta_{jk}^\rV+\omega_jt)dt\Big\}\bigg)^2\bigg] \nonumber \\
	    &= \frac{\lambda_j}{1-\delta_j}\bigg[ N_\ra f_\rc^2 \Big(\frac{G(\phi_j-\psi)}{D_j^2} + \omega_j^2t_{\text{\rm rms}}^2\Big) + \frac{N_\ra f_\rc\omega_j^2}{\pi\int |s_0(t)|^2dt}\Big(\int \Im\{s_0'(t)s_0^*(t)t^2\}dt - \frac{(\int t|s_0(t)|^2dt)(\int \Im\{s_0'(t)s_0^*(t)t\}dt)}{\int |s_0(t)|^2dt}\Big)  \nonumber \\
	    &\qquad + \frac{1}{4\pi^2}\sum_{k=1}^{N_\ra}\frac{\int |s_0'(t)(\theta_{jk}^\rV + \omega_jt)|^2dt}{\int |s_0(t)|^2dt} + \sum_{k=1}^{N_\ra}\frac{f_\rc\omega_j\theta_{jk}^\rV}{\pi}\cdot \frac{\int \Im\{s_0'(t)s_0^*(t)t\}dt}{\int |s_0(t)|^2dt} - \frac{N_\ra\omega_j^2}{4\pi^2}\cdot\Big(\frac{\int \Im \{s_0'(t)s_0^*(t)t\}dt}{\int |s_0(t)|^2dt}\Big)^2\bigg]
	\end{align}
	\setcounter{equation}{\value{mytempeqncnt}}
	\vspace*{4pt}
\end{figure*}

\subsubsection{Derivation of $A_{1j}$}
By the variable substitution $\tilde{t}=t-\tau_{jk}^{(1)}(t)$, it is obvious that
\begin{align}\label{eq:A_1j_final}\nonumber
  A_{1j} &= \frac{1}{c^2}\left[{\bf F_r}({\bf r}|\bm{\theta};\tau_j,\tau_j) \right.\\\nonumber
  &\left.\qquad - {\bf F_r}({\bf r}|\bm{\theta};\tau_j,\xi_j){\bf F_r}({\bf r}|\bm{\theta};\xi_j,\xi_j)^{-1}{\bf F_r}({\bf r}|\bm{\theta};\tau_j,\xi_j)^{\text{T}}\right] \\
  &= \frac{\lambda_j}{1-\delta_j}\bigg[\sum_{k=1}^{N_\text{\rm a}} (\beta^2 + f_\text{\rm c}^2) - \frac{(N_\ra f_\rc)^2}{N_\ra}\bigg]
  = \frac{\lambda_jN_\ra\beta^2}{1-\delta_j}.
\end{align}
\subsubsection{Derivation of $A_{2j}$}
Using the same variable substitution method, $A_{2j}$ can be expressed as (\ref{eq:Phi22_exp}), shown at the bottom of \blue{the next} page. Note that
\addtocounter{equation}{1}

\begin{figure*}[!b]
	\normalsize
	\setcounter{mytempeqncnt}{\value{equation}}
	\hrulefill
	\vspace*{1mm}
	\setcounter{equation}{86}
	\begin{align}\label{eq:A_3j}
	A_{3j} &= \frac{1}{c^2\tau_j}\left[{\bf F_r}({\bf r}|\bm{\theta};\tau_j,\phi_j) - {\bf F_r}({\bf r}|\bm{\theta};\tau_j,\xi_j){\bf F_r}({\bf r}|\bm{\theta};\xi_j,\xi_j)^{-1}{\bf F_r}({\bf r}|\bm{\theta};\phi_j,\xi_j)^{\text{T}}\right] \nonumber \\
	&= \frac{2|\alpha_j|^2}{c^2(1-\delta_j)N_0}\bigg[\sum_{k=1}^{N_\ra}\int |s'(t)|^2(\theta_{jk}^\rV+\omega_jt)dt
	- 2\pi f_\rc\sum_{k=1}^{N_\ra}\Re\Big\{\int (js(t))^* \cdot s'(t)(\theta_{jk}^\rV+\omega_jt)dt\Big\}\bigg] \nonumber \\
	&= \frac{\lambda_j}{1-\delta_j}\bigg[\beta^2\sum_{k=1}^{N_\ra} \theta_{jk}^\rV + \frac{N_\ra \omega_j}{\int |s_0(t)|^2dt}\Big(\int t|s_0'(t)|^2dt + 2\pi f_\rc \int \Im\{s_0'(t)s_0^*(t)t\}dt\Big) \bigg]
	\end{align}
	\setcounter{equation}{\value{mytempeqncnt}}
	\vspace*{4pt}
\end{figure*}

\begin{align}
  \int |s_0'(t)|^2 dt & = \int |2\pi fS_0(f)|^2 df   \nonumber\\
  & = 4\pi^2\beta^2 \int |S_0(f)|^2 df \nonumber\\
  & = 4\pi^2\beta^2 \int |s_0(t)|^2dt 
\end{align}
  \begin{align}\label{eq:Phi22_A2k_bound}
  \int t^2|s_0'(t)|^2dt 
  & \le \int 2\left(|(ts_0(t))'|^2 + |s_0(t)|^2\right)dt \nonumber \\
  & = 2\int |fS_0'(f)|^2 df + 2\int |s_0(t)|^2dt \nonumber \\
  & \le 2B^2 \int |S_0'(f)|^2 df + \frac{2}{\sigma^2}\int |ts_0(t)|^2 dt \nonumber \\
  & \le (8\pi^2B^2 + 32\pi^2\beta^2)\int |ts_0(t)|^2 dt \nonumber  \\
  & \ll 4\pi^2 f_\rc^2 \int |ts_0(t)|^2 dt
\end{align}
where
\begin{align}
  \sigma \triangleq \left[\frac{\int |ts_0(t)|^2 dt}{\int |s_0(t)|^2dt}\right]^{\frac{1}{2}}
\end{align}
and we have used the uncertainty principle $\beta\sigma \ge 1/4\pi$ and the assumption $\beta\le B\ll f_\rc$. Then by the Cauchy-Schwarz inequality, we further have
\begin{align}\nonumber
  \int t|s_0(t)|^2 dt
  &\le \sqrt{\left(\int t^2|s_0(t)|^2dt\right)\left(\int |s_0(t)|^2dt\right)}\\
  &= \sigma\int |s_0(t)|^2dt\\\nonumber
   \int \Im\{s_0'(t)s_0^*(t)t\}dt
    &\le \sqrt{\left(\int t^2|s_0'(t)|^2dt\right)\left(\int |s_0^*(t)|^2dt\right)} \\
   &\ll f_\rc\sigma \int |s_0(t)|^2dt\\\nonumber
  \int \Im\{s_0'(t)s_0^*(t)t^2\}dt
   &\le \sqrt{\left(\int t^2|s_0'(t)|^2dt\right)\left(\int t^2|s_0^*(t)|^2dt\right)} \\
  &\ll f_\rc\sigma^2 \int |s_0(t)|^2dt
\end{align}
and thus conclude that all other terms in (\ref{eq:Phi22_exp}) are negligible compared with the first term. In other words, we have
\begin{align}\label{eq:A_2j_final}
   A_{2j} \cong \frac{\lambda_jN_\ra f_\rc^2}{1-\delta_j}\Big(\frac{G(\phi_j-\psi)}{D_j^2} + \omega_j^2t_{\text{\rm rms}}^2\Big).
\end{align}

\subsubsection{Derivation of $A_{3j}$}
$A_{3j}$ can be expressed in (\ref{eq:A_3j}), shown at the bottom of \blue{this} page. By Cauchy-Schwarz inequality,
\addtocounter{equation}{1}
\begin{align}
  \int t|s_0'(t)|^2 dt  &\le \sqrt{\bigg(\int t^2|s_0'(t)|^2dt\bigg)\bigg(\int |s_0'(t)|^2dt\bigg)} \nonumber \\
  & \ll \beta f_\rc\sigma\int |s_0(t)|^2dt
\end{align}
thus we conclude that
\begin{align}
  &\frac{\lambda_j}{1-\delta_j}\bigg[\beta^2\sum_{k=1}^{N_\ra} \theta_{jk}^\rV + \frac{N_\ra \omega_j}{\int |s_0(t)|^2dt}\int t|s_0'(t)|^2dt\bigg] \nonumber\\
  & \qquad\ll \sqrt{A_1A_2}.
\end{align}

For the last term in the expression of $A_3$, suppose that $S_0(f)=|S_0(f)|\exp(j\phi(f))$, it can be shown that
\begin{align}\label{eq:Phi12_A6k_lemma}\nonumber
  \Im\bigg\{\int t s_0^*(t)s_0'(\tau)dt\bigg\} &=
  \Im\bigg\{\int fS_0^*(f)S_0'(f)df\bigg\} \\
  &= \int f|S_0(f)|^2\phi'(f)df = 0
\end{align}
where the last equality holds due to the balanced phase assumption. Hence, we conclude that $A_3\ll \sqrt{A_1A_2}$, and thus
\begin{align}\label{eq:Phi12_A6k}\nonumber
A_{1j}{\bf q}_{1j}{\bf q}_{1j}^{\text{T}} + A_{2j}{\bf q}_{2j}{\bf q}_{2j}^{\text{T}} &\succeq \sqrt{A_{1j}A_{2j}}\left({\bf q}_{1j}{\bf q}_{2j}^{\text{T}}+{\bf q}_{2j}{\bf q}_{1j}^{\text{T}}\right)\\ &\gg A_{3j}\left({\bf q}_{1j}{\bf q}_{2j}^{\text{T}}+{\bf q}_{2j}{\bf q}_{1j}^{\text{T}}\right).
\end{align}
\quad In conclusion, we have
\begin{align}\label{eq:Phi12_final}
{\bf J}_{\text{\rm e}}({\bf p}) &= \sum_{j=1}^{N_\text{\rm b}} \frac{\lambda_jN_\ra}{1-\delta_j} \bigg[ \beta^2{\bf J}_{\text{\rm r}}(\phi_j) \nonumber \\
&\qquad + f_\rc^2\Big(\frac{G(\phi_j-\psi)}{D_j^2} + \omega_j^2t_{\text{\rm rms}}^2\Big){\bf J}_{\text{\rm r}}(\phi_j+\frac{\pi}{2})\bigg]
\end{align}
which is the desired result.

\section{Proof of Geometric Properties}
\subsection{Proof of Proposition \ref{Pro_Geo}}\label{PF_Pro_Geo}
{Based on (\ref{eq:SAAF_equiv}),} we have
\begin{align}\label{eq:A_C_D_2}
  \frac{1}{2\pi}\int_0^{2\pi}G(\theta)\text{d}\theta = \frac{\widetilde{{\bf x}^2} + \widetilde{{\bf y}^2}}{2}.
\end{align}
By the definition of the infimum, for any $\epsilon>0$, the array can be covered by a circle centered at some $(x_{\text{c}},y_{\text{c}})$ with diameter $D+\epsilon$, then
\begin{align}\label{eq:A_C_D_3}\nonumber
  \widetilde{{\bf x}^2} + \widetilde{{\bf y}^2} &= \frac{1}{N_{\text{a}}}\sum_{k=1}^{N_{\text{a}}}\left[
  (x_k-\bar{x})^2 + (y_k-\bar{y})^2\right]\\\nonumber
  &\le \frac{1}{N_{\text{a}}}\sum_{k=1}^{N_{\text{a}}}\left[
  (x_k-x_{\text{c}})^2 + (y_k-y_{\text{c}})^2\right]\\
  &\le  \Big(\frac{D+\epsilon}{2}\Big)^2
\end{align}
where the first inequality utilizes the following fact
\begin{align}\label{eq:A_C_D_4}
  (\bar{x},\bar{y}) = \arg\min_{(x_{\text{c}},y_{\text{c}})} \sum_{k=1}^{N_{\text{a}}}\left[
  (x_k-x_{\text{c}})^2 + (y_k-y_{\text{c}})^2\right].
\end{align}
Then the inequality to be proved is the combination of (\ref{eq:A_C_D_2}) and (\ref{eq:A_C_D_3}) by letting $\epsilon\rightarrow0^+$. In particular, for UCOA, $G(\theta)$ is invariant with $\theta$, and all equalities in (\ref{eq:A_C_D_3}) hold, which completes the proof for the second part.

\subsection{Proof of Theorem \ref{TH_array_geo}}\label{PF_TH_array_geo}
{Define $T_{\text{ES}}: \mathbb{S}_{++}^n \mapsto \mathbb{R}$ as the mapping from EFIM for the position to SPEB, and $T_{\text{ES}}({\bf A})=\text{tr}\{{\bf A}^{-1}\}$ for ${\bf A}\in \mathbb{S}_{++}^n$. First we show that $T_{\text{ES}}(\cdot)$ is convex and monotonically non-increasing in terms of the L\"{o}wner semiorder $\succeq$. Suppose ${\bf A}\succeq{\bf B}$ are two positive definite matrices, and denote by $\{\lambda_k\}, \{\mu_k\}$ their eigenvalues in descending order. The Courant-Fischer-Weyl min-max principle\cite{reed1978methods} yields
\begin{align}\label{eq:A_C_B_3}\nonumber
  \lambda_k &= \inf_{{{\bf C}}\in \mathbb{C}^{2\times (k-1)}}\sup_{{\bf Cx}={\bf 0},\|{\bf{x}}\|=1}{\bf x^TA}{\bf{x}}\\
  &\ge \inf_{{\bf{C}}\in \mathbb{C}^{2\times (k-1)}}\sup_{{\bf Cx}={\bf 0},\|{\bf{x}}\|=1}{\bf x^TB}{\bf{x}} = \mu_k
\end{align}
by taking supremum and infimum successively. Hence
\begin{align}\label{eq:A_C_B_4}
  T_{\text{ES}}({\bf A}) = \text{tr}\{{\bf A}^{-1}\} = \sum \lambda_k^{-1}
  \le \sum \mu_k^{-1} = T_{\text{ES}}({\bf B})
\end{align}
gives the monotonicity of $T_{\text{ES}}(\cdot)$. The convexity follows from the fact that $T_{\text{ES}}(\cdot)$ is a compound function of a convex function $g: \mathbb{S}_{++}^n\mapsto\mathbb{S}_{++}^n$ with $g({\bf{A}})={\bf{A}}^{-1}$ and a linear function $h: \mathbb{S}_{++}^n\mapsto \mathbb{R}$ with $h({\bf{A}})=\text{tr}\{{\bf{A}}\}$}.

{Then by Jensen's inequality, we have
\begin{align}\label{eq:T5_6}\nonumber
  \frac{1}{2\pi}\int_0^{2\pi}\text{SPEB}(\psi)d\psi &= \frac{1}{2\pi}\int_0^{2\pi}T_{\text{ES}}({\bf{J}}_{\text{e}}({\bf{p}},\psi))d\psi
  \\&\ge \nonumber T_{\text{ES}}\Big(\frac{1}{2\pi}\int_0^{2\pi}{\bf{J}}_{\text{e}}({\bf{p}},\psi)d\psi\Big)\\\nonumber
  &\ge T_{\text{ES}}({\bf{J}}_{\text{e}}^{\text{\rm UCOA}}({\bf p})) \\
  &= \text{SPEB}^{\text{UCOA}}
\end{align}
where we have used the monotonicity of $T_{\text{ES}}(\cdot)$ and Proposition \ref{Pro_Geo} in the last inequality. Since the Bayes estimator takes constant value in the support of the prior, we conclude that UCOA is also optimal in the minimax sense \cite{lehmann1998theory}, which completes the proof.}

\subsection{Proof of Theorem \ref{TH_anchor_geo}}\label{PF_TH_anchor_geo}
Defining $r_i \triangleq \lambda_i\beta^2, s_i\triangleq \lambda_if_{\text{c}}^2D_i^{-2}G^{\text{UOA}}$, the orientation-known EFIM for the position is given by
\begin{align}\label{eq:86}
  {\bf{J}}_{\text{e}}({\bf{p}}) = \sum_{j\in\mathcal{N}_{\text{L}}}\left(r_j {\bf J}_{\text{\rm r}}(\phi_j) + s_j{\bf J}_{\text{\rm r}}(\phi_j+\frac{\pi}{2})\right).
\end{align}
Hence the SPEB is expressed as
\begin{align}\label{eq:88}
  \text{SPEB}
  = \frac{2\sum_{k\in \mathcal{N}_{\text{L}}}(r_k+s_k)}{\sum_{i,i'\in \mathcal{N}_{\text{L}}}(u_iu_{i'}\sin^2(\phi_i-\phi_{i'}) + s_ir_{i'} + r_is_{i'})}
\end{align}
where $u_i\triangleq r_i-s_i$. Then minimizing SPEB is equivalent to maximizing the denominator
\begin{align}\label{eq:89}
 &\sum_{i,i'\in \mathcal{N}_{\text{L}}}u_iu_{i'}\sin^2(\phi_i-\phi_{i'}) \nonumber\\
 &\qquad = \frac{1}{2}\Big[\Big|\sum_{i\in \mathcal{N}_{\text{L}}}u_i\Big|^2 -\Big|\sum_{i\in \mathcal{N}_{\text{L}}}u_i\exp(j2\phi_{i})\Big|^2\Big].
\end{align}

As a direct corollary, the minimum SPEB requires $\phi_i$'s to satisfy
$
  \sum_{i\in \mathcal{N}_{\text{L}}}u_i\exp(j2\phi_i) = 0
$,
which completes the proof for orientation-known case. In the orientation-unknown case, it follows from Corollary \ref{cor_EFIM_ori_unknown} that
\begin{align}\label{eq:96}
  \sum_{i\in\mathcal{N}_{\text{L}}}\frac{\lambda_i}{D_i}\exp(j\phi_i) = 0
\end{align}
provides a further criterion.

\bibliographystyle{IEEEtran}
\bibliography{IEEEabrv,StringDefinitions,BiblioCV,WGroup,Reference}

\begin{thebibliography}{10}
\providecommand{\url}[1]{#1}
\csname url@samestyle\endcsname
\providecommand{\newblock}{\relax}
\providecommand{\bibinfo}[2]{#2}
\providecommand{\BIBentrySTDinterwordspacing}{\spaceskip=0pt\relax}
\providecommand{\BIBentryALTinterwordstretchfactor}{4}
\providecommand{\BIBentryALTinterwordspacing}{\spaceskip=\fontdimen2\font plus
\BIBentryALTinterwordstretchfactor\fontdimen3\font minus
  \fontdimen4\font\relax}
\providecommand{\BIBforeignlanguage}[2]{{%
\expandafter\ifx\csname l@#1\endcsname\relax
\typeout{** WARNING: IEEEtran.bst: No hyphenation pattern has been}%
\typeout{** loaded for the language `#1'. Using the pattern for}%
\typeout{** the default language instead.}%
\else
\language=\csname l@#1\endcsname
\fi
#2}}
\providecommand{\BIBdecl}{\relax}
\BIBdecl

\bibitem{WinConMazSheGifDarChi:J11}
M.~Z. Win, A.~Conti, S.~Mazuelas, Y.~Shen, W.~M. Gifford, D.~Dardari, and
  M.~Chiani, ``Network localization and navigation via cooperation,''
  \emph{{IEEE} Commun. Mag.}, vol.~49, no.~5, pp. 56--62, May 2011.

\bibitem{GezTiaGiaKobMolPooSah:05}
S.~Gezici, Z.~Tian, G.~B. Giannakis, H.~Kobayashi, A.~F. Molisch, H.~V. Poor,
  and Z.~Sahinoglu, ``Localization via ultra-wideband radios: A look at
  positioning aspects for future sensor networks,'' \emph{{IEEE} Signal
  Process. Mag.}, vol.~22, no.~4, pp. 70--84, Jul. 2005.

\bibitem{SheWin:J10a}
Y.~Shen and M.~Z. Win, ``Fundamental limits of wideband localization --
  {Part~I}: A general framework,'' \emph{{IEEE} Trans. Inf. Theory}, vol.~56,
  no.~10, pp. 4956--4980, Oct. 2010.

\bibitem{SayTarKha:05}
A.~H. Sayed, A.~Tarighat, and N.~Khajehnouri, ``Network-based wireless
  location: Challenges faced in developing techniques for accurate wireless
  location information,'' \emph{{IEEE} Signal Process. Mag.}, vol.~22, no.~4,
  pp. 24--40, Jul. 2005.

\bibitem{PahLiMak:02}
K.~Pahlavan, X.~Li, and J.-P. M{\"a}kel{\"a}, ``Indoor geolocation science and
  technology,'' \emph{{IEEE} Commun. Mag.}, vol.~40, no.~2, pp. 112--118, Feb.
  2002.

\bibitem{CafStu:98}
J.~J. Caffery and G.~L. St\"{u}ber, ``Overview of radiolocation in {CDMA}
  cellular systems,'' \emph{{IEEE} Commun. Mag.}, vol.~36, no.~4, pp. 38--45,
  Apr. 1998.

\bibitem{WymLieWin:J09}
H.~Wymeersch, J.~Lien, and M.~Z. Win, ``Cooperative localization in wireless
  networks,'' \emph{Proc. {IEEE}}, vol.~97, no.~2, pp. 427--450, Feb. 2009.

\bibitem{PatAshKypHerMosCor:05}
N.~Patwari, J.~N. Ash, S.~Kyperountas, A.~O. Hero, R.~L. Moses, and N.~S.
  Correal, ``Locating the nodes: Cooperative localization in wireless sensor
  networks,'' \emph{{IEEE} Signal Process. Mag.}, vol.~22, no.~4, pp. 54--69,
  Jul. 2005.

\bibitem{DarConFerGioWin:J09}
D.~Dardari, A.~Conti, U.~J. Ferner, A.~Giorgetti, and M.~Z. Win, ``Ranging with
  ultrawide bandwidth signals in multipath environments,'' \emph{Proc. {IEEE}},
  vol.~97, no.~2, pp. 404--426, Feb. 2009.

\bibitem{KhaKarMou:09}
U.~A. Khan, S.~Kar, and J.~M.~F. Moura, ``Distributed sensor localization in
  random environments using minimal number of anchor nodes,'' \emph{{IEEE}
  Trans. Signal Process.}, vol.~57, no.~5, pp. 2000--2016, May 2009.

\bibitem{TOA1}
E.~R. Robinson and A.~H. Quazi, ``{Effect of sound-speed profile on
  differential time-delay estimation},'' \emph{J. Acoust. Society America},
  vol.~77, no.~3, pp. 1086--1090, Sept. 1985.

\bibitem{TOA2}
C.~Knapp and G.~C. Carter, ``{The generalized correlation method for estimation
  of time delay},'' \emph{IEEE Trans. Acoust. Speech and Signal Process.},
  vol.~24, no.~4, pp. 320--327, Aug. 1976.

\bibitem{JouDarWin:J08}
D.~B. Jourdan, D.~Dardari, and M.~Z. Win, ``Position error bound for {UWB}
  localization in dense cluttered environments,'' \emph{{IEEE} Trans. Aerosp.
  Electron. Syst.}, vol.~44, no.~2, pp. 613--628, Apr. 2008.

\bibitem{ConGueDarDecWin:J12}
A.~Conti, M.~Guerra, D.~Dardari, N.~Decarli, and M.~Z. Win, ``Network
  experimentation for cooperative localization,'' \emph{{IEEE} J. Sel. Areas
  Commun.}, vol.~30, no.~2, pp. 467--475, Feb. 2012.

\bibitem{TDOA1}
J.~J. Caffery, \emph{{Wireless Location in CDMA Cellular Radio Systems}}.\hskip
  1em plus 0.5em minus 0.4em\relax Norwell, MA, USA: Kluwer Academic
  Publishers, 1999.

\bibitem{TDOA2}
T.~S. Rappaport, J.~Reed, and B.~D. Woerner, ``{Position location using
  wireless communications on highways of the future},'' \emph{IEEE Commun.
  Mag.}, vol.~34, no.~10, pp. 33--41, Oct. 1996.

\bibitem{TDOA3}
Y.~Qi, H.~Kobayashi, and H.~Suda, ``{Analysis of wireless geolocation in a
  non-line-of-sight environment},'' \emph{IEEE Trans. Wireless Commun.},
  vol.~5, no.~3, pp. 672--681, Mar. 2006.

\bibitem{AOA2}
J.~Xu, M.~D. Ma, and C.~L. Law, ``{AOA cooperative position localization},'' in
  \emph{Proc. IEEE Global Telecomm. Conf.}, New Orleans, LA, USA, Nov. 2008,
  pp. 1--5.

\bibitem{study1}
A.~Mallat, J.~Louveaux, and L.~Vandendorpe, ``{UWB based positioning: Cramer
  Rao bound for angle of arrival and comparison with time of arrival},'' in
  \emph{Symp. Commun. Veh. Technol.}, Liege, Belgium, Nov. 2006, pp. 65--68.

\bibitem{AOA1}
D.~Niculescu and B.~Nath, ``{Ad hoc positioning system (APS) using AOA},'' in
  \emph{Annual IEEE Conf. Computer Commun.}, vol.~3, San Francisco, CA, USA,
  Apr. 2003, pp. 1734--1743.

\bibitem{AOA3}
B.~D. Van~Veen and K.~M. Buckley, ``{Beamforming: a versatile approach to
  spatial filtering},'' \emph{IEEE ASSP Mag.}, vol.~5, no.~2, pp. 4--24, Apr.
  1988.

\bibitem{RSS1}
H.~Hashemi, ``{The indoor radio propagation channel},'' \emph{Proc. IEEE},
  vol.~81, no.~7, pp. 943--968, Jul. 1993.

\bibitem{RSS2}
F.~Evennou and F.~Marx, ``{Advanced integration of WIFI and inertial navigation
  systems for indoor mobile positioning},'' \emph{EURASIP J. Appl. Signal
  Process.}, vol. 2006, pp. 164--164, Jan. 2006.

\bibitem{RSS3}
N.~Patwari, R.~O'Dea, and Y.~Wang, ``{Relative location in wireless
  networks},'' in \emph{IEEE Veh. Technol. Conf.}, vol.~2, Atlantic City, NJ,
  USA, May 2001, pp. 1149--1153.

\bibitem{BasicIntro}
A.~Roxin, J.~Gaber, M.~Wack, and A.~Nait-Sidi-Moh, ``Survey of wireless
  geolocation techniques,'' in \emph{IEEE Globecom Workshops}, Washington, DC,
  USA, Nov. 2007, pp. 1--9.

\bibitem{TOA-AOA}
P.~Deng and P.~Fan, ``{An AOA assisted TOA positioning system},'' in
  \emph{Proc. Commun. Technol. Conf.}, vol.~2, Beijing, China, Aug. 2000, pp.
  1501--1504.

\bibitem{SheDaiWin:J14}
Y.~Shen, W.~Dai, and M.~Z. Win, ``Power optimization for network
  localization,'' \emph{{IEEE/ACM} Trans. Netw.}, vol.~22, no.~4, pp.
  1337--1350, Aug. 2014.

\bibitem{WanLeuHua:09}
T.~Wang, G.~Leus, and L.~Huang, ``Ranging energy optimization for robust sensor
  positioning based on semidefinite programming,'' \emph{{IEEE} Trans. Signal
  Process.}, vol.~57, no.~12, pp. 4777--4787, Dec. 2009.

\bibitem{GarHaiCouLop:14}
N.~Garcia, A.~M. Haimovich, M.~Coulon, and M.~Lops, ``Resource allocation in
  {MIMO} radar with multiple targets for non-coherent localization,''
  \emph{{IEEE} Trans. Signal Process.}, vol.~62, no.~10, pp. 2656--2666, Oct.
  2014.

\bibitem{DaiSheWin:J15}
W.~Dai, Y.~Shen, and M.~Z. Win, ``Distributed power allocation for cooperative
  wireless network localization,'' \emph{{IEEE} J. Sel. Areas Commun.},
  vol.~33, no.~1, pp. 28--40, Jan. 2015.

\bibitem{DaiSheWin:J15a}
------, ``Energy-efficient network navigation algorithms,'' \emph{{IEEE} J.
  Sel. Areas Commun.}, vol.~33, no.~7, pp. 1418--1430, Jul. 2015.

\bibitem{GodPetPoo:11}
H.~Godrich, A.~P. Petropulu, and H.~V. Poor, ``Power allocation strategies for
  target localization in distributed multiple-radar architectures,''
  \emph{{IEEE} Trans. Signal Process.}, vol.~59, no.~7, pp. 3226--3240, Jul.
  2011.

\bibitem{JeoSimHaiKan:14}
S.~Jeong, O.~Simeone, A.~Haimovich, and J.~Kang, ``Beamforming design for joint
  localization and data transmission in distributed antenna system,''
  \emph{{IEEE} Trans. Veh. Technol.}, 2014, to appear.

\bibitem{SheWymWin:J10}
Y.~Shen, H.~Wymeersch, and M.~Z. Win, ``Fundamental limits of wideband
  localization -- {Part~II}: Cooperative networks,'' \emph{{IEEE} Trans. Inf.
  Theory}, vol.~56, no.~10, pp. 4981--5000, Oct. 2010.

\bibitem{D2}
Y.~Qi, H.~Kobayashi, and H.~Suda, ``On time-of-arrival positioning in a
  multipath environment,'' \emph{IEEE Trans. Veh. Technol.}, vol.~55, no.~5,
  pp. 1516--1526, Sept. 2006.

\bibitem{van2000asymptotic}
A.~W. Van~der Vaart, \emph{Asymptotic statistics}.\hskip 1em plus 0.5em minus
  0.4em\relax Cambridge university press, 2000, vol.~3.

\bibitem{hajek1970characterization}
J.~H{\'a}jek, ``A characterization of limiting distributions of regular
  estimates,'' \emph{Zeitschrift f{\"u}r Wahrscheinlichkeitstheorie und
  verwandte Gebiete}, vol.~14, no.~4, pp. 323--330, 1970.

\bibitem{hajek1972local}
------, ``Local asymptotic minimax and admissibility in estimation,'' in
  \emph{Proc. Sixth Berkeley Symp. Math. Statist. Probab}, vol.~1.\hskip 1em
  plus 0.5em minus 0.4em\relax Univ. of California Press, Berkeley, 1972, pp.
  175--194.

\bibitem{Barakin}
R.~McAulay and E.~Hofstetter, ``Barankin bounds on parameter estimation,''
  \emph{IEEE Trans. Inf. Theory}, vol.~17, no.~6, pp. 669--676, Nov. 1971.

\bibitem{MonMazVitWin:J13}
F.~Montorsi, S.~Mazuelas, G.~M. Vitetta, and M.~Z. Win, ``On the performance
  limits of map-aware localization,'' \emph{{IEEE} Trans. Inf. Theory},
  vol.~59, no.~8, pp. 5023--5038, Aug. 2013.

\bibitem{RADAR}
A.~W. Rihaczek, \emph{Principles of High-Resolution Radar}.\hskip 1em plus
  0.5em minus 0.4em\relax McGraw-Hill, Jan. 1969.

\bibitem{envelope}
Z.~Gu and E.~Gunawan, ``{Cramer-Rao bound for joint direction of arrival, time
  delay estimation in DS-CDMA systems},'' in \emph{Proc. Military Commun.
  Conf.}, vol.~1, Los Angeles, CA, USA, Oct. 2000, pp. 614--618.

\bibitem{rice2000narrowband}
M.~Rice, R.~Dye, and K.~Welling, ``{Narrowband channel model for aeronautical
  telemetry},'' \emph{IEEE Trans. Aerosp. Electron. Syst.}, vol.~36, no.~4, pp.
  1371--1376, Oct. 2000.

\bibitem{bekkerman2006target}
I.~Bekkerman and J.~Tabrikian, ``Target detection and localization using mimo
  radars and sonars,'' \emph{IEEE Trans. Signal Process.}, vol.~54, no.~10, pp.
  3873--3883, Oct. 2006.

\bibitem{tichavsky1998posterior}
P.~Tichavsk{\`y}, C.~H. Muravchik, and A.~Nehorai, ``Posterior {Cram{\'e}r-Rao}
  bounds for discrete-time nonlinear filtering,'' \emph{IEEE Trans. Signal
  Process.}, vol.~46, no.~5, pp. 1386--1396, May 1998.

\bibitem{meng2009computationally}
H.~Meng, M.~Hernandez, Y.~Liu, and X.~Wang, ``Computationally efficient {PCRLB}
  for tracking in cluttered environments: measurement existence conditioning
  approach,'' \emph{IET Signal Process.}, vol.~3, no.~2, pp. 133--149, Mar.
  2009.

\bibitem{SheMazWin:J12}
Y.~Shen, S.~Mazuelas, and M.~Z. Win, ``Network navigation: {T}heory and
  interpretation,'' \emph{{IEEE} J. Sel. Areas Commun.}, vol.~30, no.~9, pp.
  1823--1834, Oct. 2012.

\bibitem{mckay1997geometry}
J.~McKay and M.~Pachter, ``{Geometry optimization for GPS navigation},'' in
  \emph{Proc. IEEE Conf. Decision and Control}, vol.~5, San Diego, CA, USA,
  Dec. 1997, pp. 4695--4699.

\bibitem{hegazy2003sensor}
T.~Hegazy and G.~Vachtsevanos, ``Sensor placement for isotropic source
  localization,'' in \emph{Information Processing in Sensor Networks}.\hskip
  1em plus 0.5em minus 0.4em\relax Springer, 2003, pp. 432--441.

\bibitem{caffery2000new}
J.~J. Caffery~Jr, ``{A new approach to the geometry of TOA location},'' in
  \emph{Proc. IEEE Semiannual Veh. Technol. Conf.}, vol.~4, Boston, MA, USA,
  Sept. 2000, pp. 1943--1949.

\bibitem{bishop2007optimality}
A.~N. Bishop, B.~Fidan, B.~D. Anderson, P.~N. Pathirana, and K.~Dogan{\c{c}}ay,
  ``Optimality analysis of sensor-target geometries in passive localization:
  Part 2-time-of-arrival based localization,'' in \emph{Proc. Int. Conf.
  Intell. Sensors Sensor Networks Inf. Process. (3rd ISSNIP)}, Melbourne, Qld.,
  Australia, Dec. 2007.

\bibitem{martinez2006optimal}
S.~Mart{\'\i}Nez and F.~Bullo, ``Optimal sensor placement and motion
  coordination for target tracking,'' \emph{Automatica}, vol.~42, no.~4, pp.
  661--668, Apr. 2006.

\bibitem{abel1990optimal}
J.~Abel, ``Optimal sensor placement for passive source localization,'' in
  \emph{Proc. IEEE Int. Conf. Acoustics, Speech, and Signal Processing},
  Albuquerque, NM, USA, Apr. 1990, pp. 2927--2930.

\bibitem{godrich2010target}
H.~Godrich, A.~M. Haimovich, and R.~S. Blum, ``{Target localization accuracy
  gain in MIMO radar-based systems},'' \emph{IEEE Trans. Inf. Theory}, vol.~56,
  no.~6, pp. 2783--2803, Jun. 2010.

\bibitem{Rappaport1996Wireless}
T.~S. Rappaport, \emph{Wireless communications: principles and practice}.\hskip
  1em plus 0.5em minus 0.4em\relax Prentice Hall PTR, 1996.

\bibitem{vanTrees1968detection}
H.~L.~V. Trees, \emph{Detection, Estimation and Modulation Theory}.\hskip 1em
  plus 0.5em minus 0.4em\relax New York: Wiley, 1968, vol.~1.

\bibitem{cam1986asymptotic}
L.~Le~Cam, \emph{Asymptotic methods in statistical decision theory}.\hskip 1em
  plus 0.5em minus 0.4em\relax Springer, 1986.

\bibitem{Zhu1992Calculation}
J.~Zhu, ``Calculation of geometric dilution of precision,'' \emph{IEEE Trans.
  Aeros. Electron. Syst.}, vol.~28, Jul. 1992.

\bibitem{wang2012efficient}
D.~Wang, H.~Leung, M.~Fattouche, and F.~Ghannouchi, ``Efficient spectrum
  allocation and time of arrival based localization in cognitive networks,''
  \emph{Wireless Per. Commun.}, vol.~66, pp. 813--831, Oct. 2012.

\bibitem{reed1978methods}
M.~Reed and B.~Simon, \emph{Methods of modern mathematical physics: Analysis of
  operators}.\hskip 1em plus 0.5em minus 0.4em\relax Academic Press, 1978.

\bibitem{lehmann1998theory}
E.~L. Lehmann and G.~Casella, \emph{Theory of point estimation}.\hskip 1em plus
  0.5em minus 0.4em\relax Springer, 1998, vol.~31.

\end{thebibliography}

\begin{IEEEbiographynophoto}{Yanjun Han}
	(S'14) received his B.Eng. degree with the highest honor in electronic engineering from Tsinghua University, Beijing, China in 2015. He is currently working towards the Ph.D. degree in the Department of Electrical Engineering at Stanford University. His research interests include information theory and statistics, with applications in communications, data compression, and learning.
\end{IEEEbiographynophoto}

\begin{IEEEbiographynophoto}{Yuan Shen}
	(S'05-M'14) received the Ph.D.\ degree and the S.M.\ degree in electrical engineering and computer science from the Massachusetts Institute of Technology (MIT), Cambridge, MA, USA, in 2014 and 2008, respectively, and the B.E.\ degree (with highest honor) in electronic engineering from Tsinghua University, Beijing, China, in 2005.

	He is an Associate Professor with the Department of Electronic Engineering at Tsinghua University. Prior to joining Tsinghua University, he was a Research Assistant and then Postdoctoral Research Associate with the Laboratory for Information and Decision Systems (LIDS) at MIT in 2005-2014. He was with the Wireless Communications Laboratory at The Chinese University of Hong Kong in summer 2010, the Hewlett-Packard Labs in winter 2009, and the Corporate R\&D at Qualcomm Inc.\ in summer 2008. His research interests include statistical inference, network science, communication theory, information theory, and optimization. His current research focuses on network localization and navigation, inference techniques, resource allocation, and intrinsic wireless secrecy.
	
	Dr.\ Shen was a recipient of the Qiu Shi Outstanding Young Scholar Award (2015), the China's Youth 1000-Talent Program (2014), the Marconi Society Paul Baran Young Scholar Award (2010), the MIT EECS Ernst A. Guillemin Best S.M.\ Thesis Award (1st place) (2008), the Qualcomm Roberto Padovani Scholarship (2008), and the MIT Walter A. Rosenblith Presidential Fellowship (2005). His papers received the IEEE Communications Society Fred W. Ellersick Prize (2012) and three Best Paper Awards from the IEEE Globecom (2011), the IEEE ICUWB (2011), and the IEEE WCNC (2007). He is elected Secretary (2015-2017) for the Radio Communications Committee of the IEEE Communications Society. He serves as symposium Co-Chair of the Technical Program Committee (TPC) for the IEEE Globecom (2016) and the European Signal Processing Conference (EUSIPCO) (2016). He also serves as Editor for the {\scshape IEEE Communications Letters} since 2015 and Guest-Editor for the {\scshape International Journal of Distributed Sensor Networks} (2015).
\end{IEEEbiographynophoto}

\begin{IEEEbiographynophoto}{Xiao-Ping Zhang}
	(M'97-SM'02) received the B.S. and Ph.D. degrees in electronics engineering from Tsinghua University, in 1992 and 1996, respectively, and the M.B.A. (Hons.) degree in finance, economics and entrepreneurship from the University of Chicago Booth School of Business, Chicago, IL.
	
	Since Fall 2000, he has been with the Department of Electrical and Computer Engineering, Ryerson University, where he is now Professor, Director of Communication and Signal Processing Applications Laboratory (CASPAL). He has served as Program Director of Graduate Studies. He is cross appointed to the Finance Department at the Ted Rogers School of Management at Ryerson University. His research interests include statistical signal processing, multimedia content analysis, sensor networks and electronic systems, computational intelligence, and applications in big data, finance, and marketing. He is a frequent consultant for biotech companies and investment firms. He is cofounder and CEO for EidoSearch, an Ontario based company offering a content-based search and analysis engine for financial data. 
	
	Dr. Zhang is a Registered Professional Engineer in Ontario, Canada, and a member of the Beta Gamma Sigma Honor Society. He is the General Chair of MMSP'15. He was the publicity chair of ICME'06 and the program chair of ICIC'05 and ICIC'10. He served as a guest editor of Multimedia Tools and Applications, and the International Journal of Semantic Computing. He is a tutorial speaker in ACMMM2011, ISCAS2013, ICIP2013, and ICASSP2014. He is an Associate Editor of the \textsc{IEEE Transactions on Signal Processing}, the \textsc{IEEE Transactions on Image Processing}, the \textsc{IEEE Transactions on Multimedia}, the \textsc{IEEE transactions on Circuits and Systems for Video Technology}, the \textsc{IEEE Signal Processing Letters} and the \emph{Journal of Multimedia}.
\end{IEEEbiographynophoto}

\begin{IEEEbiographynophoto}{Moe Z. Win} (S'85-M'87-SM'97-F'04)
received both the Ph.D. in Electrical Engineering and the M.S. in Applied Mathematics as a Presidential Fellow at the University of Southern California (USC) in 1998. He received the M.S. in Electrical Engineering from USC in 1989 and the B.S. ({\em magna cum laude}) in Electrical Engineering from Texas A\&M University in 1987. 

He is a Professor at the Massachusetts Institute of Technology (MIT) and the founding director of the Wireless Communication and Network Sciences Laboratory. Prior to joining MIT, he was with AT\&T Research Laboratories for five years and with the Jet Propulsion Laboratory for seven years. His research encompasses fundamental theories, algorithm design, and experimentation for a broad range of real-world problems. His current research topics include network localization and navigation, network interference exploitation, intrinsic wireless secrecy, adaptive diversity techniques, and ultra-wideband systems.

Professor Win is an elected Fellow of the AAAS, the IEEE, and the IET, and was an IEEE Distinguished Lecturer. He was honored with two IEEE Technical Field Awards: the IEEE Kiyo Tomiyasu Award (2011) and the IEEE Eric E. Sumner Award (2006, jointly with R. A. Scholtz). Together with students and colleagues, his papers have received numerous awards, including the IEEE Communications Society’s Stephen O. Rice Prize (2012), the IEEE Aerospace and Electronic Systems Society’s M. Barry Carlton Award (2011), the IEEE Communications Society’s Guglielmo Marconi Prize Paper Award (2008), and the IEEE Antennas and Propagation Society’s Sergei A. Schelkunoff Transactions Prize Paper Award (2003). Highlights of his international scholarly initiatives are the Copernicus Fellowship (2011), the Royal Academy of Engineering Distinguished Visiting Fellowship (2009), and the Fulbright Fellowship (2004). Other recognitions include the International Prize for Communications Cristoforo Colombo (2013), the {\em Laurea Honoris Causa} from the University of Ferrara (2008), the Technical Recognition Award of the IEEE ComSoc Radio Communications Committee (2008), and the U.S. Presidential Early Career Award for Scientists and Engineers (2004).

Dr. Win was an elected Member-at-Large on the IEEE Communications Society Board of Governors (2011-2013). He was the Chair (2004-2006) and Secretary (2002-2004) for the Radio Communications Committee of the IEEE Communications Society. Over the last decade, he has organized and chaired numerous international conferences. He is currently an Editor-at-Large for the {\scshape IEEE Wireless Communications Letters}. He served as Editor (2006-2012) for the {\scshape IEEE Transactions on Wireless Communications}, and as Area Editor (2003-2006) and Editor (1998-2006) for the {\scshape IEEE Transactions on Communications}. He was Guest-Editor for the {\scshape Proceedings of the IEEE} (2009) and for the {\scshape IEEE Journal on Selected Areas in Communications} (2002).
\end{IEEEbiographynophoto}

\begin{IEEEbiographynophoto}{Huadong Meng}(S'01-M'04-SM'14) 
received the B.Eng. degree and Ph.D. degree in electronic engineering, both from Tsinghua University, Beijing, China, in 1999 and 2004 respectively.

In 2015, he joined University of California, Berkeley, where he is currently a Visiting Associate Researcher at California PATH. During 2008 to 2015, he was an Associate Professor in the Department of Electronic Engineering of Tsinghua University. His current research interests include statistical signal processing, intelligent transportation systems, wireless localization, and target tracking.

Dr. Meng was a Technical Program Committee (TPC) member of the IET International Radar Conference 2013 and 2015. He also serves as an Editorial Board member for \emph{Remote Sensing Information}.
\end{IEEEbiographynophoto}

\end{document}